\documentclass[iop,apj,numberedappendix,appendixfloats]{emulateapj}
\usepackage{graphicx,natbib,amsmath, longtable, apjfonts}

\newcommand{\dtwo}{{$_2$}}
\newcommand{\dthree}{{$_3$}}
\newcommand{\jon}{{$^+$}}
\newcommand{\ijon}{{$^-$}}
\newcommand{\ra}{$\rightarrow$}

\shorttitle{New extended deuterium fractionation model}
\shortauthors{Albertsson et al.}

\begin{document}

\title{New extended deuterium fractionation model: assessment at dense ISM conditions and sensitivity analysis}

\author{T. Albertsson\altaffilmark{1}\altaffilmark{,*}, D. A. Semenov\altaffilmark{1}, A. I. Vasyunin\altaffilmark{2}, Th. Henning\altaffilmark{1}, and E. Herbst\altaffilmark{2}}
\affil{(1) Max-Planck-Institut f\"ur Astronomie, K\"onigstuhl 17, 69117 Heidelberg, Germany \\
(2) Departments of Chemistry and Astronomy, University of Virginia, Charlottesville, VA 22904, USA\\
* Visiting scientist at University of Virginia, Charlottesville}

\begin{abstract}
Observations of deuterated species are {useful in probing the temperature, ionization level, evolutionary stage, chemistry,} and thermal history of astrophysical environments. The analysis of data { from ALMA and other new telescopes} requires an elaborate model of deuterium fractionation. This paper presents a publicly available chemical network with multi-deuterated species and an extended, up-to-date set of gas-phase and surface reactions. To test this { network}, we simulate deuterium fractionation in diverse interstellar sources. Two cases of initial abundances are considered: i) atomic except for H$_2$ and HD, and ii) molecular from a prestellar core. We reproduce the observed D/H ratios of many deuterated molecules, and sort the species according to their sensitivity to temperature gradients and initial abundances. We find that many multiply-deuterated species produced at 10~K retain enhanced D/H ratios at temperatures $ \la 100$~K. We study how recent updates to reaction rates affect calculated D/H ratios, and perform a detailed sensitivity analysis of the uncertainties of the gas-phase reaction rates in the network. {We find that uncertainties are generally lower in dark cloud environments than in warm IRDCs and that uncertainties increase with the size of the molecule and number of D-atoms.} A set of the most problematic reactions is presented. We list potentially observable deuterated species predicted to be abundant in low- and high-mass star-formation regions.
\end{abstract}


\section{Introduction}
The life cycle of molecules covers a wide range of environments, starting from the sparse interstellar medium (ISM), which eventually evolves into stars and planets. As molecular hydrogen cannot be easily observed in the cold interstellar medium, other molecular tracers are employed to probe the relevant physical conditions and chemical composition. More than 170 molecules have been observed in the interstellar medium to date\footnote{\url{http://www.astrochymist.org/astrochymist$\_$mole.html}}, ranging {from diatomic species} to the fullerenes C$_{60}$ and C$_{70}$ \citep{Cami_ea2010} and including deuterated species. A variety of deuterated species have been detected in various astrophysical environments, including molecular clouds: DCO\jon~\citep{2009A&A...507..347V, 1977ApJ...217L.165G}, DNC~\citep{2009A&A...507..347V, 1978ApJ...225L..75T}, H\dtwo D\jon~\citep{2011A&A...526A..31P, 1999ApJ...521L..67S}, HDCO \citep{1985ApJ...299..947L}, D$_{2}$CO \citep{1990ApJ...362L..29T}, HD\dtwo\jon~\citep{2011A&A...526A..31P}, HDO \citep{1973ApL....15...17P}; {pre-stellar cores: D\dtwo CO \citep{2004BaltA..13..402B}, H\dtwo D\jon~\citep{2003A&A...403L..37C, 2006ApJ...645.1198V, 2008A&A...492..703C}, HD\dtwo\jon~\citep{2012A&A...547A..33V}, N\dtwo D\jon~\citep{2012A&A...538A.137M}, NHD\dtwo~\citep{2000A&A...354L..63R};} hot cores/corinos: D\dtwo CO, HDCO~\citep{2011A&A...527A..39B}, DCOOCH\dthree~\citep{2010A&A...517A..17D, 2010ApJ...714.1120M}, HD\dtwo\jon~\citep{2004ApJ...606L.127V}, HDO, NH\dtwo D~\citep{1990A&A...228..447J}; warm protostellar envelopes: DCO\jon, HDCO~\citep{2009A&A...508..737P}, HDO~\citep{2010ApJ...725L.172J, 2011A&A...527A..19L}, OD~\citep{2012A&A...542L...5P}; protoplanetary disks: DCN, DCO\jon~\citep{2003A&A...400L...1V, 2006A&A...448L...5G, 2008ApJ...681.1396Q} and comets: CH\dthree D~\citep{2009ApJ...699.1563B, 2012ApJ...750..102G}, HDCO~\citep{2008cosp...37.1640K}, HDO~\citep{2009ApJ...690L...5V,2011Natur.478..218H,2012ApJ...750..102G}. For comprehensive reviews, we refer to \citet{2003SSRv..106...61R} and \citet{2007prpl.conf..751B}. The study of deuterium chemistry has proven useful to constrain the ionization fraction, density and thermal history of the ISM and protoplanetary disks \citep[e.g.,][]{1981A&A....93..189G, 2001A&A...371.1107A, 2005ApJ...619..379C,2007ApJ...660..441W, 2012ApJ...749..162O}. 

Still, many more deuterated species of key importance remain to be detected. Upon the completion of the high-sensitive, high-resolution Atacama Large Millimeter Array (ALMA), we will be able to, for the first time, detect and spatially resolve emission lines of numerous new complex and rare-isotope molecules. In order to analyze these rich observational data, new astrochemistry models including isotope-exchange reactions and state-to-state processes will be required. The main goal of the present paper is to present and provide a new extended, public deuterium fractionation model, and to explore its validity and accuracy.

The implementation of deuterium chemistry is a challenging task though because of a limited number of accurately determined rate coefficients of relevant reactions, and the sheer number of hydrogen-dominated reactions in astrochemical networks. Previous studies used the available limited set of reaction data, substituted with ``educated guesses'' for missing reaction rates, and cloned data from similar reactions involving hydrogen-bearing species \citep[e.g.,][]{1982A&A...111...76H,1988MNRAS.231..409B,1996MNRAS.280.1046R, 1997ApJ...482L.203C, 2001ApJS..136..579T, 2003ApJ...593..906A,2012ApJ...760...40A, 2013arXiv1304.4031S}. In many cases only mono- and double-deuterated species were considered. We follow the general approach, but abandon the restriction on the total number of deuterons in chemical species. 

The redistribution of elemental deuterium, initially locked mainly in HD, is initiated by fast ion-molecule reactions with polyatomic ions, such as H\dthree\jon~(and H$_2$D$^+$, HD$_2^+$, D\dthree\jon). Due to zero-point vibrational energy differences between modes with D and H, and the lack of a ground rotational state for H$_{3}^{+}$, the backward reactions between H$_2$ and H$_3^+$ isotopologues are endothermic, with barriers of $\sim 100-300$~K, leading to the initial enrichment of abundances of the H$_3^+$ isotopologues at $\la 20-30$~K. In a similar way, other ions such as CH$_2$D$^+$ and C$_2$HD$^+$ allow deuterium fractionation to proceed effectively at warmer temperatures of $\la 30-80$~K because of larger endoergicities for the backward reactions \citep[e.g.,][]{1989ApJ...340..906M, 2005A&A...438..585R, 2009A&A...508..737P}. These deuterated ions react further with abundant molecules such as CO and N$_2$, transferring deuterium atoms to new molecules. These findings have been proven both observationally and theoretically \citep[see, e.g.,][]{2002A&A...389L...6B, 2003ApJ...585L..55B, 2003ApJ...591L..41R, 2006RSPTA.364.3101V, 2006RSPTA.364.3063R, 2008A&A...492..703C}. The initial gas-phase deuterium enrichment of H$_3^+$ is even more pronounced in cold, dense regions, where some destructive neutral species{, especially CO,} become severely depleted onto dust grain surfaces. The dissociative recombination of abundant H$_3^+$ isotopologues leads to a high flux of D atoms ($\la 10-30\%$ compared to H) sticking to dust grains, which further react with surface species such as CO forming abundant multi-deuterated complex (organic) ices such as formaldehyde and methanol. These molecules can later desorb into the gas-phase due to non-thermal desorption or due to the gradual warm-up of the environment by a forming protostar.

The physical environment plays an important role in deuterium fractionation. In one of the first theoretical studies of deuterium fractionation, \citet{1973ApJ...180L..89S} estimated the D/H ratio for HCN in Orion, albeit erroneously assuming that fractionation is driven solely by neutral-neutral processes involving HD and atomic D. The ion-molecule fractionation route was first proposed by \citet{1974ApJ...188...35W} and \citet{1976RvMP...48..513W}, who used it to constrain the interstellar D/H elemental ratio. \citet{1977ApJ...217L.165G} used observations of DCO$^+$ to constrain the electron abundance in dark clouds, while \citet{1982A&A...111...76H} showed that the DCO$^+$/HCO$^+$ ratio can be used as a sensitive measure of the gas kinetic temperature in the ISM. \citet{1984ApJ...287L..47D} studied the D/H ratio of HCO$^+$, considering cold and warm ISM conditions, and illustrated the role of depletion (and thus the density of the environment) for deuterium fractionation processes. 

The first complete gas-phase model of deuterium chemistry in a dense cold cloud was undertaken by \citet{1989ApJ...340..906M}, while \citet{1989MNRAS.240P..25B} explored grain-surface deuteration processes. More recently, \citet{2000A&A...361..388R} studied deuterium chemistry over a wide range of physical parameters, by varying density, temperature, initial abundances and freeze-out. They found that if freeze-out is present, molecular D/H ratios can become very high; e.g., $\ga 1-10\%$, and gas-phase chemistry can produce abundant mono- and multi-deuterated molecules. 

Not only are the current physical properties of the environment important for chemical evolution, but also is the evolutionary history. \citet{2012ApJ...748L...3T} have considered a two-stage model to study deuterium chemistry in prestellar cores, with a gas-phase steady-state phase followed by the formation and evolution of grain mantles and surface deuterium fractionation. With such a simple approach they have reproduced high observed abundances of the isotopologues of formaldehyde and methanol. Taquet et al. have concluded that D and H abstraction and substitution reactions on dust surfaces are crucial for attaining the observed high D/H ratios. The role of abstraction reactions for deuterium fractionation has been intensively investigated in the laboratory \citep[see e.g.,][]{2005ApJ...624L..29N,2009ApJ...702..291H,2012ApJ...757..185H}. 

\citet{2011ApJ...741L..34C} have studied the chemistry of formaldehyde and water by modeling the formation of ices in translucent clouds, and later following the chemical evolution as the cloud collapses to eventually form a Class~I protostellar object. Their results show that the degree of deuteration of formaldehyde is sensitive to the initial D/H ratios of gaseous molecules attained before the collapse phase, while the degree of deuteration of water depends strongly on the dust temperature during the water ice formation. Intriguingly, \citet{2012A&A...539A.132C} have observed deuterated water vapor in the low-mass protostar IRAS 16293-2422 and found that the water D/H ratio is lower than for other deuterated species detected in the same source. This observational trend continues toward more evolved hot cores/corinos \citep[e.g.,][]{2004A&A...416..159P, 2012A&A...541L..12B} and suggests that the water may have formed relatively early, in a warm dilute ISM environment, while the depletion of CO at a later, cold and dense dense core stage allows for efficient surface synthesis of highly deuterium-enriched complex ices. An alternative explanation could be efficient abstraction and substitution reactions of H atoms by D atoms for organic ices like formaldehyde and methanol during cold prestellar cloud phase, which would not be as effective for water ice. A detailed one-dimensional chemical-hydrodynamical model of the prestellar core collapse and the formation of a protostar, coupled to the gas-grain chemistry and deuterium fractionation, has been developed by \cite{2012ApJ...760...40A}. They have found that due to initially high D/H ratios accumulated in the cold phase, large (organic) molecules and carbon chains remain strongly deuterated even at later, warmer conditions.
 
While physical properties, such as temperature and density, and surface chemistry can have a significant effect on deuterium fractionation, smaller effects can derive from other global properties such as metallicity and ionization fields. In order to understand the influence of metallicity and ionization one needs to study deuterated species on a more global scale. \citet{2010ApJ...725..214B} have conducted an observational survey of deuterated species in extragalactic star-forming regions and studied the influence of density, temperature, far-UV radiation field, cosmic-ray ionization and metallicity on the D/H ratios for $\sim$20 deuterated species. Without modeling any particular source, they have compared the predicted column densities with those derived from the current limited set of observational data in external galaxies and found an overall reasonable agreement. \citet{2010ApJ...725..214B} have provided a list of key deuterated species in extragalactic environments to be searched for with ALMA. 

ALMA is a truly revolutionary observational facility not only for extragalactic and cosmological studies, but also for observations of the Milky Way ISM, the analysis of which requires better chemical tools. In this paper we present a new up-to-date, extended, multi-deuterated chemical network and assess its reliability by modeling the deuterium fractionation in various phases of the ISM and comparing the results with observed D/H ratios of a variety of mono-, doubly-, and triply-deuterated species in distinct astrophysical environments. A list of the most promising, deuterated species potentially detectable with ALMA in the local Milky Way ISM is provided. Also, we report a detailed sensitivity analysis to understand and to quantify the intrinsic error bars in calculated abundances of deuterated species in several representative astrophysical environments. We isolate the most problematic gas-phase reactions with uncertain rate coefficients to be studied in the laboratory or theoretically, and quantify the associated uncertainties in modeled abundances. The User Manual and the new deuterium network are freely available on the Internet\footnote{\url{http://mpia.de/PSF/codes.php}}.

The remainder of the paper is structured as follows. In Section~\ref{sec:model} we present our new deuterium chemistry model. We give a detailed description of the construction of our deuterium network and our choice of the relevant reaction rates and branching ratios. We also discuss error propagation based on uncertainties in rate coefficients. In Section~\ref{sec:results} this model is used to calculate abundances and D/H ratios under a wide range of physical conditions. We discern general trends in D/H ratios with temperature, density, and initial abundances, and divide the species according to whether temperature or initial abundance influences D/H ratios more strongly. A sensitivity analysis is performed to quantify intrinsic uncertainties in modeled abundances due to uncertainties in reaction rate data. A list of the most problematic reactions for deuterium chemistry is presented. We then discuss our results and compare them with recent observations and theoretical studies in Section~\ref{sec:discussion}. Conclusions follow in Section~\ref{sec:conclusion}. Appendices are only available as online material. Appendix A contains updated and added reactions, Appendix B shows dominant reaction pathways for selected deuterated species, and Appendix C lists deuterium fractionation rate coefficients in cold dark environments.

\section{Model}
\label{sec:model}
\subsection{Parameter space}
\label{sec:phys_model}
In this work we are primarily concerned with providing a new extended deuterium network and assessing its capacity to model chemistry under both static cold and warm conditions in the ISM. We do not consider evolutionary models such as those pertaining to low-mass and high-mass star formation separately, although the static conditions we consider arise from evolutionary processes. The evolution of the interstellar medium begins from fragmentation of turbulent, mainly atomic clouds {with kinetic temperatures} up to $\sim$100~K {and densities} of $\sim$~10--100~cm$^{-3}$. {The denser clumps evolve into starless molecular cores \citep{2007ARA&A..45..339B} with} temperatures of ~8--15~K {and densities of} $\sim 10^4-10^6$ cm$^{-3}$ \citep{2006ARA&A..44..367S, 2009sfa..book..254A, 2010ApJS..188..139L, 2012A&A...547A..11N}. Some of these gravitationally-bound cores may begin contracting, first isothermally, and then with increasing internal densities and temperatures. Then {a central hydrostatic object forms, which starts heating} up the surrounding gas. Protostars with a mass greater than 8 solar masses are generally referred to as ÒHigh-mass protostellar objects (HMPOs)Ó. The collapsing envelope material can then reach temperatures of several hundred K closer to the central star, and peak densities of $\sim 10^{8}$ cm$^{-3}$, conditions which define a ``hot core'' or, for low-mass protostars, a hot ``corino'' \citep{2006PNAS..10312249V}. In this paper we concentrate on the evolutionary stages ranging from a cold molecular cloud to the warm envelopes of protostars. We choose a wide parameter space with a grid of 1$\,$000 points covering temperatures between 5--150~K and densities of 10$^3$ - 10$^{10}$~cm$^{-3}$, and assume the standard ISM dust and a fixed A$_{\rm V}$~=~10~mag, {meaning that the photochemistry is only driven by secondary UV photons}. Assuming a fixed A$_{\rm V}$ reduces the problem to two dimensions, which is easier to analyze and visualize.

\subsection{Chemical model}
\label{sec:InitChem}
We have utilized the gas-grain chemical model ``ALCHEMIC'' developed by \citet{2010A&A...522A..42S}, where a detailed description of the code and its performance is presented. The code is optimized for modeling the time-dependent evolution of large chemical networks, including both gas-phase and surface species. In this paper we added a large set of reactions involving deuterated species. A few features of the ``ALCHEMIC" model are summarized below.

The self-shielding of H$_2$ from photodissociation was calculated using Equation~(37) from \citet{1998ARA&A..36..317V}. The shielding of CO by dust grains, H$_2$, and its self-shielding was calculated using the precomputed table of \citet[][Table~11]{1996A&A...311..690L}. We consider cosmic rays (CRP) as the only external ionizing source, using {a} CRP ionization rate for H$_{2}$, $\zeta_{\rm CR}$~=~1.3~$\times 10^{-17}$~s$^{-1}$ \citep{1973ApJ...185..505H}, {appropriate for molecular cloud environments and which has been utilized in several previous studies \citep[such as][]{2006A&A...459..813W, 2008ApJ...672..629V, 2012MNRAS.426..354D}}. The gas-grain interactions include dissociative recombination and neutralization of ions on charged grains, sticking of neutral species and electrons to uniformly-sized 0.1 $\mu$m dust grains with a sticking coefficient of 1 and release of ices by thermal, CRP-, and UV-induced desorption, such that at high temperatures the surface population will be low as thermal desorption takes over. We do not allow H$_2$ and its isotopologues to stick to grains. We assume a UV photodesorption yield of $10^{-3}$ \citep[e.g.,][]{2009A&A...504..891O, 2009A&A...496..281O}. With our fixed visual extinction, the photon field derives from secondary electron excitation of molecular hydrogen followed by fluorescence.
 
We assume that each $0.1\mu$m spherical silicate grain provides $\approx $1.88 $\times$10$^6$ sites \citep[][]{2001ApJ...553..595B} for surface recombination that proceeds solely through the classical Langmuir-Hinshelwood mechanism \citep[e.g.][]{1992ApJS...82..167H}. The grain surface topology, the presence of high- and low-energy binding sites, grain sizes and shapes are all separate parameters that may severely impact the chemistry. An accurate study of this impact will require a detailed treatment of the microscopic physics of molecules on various solid surfaces, which is far beyond the scope of the present study. For further reading we recommend papers by \citet{2006MNRAS.365..801P, 2009A&A...508..275C, 2013ApJ...762...86V, 2013ApJ...769...34V}, where some of these issues are already addressed.

Upon a surface recombination, we assume there is a 5\% probability for the products to leave the grain due to the conversion of some of the exothermicity of reaction into breaking the surface-adsorbate bond \citep{2007A&A...467.1103G}. We do not find significant differences {(less than a factor of 2)} in D/H ratios and abundances of essential species, such as H\dthree\jon, water, ammonia from varying this probability between {1$-$10}\%. 
{However, we found a significant variation in ice abundances of formaldehyde and methanol of up to a factor of 6 at lower temperatures ($\lesssim 25$ K) {when considering higher desorption probabilities $\gtrsim$ 5\%}. Interestingly, ice abundances increase with the desorption probability and we find that this is due to a much more efficient formation of formaldehyde and methanol at intermediate times. Due to more intense gas-grain interactions precursor species of H$_2$CO and CH$_3$OH are able to form more readily in the gas phase and later stick to grains. Consequently, formaldehyde and methanol are formed faster via surface processes.} 
Following experimental studies on the formation of molecular hydrogen on dust grains by \citet{1999ApJ...522..305K}, we adopt the standard rate equation approach to the surface and ice chemistry without quantum-mechanical tunneling through the potential walls of the surface sites. We also do not consider competition kinetics between activation and diffusive barriers \citep{2011ApJ...735...15G}.

A typical run, with relative and absolute tolerances of 10$^{-5}$ and 10$^{-25}$, utilizing the original gas-grain network without deuterium chemistry ($\sim$7\,000 reactions, $\sim 700$ species) takes 1--5~s for 1 Myr of evolution with a Xeon 3.0~GHz CPU. With our new, almost tenfold larger deuterium network, the same run takes approximately an order of magnitude longer to calculate. The linear dependence of the CPU time vs. species number in the model is due to the advanced numerical scheme implemented in the ALCHEMIC code, which generates and uses sparse Jacobi matrices.

\begin{table*}
\centering
\caption{Initial abundances for the ``Primordial''  model with respect to $n_{\rm H}$. \label{tab:IAp}}
\begin{tabular}{r|ccccccc}
\hline
\tablewidth{0.95\textwidth}
\tabletypesize{\footnotesize}
Species~~	&	H$_2$			&	H				&	HD				&	He				&	C				&	N				&	O			\\
		&	0.499 			& 2.00 $\times$10$^{-3}$ 		& 1.50 $\times$10$^{-5}$ 		& 9.75 $\times$10$^{-2}$ 		& 7.86 $\times$10$^{-5}$ 		& 2.47 $\times$10$^{-5}$ 		& 1.80 $\times$10$^{-4}$ \\[3pt]
Species~~		&	S 				&	Si				&	Na				&	Mg				&	Fe				&	P 				&	Cl			\\
		&	9.14 $\times$10$^{-8}$	&	9.74 $\times$10$^{-9}$	&	2.25 $\times$10$^{-9}$	&	1.09 $\times$10$^{-8}$	&	2.74 $\times$10$^{-9}$	&	2.16 $\times$10$^{-10}$	&	1.00 $\times$10$^{-9}$	\\
		\hline
\end{tabular} 
\end{table*}

\begin{table*}
\centering
\caption{Initial abundances of major species for the ``Evolution'' model with respect to $n_{\rm H}$. \label{tab:IAe}}
\begin{tabular}{r|cccccccc}
\hline
\tablewidth{0.95\textwidth}
Species~~	&	H$_2$			&	He				&	H				&	HD				&	C				&	N				&	O			\\[1pt]
		&	0.500 			& 9.76 $\times 10^{-2}$ 	& 2.32 $\times 10^{-4}$ 	& 	9.57 $\times 10^{-6}$& 1.02 $\times 10^{-8}$ 	& 8.75 $\times 10^{-8}$ 	& 1.45 $\times 10^{-6}$ \\[3pt]
Species~~	&	S 				&	Si				&	Na				&	Mg				&	Fe				&	P 				&	Cl			\\
		&	2.00 $\times$10$^{-9}$	& 7.00 $\times$10$^{-11}$		& 6.37 $\times$10$^{-11}$		& 3.70 $\times$10$^{-10}$		&	6.07 $\times$10$^{-11}$	&	5.72 $\times$10$^{-12}$	&	1.64 $\times$10$^{-10}$	\\[3pt]
Species~~	&	H$_{2}$O (ice)			&	CO (ice)			&	CO				&	CH$_{4}$ (ice)			&	NH$_{3}$ (ice)		&	O$_{2}$			&	N$_{2}$	\\[1pt]
		& 9.90 $\times 10^{-5}$	& 3.91 $\times 10^{-5}$	& 1.85 $\times 10^{-5}$	& 1.66 $\times 10^{-5}$	& 1.30 $\times 10^{-5}$	& 7.04 $\times 10^{-6}$	& 3.78 $\times 10^{-6}$ \\[3pt]
Species~~	&	O$_{2}$ (ice)		&	D				&	N$_{2}$ (ice)		&	HDO (ice)			& C$_{3}$H$_{2}$ (ice)	&	HNO (ice)			&	D$_{2}$	\\[1pt]
		& 2.29 $\times 10^{-6}$	& 1.81 $\times 10^{-6}$	& 1.26 $\times 10^{-6}$	& 1.08 $\times 10^{-6}$	& 7.68 $\times 10^{-7}$	& 7.25 $\times 10^{-7}$	& 7.07 $\times 10^{-7}$	\\
		\hline
\end{tabular} 
\end{table*}

\subsection{Initial abundances}
\label{sec:initabund}
As input data, reaction rate coefficients and physical properties need to be specified, as do initial abundances. We have chosen to implement two different initial abundance sets and calculate the chemical evolution with the new deuterium network for 1 Myr. 

For the first set, hereafter referred to as the  ``Primordial''  model, we utilized the ``low metals" abundances of \citet{1982ApJS...48..321G} and \citet{1998A&A...334.1047L}. Initially all deuterium is located in HD, with D/H $= 1.5 \times 10^{-5}$ \citep{1998ApJ...509....1S, 2003SSRv..106...49L}, see Table~\ref{tab:IAp}. The abundances in the second set, the ``Evolution'' model, were calculated with our deuterium chemistry model, assuming a TMC1-like environment: $T = 10$~K and $n_{\rm H}=10^4$ cm$^{-3}$, at $t= 1$~Myr. Under such conditions elemental deuterium from HD is efficiently redistributed to other molecules, leading to their high initial D/H fractionation. These final abundances at 1~Myr are used as input in the ``Evolution'' model (see Table~\ref{tab:IAe}). The ``Evolution'' model serves as a simple example of a two-stage chemical model with physical conditions that can change dramatically at 1 Myr, unless the evolutionary model is run strictly under TMC-1 conditions. 

\subsection{Deuterium fractionation chemistry}
\label{sec:deut_network}
As a first step toward creating a consistent network with deuterium fractionation, we undertook a thorough search in the literature for updates to the reaction rates of the original non-deuterated network. We utilized the latest osu.2009 gas-phase chemical network and incorporated all essential updates as of December 2012, adopted from \citet{2004ApJ...611..605H}, \citet{2010A&A...524A..39C}, \citet{2010A&A...514A..83H}, \citet{2010SSRv..156...13W}, \citet{2011ApJ...728...71L}, as well as those reported in the KInetic Database for Astrochemistry (KIDA)\footnote{{http://kida.obs.u-bordeaux1.fr/} as of [2012-12-26]}. Further, to allow for the synthesis of the few complex molecules in our network such as methanol (CH$_3$OH), methyl formate (HCOOCH$_3$), and dimethyl ether (CH$_3$OCH$_3$), an extended list of surface reactions and photodissociation of ices was adopted from \citet{2006A&A...457..927G}. Several tens of gas-phase photoreaction rates were updated using the new calculations of \citet{2006FaDi..133..231V}\footnote{\url{http://www.strw.leidenuniv.nl/\textasciitilde ewine/photo/}}. 

Next, we applied a cloning routine to this updated network \citep[as described in][]{1996MNRAS.280.1046R}, and added all additional primal isotope exchange reactions for H$_3^+$ as well as CH$_3^+$ and C$_{2}$H$_{2}^+$ from \citet{2000A&A...361..388R, 2002P&SS...50.1275G, 2004A&A...424..905R, 2005A&A...438..585R}. In this cloning routine all reactions bearing hydrogen atoms are considered to have deuterated analogues, and ``cloned" accordingly (assuming the same rate coefficient if no laboratory data are available). In cases where the position of the deuterium atom is ambiguous, we apply a statistical branching approach. In the resulting network we do not yet distinguish between the ortho/para states of molecules, and leave this for a separate paper. 

A typical example of the outcome of the cloning procedure is presented for the reaction between C$^+$ and CH$_3$: 
\begin{equation}
{\rm C^+ + CH_3} \rightarrow {\rm C_2H^+ + H_2} \Rightarrow \left\{
\begin{array}{lcl}

{\rm C^+ + CHD_2} \rightarrow {\rm C_2D^+ + HD } \\
{\rm C^+ + CHD_2} \rightarrow {\rm C_2H^+ + D_2 } 

\label{eq:CloneScheme}
\end{array}
\right.
\end{equation}

The single ion-molecule reaction of C$^+$ with CH$_3$ is cloned into two separate channels for CHD$_2$. Moreover, the branching of these two new channels is not equal. To visualize this, we label the two deuterium atoms in CHD\dtwo~{D$_{\rm a}$ and D$_{\rm b}$}. For the first reaction, which forms C\dtwo D\jon,~{D$_{\rm a}$} can be placed on either product and {D$_{\rm b}$} on the other, hence we have two possibilities: C\dtwo {D$_{\rm a}$}\jon + H{D$_{\rm b}$} or C\dtwo {D$_{\rm b}$}\jon + H{D$_{\rm a}$}. For the second channel, which forms C\dtwo H\jon, both deuterons have to be placed on D\dtwo~and we only have one possibility. This analysis assumes that the deuterons on D$_{2}$ are indistinguishable, which is in agreement with the Pauli Exclusion Principle. Alternatively, we could initially assume that they are distinguishable, but because half of the D\dtwo~rotational-nuclear spin states are missing, the simple argument about 2/3 and 1/3 branching ratios remains valid. 

To limit the size of the network we have restricted the cloning process to avoid any -OH endgroups. Observations of deuterated species suggest that fractionation of species with -OD endgroups is less important in low-mass protostars, but may still be important for high-mass protostars. For example, \citet{2006A&A...453..949P} conducted a survey of deuterated formaldehyde and methanol in a sample of seven low-mass class 0 protostars, and found CH$_3$OD / CH$_2$DOH $\lesssim$ 0.1. A hypothesis of rapid conversion of CH\dthree OD into CH\dthree OH in the gas-phase due to protonation reactions that would affect only species for which deuterium is bound to the electronegative oxygen has been suggested by \citet{1997ApJ...482L.203C} and \citet{2004A&A...421.1101O}. We conducted a small study using a version of our deuterium network where -OH endgroups were cloned and found no significant changes in the resulting time-dependent molecular abundances.

{Full tables} of added and updated reactions are {found in Tables~\ref{tab:netwOrg} and \ref{tab:netwDeut} in} Appendix A of the online material. The resulting chemical network consists of $\sim 55\,000$ reactions connected by $\ga 1\,900$ species, to our knowledge the most extended network for deuterium chemistry to date\footnote{Publicly available at: \url{http://mpia.de/PSF/codes.php}}. 

\subsection{Analysis of reaction updates}

\begin{table}
\centering
\caption{Species showing variations in D/H ratios by more than a factor of 5 due to the updates in the reaction network.$^{a}$ \label{tab:UpdateCompares}}
\tablewidth{0.2\textwidth}
\begin{tabular}{cc|cc}
\hline
Species			&	$R({\rm D/H})$	&	Species			&	$R({\rm D/H})$	\\
\hline
HD\dtwo O\jon		&	24.3			&	CHD\dthree		&	8.5		\\
CH\dtwo D\dtwo	&	8.5			&	CH\dthree D		&	8.1		\\
CH$_{4}$D\jon		&	7.9			&	C\dtwo H\dtwo D\dtwo&	7.3		\\
CD$_{4}$			&	7.0			&	H\dtwo DO\jon		&	5.9		\\
HD\dtwo CS\jon	&	5.4			&	D\dtwo CS		&	5.3		\\
\hline\\[-7.0pt]
\multicolumn{4}{l}{$^{a}$ Includes only species with fractional abundances $> 10^{-25}$}
\end{tabular}
\end{table}

Given the large size of the network with uncertainties in the adopted rate coefficients, reaction barriers, and branching ratios, it is educational to estimate how these uncertainties propagate in time-dependent modeled abundances. Before performing a detailed sensitivity analysis, as in our study of disk chemistry uncertainties \citep{2008ApJ...672..629V}, it is of interest to characterize the influence of the reaction rate updates on the calculated abundances and the D/H ratios. This may help us to highlight the significance of recent laboratory astrochemistry activities, both for deuterated and un-deuterated species, in providing more accurate astrochemical data to the community. 

First, we studied the effects of introducing deuterium chemistry into our model on abundances of un-deuterated species by comparing abundances throughout the parameter space to a non-deuterated version of the network. We found that species with relative abundance $> 10^{-25}$ show mean values in abundance variations between the two networks within a factor {0.95 - 1.05}. {Since we} did not find any particularly large variations in abundances for H-bearing species, we conclude that the results from our updated analysis are a pure effect of updated reaction rates and not caused by the additional pathways created by the cloning routine. 

{In order to separate the effect that recent updates have had on abundances}, we generated an additional network by cloning an outdated network restricted to the reaction rate updates up until 2005. We then studied the impact of updated reaction rate coefficients by comparing the calculated time-dependent abundances between the ``old'' chemical network and the ``new'' network in the 2D-parameter space discussed in Sect.~\ref{sec:phys_model}. In addition to the D/H abundance ratios, we will emphasize the differences in these ratios between the models, which we calculated by dividing the respective D/H ratios in the updated 2012 network by those from the outdated 2005 network, and will denote this ratio as $R({\rm D/H})$. The results have been obtained with the ``Primordial'' model only. In this comparison, we have excluded minor species with relative abundances below $10^{-25}$. It should be noted that the $R({\rm D/H})$ ratios may remain unchanged when absolute abundances of species and their isotopologues in the updated and outdated networks increase in unison.

We list in Table~\ref{tab:UpdateCompares} the arithmetic mean value {calculated over the parameter grid} ($T=5-150$~K, $n_{\rm H}=10^{3} -10^{10}$~cm$^{-3}$) of $R$(D/H) ratios for all species with fractional abundances $\ge 10^{-25}$ for which the mean value of $R$(D/H) ratios have changed by more than a factor of 5. Among the listed species, we find light hydrocarbons (e.g., CHD\dthree, CH\dtwo D\dtwo), ions (CH$_{4}$D\jon, HD\dtwo O\jon), and simple organic molecules (e.g. DCOOH, D$_2$CO), as well as key molecules such as doubly-deuterated water and ammonia. Multi-deuterated species appear to be more affected by the updates than their singly-deuterated analogues, as there are more intermediate pathways involved in their chemistry, {as is most evident by comparing HD\dtwo O\jon~and H\dtwo DO\jon~in Table~\ref{tab:UpdateCompares}}. 

There are also several species affected by the abundances that do no show any variance in D/H ratios, i.e. both deuterated and undeuterated species are similarly affected by updates. The abundances of CH\dtwo D\jon~and CHD\dtwo\jon~provide good examples of such behavior. These species show a coherent increase (within a factor of 1.1 between un-deuterated species and isotopologues) in their gas-phase abundances at 1 Myr and at high temperatures ($\gtrsim 100$ K) and densities ($n_{\rm H} \gtrsim 10^{7}$ cm$^{-3}$). As a result, their $R({\rm D/H})$ values remain close to unity. We identified the coherent increase in abundance as originating from an update taken from KIDA in the rate coefficient for the slow radiative association reaction forming CH$_5$\jon~via CH\dthree\jon~colliding with H\dtwo. The rate coefficient of this reaction was lowered by almost two orders of magnitude, from 1.30 $\times 10^{-14}$~cm$^3$ s$^{-1}$ to 4.10 $\times$10$^{-16}$~cm$^3$ s$^{-1}$ \citep[at room temperature; see][]{2010SSRv..156...13W}. We note that the older value was based on a misinterpretation of the original literature, which used 300 K in the formula for the rate coefficient but was intended only for temperatures up to 50 K \citep{1985ApJ...291..226H}. The same D/H ratio variation is transferred to HD\dtwo O\jon~and H\dtwo DO\jon~through the ion-neutral reaction with CH$_{5}$\jon~and its isotopologues reacting with free oxygen atoms. The abundances of the CH$_4$ isotopologues derive from the dissociative recombination of CH$_5$\jon~(and its isotopologues) as well as from ion-molecule reactions with CO, so CH$_4$ is directly affected by the updated reaction rate. It then transfers its D/H ratio variation into C\dtwo H\dtwo D\dtwo~through CH\dtwo D\dtwo~reacting with CH. Another route involves the intermediary reaction between CH\dtwo D\dtwo~and S\jon~to form HD\dtwo CS\jon, which later dissociatively recombines into D\dtwo CS, which in turn reacts with CH$_{2}$D$_{2}$.

On the other hand, the deuterated analogue of this radiative association reaction does not occur; instead CH$_3^+$ + HD produces CH$_2$D$^+$ and H$_2$, and the corresponding rate constant is the same in the outdated and new networks \citep{1989ApJ...340..906M}. This slows down production of the key ion, CH$_5^+$, while deuterated isotopologues of CH$_3^+$ and CH$_5^+$ are produced with almost the same rate, consequently affecting D/H ratios of the gas-phase species listed in Table~\ref{tab:UpdateCompares}. It does not affect abundant organic species such as methanol and formaldehyde however, as they are mainly formed by surface hydrogenation of CO ice.

{We find that there is a particularly large variance in abundances for the} two sulphur-bearing species, C\dtwo S and C\dtwo S\jon, which show an increase in abundance by a factor of 187 and 26, respectively, compared with the non-deuterated network. We are not concentrating on the chemistry of sulphur-bearing molecules in this study because their chemistry is still poorly understood and often restricted to a few pathways. But the additional pathways that the cloning routine generates has a stronger effect on these two species as pathways reducing their abundances proceed much slower than their formation pathways.

\subsection{Error propagation in deuterium fractionation chemistry}
We studied the impact of uncertainties in reaction rate coefficients on the resulting chemical abundances, and how they propagate throughout the chemical evolution. Two separate environments were chosen for this study, representing dark clouds \citep[$T$ = 10 K, $n_{\rm H}$ = 10$^{4}$ cm$^{-3}$;][]{2012A&A...547A..11N} and lukewarm infrared dark clouds \citep[$T = 25$ K, $n_{\rm H} = 10^{5}$ cm$^{-3}$;][]{2012ApJ...751..105V}. In these and all subsequent runs mentioned in this paper, we use the ``Primordial'' initial conditions unless stated to the contrary. We chose to concentrate on the uncertainties of the rate coefficients of the gas-phase reactions. Including variations in surface chemistry rates is a tricky problem as these depend on surface mobility, binding and diffusion energies of reactants, and properties of the surface itself (porosity, irregularities, etc.). Recently, \citet{2012ApJ...748L...3T} have studied the importance of deuterium fractionation on dust surfaces, using a multi-layered ice mantle model and quantified some of the associated errors in the calculated abundances due to the uncertainties in the surface chemistry, so we do not repeat such a study here.

Our analysis is based on the same method as employed by \citet{2004AstL...30..566V} and \citet{2008ApJ...672..629V}. We performed computations for a large set of models, using identical physical conditions and initial abundances, and the same chemical network but with randomly varied rate coefficients within their uncertainty limits. The rate uncertainties were taken from the KIDA database. Most of the reactions with deuterated species were created by the cloning procedure, and hence have unknown uncertainties. For these reactions we used the high standard error value in KIDA\footnote{http://kida.obs.u-bordeaux1.fr/}, with a normal logarithmic error distribution of a factor of two. 

With this method, we generated 10$\,$000 networks with the new rate coefficients $k(T)$ randomly distributed as follows: 

\begin{equation}
k(T) = {\rm exp} ({\rm ln}~k_{0}(T) + {\rm ln}~F~\times~N[0,1] ),
\label{eq:ratedraw}
\end{equation}
where $k_{0}(T)$ is the measured, calculated, or estimated rate coefficient at temperature $T$, $F$ is the statistical distribution of the uncertainty, and $N[0,1]$ is a random value drawn from a standard Gaussian distribution with mean $\mu = 0$ and variance 1. The time needed for a full run consisting of 10\,000 networks for a specified temperature and density takes approximately one day of computational time on a Xeon 3.0~GHz CPU (with relative and absolute tolerances of 10$^{-4}$ and $10^{-20}$, respectively). 

The huge size of our new deuterium chemical network makes it a challenging task to find a very precise correlation between the rate uncertainties and uncertainties in the molecular abundances of a particular species. Since for many deuterated species the set of primal pathways easily exceeds several reactions, the relative contribution of each individual reaction to the final uncertainty is likely to be small, $\la 10\%$. To isolate the most problematic reactions for several key observed species, we conducted a sensitivity analysis using the same cross-correlation method as implemented by \citet{2008ApJ...672..629V}. We selected a handful of molecules, including their isotopologues and isomers; viz., H\dthree\jon, H\dtwo O, HCN, HCO\jon, and CH\dthree OH, and for each of these species we calculated time-dependent linear correlation coefficients between the abundances and the rate coefficients for all the 10\,000 network realizations and for each of our 30 logarithmically taken time steps. The linear correlation coefficients $c_{L}(i,j,t)$ at specific density, $i$, and temperature, $j$, points and at time $t$ are calculated by:
\begin{equation}
c_{L} (i,j,t) = \frac{\Sigma_{l}~(x_l^s (i,j,t) - \overline{x^s (i,j,t)} )~~(\alpha_l^j - \overline{\alpha^j}) }{~~~~~\Sigma_l~(x_l^s (i,j,t) - \overline{x^s (i,j,t)} )^2~~\Sigma_l (\alpha_l^j - \overline{\alpha^j})^2 } 
\label{eq:lincoeff}
\end{equation}
with $x_s^l (i,j,t)$ being the molecular abundance for species $s$ and iteration $l$, and $\overline{x_s^l (i,j,t)}$ and $\overline{\alpha^j}$ signifying the standard (mean) abundances and rate coefficients for species $s$, respectively. Because key reactions can vary through time evolution, we calculate and use cumulative correlation coefficients in our results, for which we integrated the absolute values of time-dependent linear correlation coefficients over the 30 logarithmic taken time steps taken over 1 Myr. In our results in the next section, we restrict discussion to the cumulative correlation coefficients, designated as $c$.

\section{Results}
\label{sec:results}

\subsection{Sensitivity analysis}
\label{sec:sensAnalys}
We have determined that 10\,000 realizations of the network may not be adequate for results of our sensitivity analysis to fully converge. For correlation coefficients $c < 0.1$ we sometimes still see small variations when we compare our results with a model containing only 9\,000 realizations. Therefore, we also ran a separate set of simulations with 20\,000 realizations for a dark cloud environment, but found the same result with only minor deviations for the reactions with low correlation coefficients, $c < 0.1$. All correlation coefficients should stop fluctuating with size as soon as the number of the network realizations exceeds the number of reactions in the network, which is $\sim 50\,000$. Running the sensitivity analysis code with so many realizations, however, would be prohibitively time consuming. Therefore, below we consider reactions with $c > 0.1$ in our discussion.

\begin{table*}
\centering
\caption{40 most problematic reactions (with cumulative correlatlation coefficients $c > 0.1$) and their associated (real) uncertainty factors. \label{tab:topProbReact}}
\tablewidth{0.95\textwidth}
\begin{tabular}{lclc|lclc}
\hline
\multicolumn{3}{l}{Reaction}		&	{Uncertainty}		&	\multicolumn{3}{l}{Reaction}		&	{Uncertainty}	\\
\hline
H\dtwo~					+	CRP		&\ra&	H\dtwo\jon~				+	e\ijon	&	2.00	&	He			+	CRP				&\ra&	He\jon~				+	e\ijon		&	2.00		\\
H\dthree\jon / H\dtwo D\jon~	+	HD		&\ra&	H\dtwo D\jon / HD\dtwo\jon~	+	H\dtwo	&	1.25	&	H\dthree\jon~	+	D\dtwo			&\ra& 	HD\dtwo\jon~			+	H\dtwo 		& 2.00	\\
H\dtwo D\jon / HD\dtwo\jon~ 	+	H\dtwo	&\ra&	H\dthree\jon / H\dtwo D\jon~	+	HD		&	2.00	&	H\dthree\jon / H\dtwo D\jon~ + D 		&\ra& 	H\dtwo D\jon / HD\dtwo\jon / D\dthree\jon~ +	H &	2.00	\\
HD\dtwo\jon~				+	D		&\ra&	D\dthree\jon~				+	H		&	2.00	&	H\dthree / H\dtwo D\jon~	+	e\ijon	&\ra&	H/D			+	H		+	H		&	2.00	\\	
H\dthree\jon~				+	CO		&\ra&	HCO\jon/HOC\jon~			+	H\dtwo	&	1.25	&	H\dtwo D\jon~	+	CO				&\ra&	DCO\jon/DOC\jon~		+	H\dtwo		&	1.25		\\
\hline
H\dthree\jon~				 +	OH/OD	&\ra&	 H\dtwo O\jon/HDO\jon~ 		+	H\dtwo	&	2.00	&	H\dtwo D\jon~	+	OH/OD			&\ra&	HDO\jon/D\dtwo O\jon~	+	H\dtwo		&	2.00	\\
HD\dtwo\jon~				+	OH/OD	&\ra&	D\dtwo O\jon~				+	H/D		&	2.00	&	H\dthree\jon~	+	H\dtwo O/HDO		&\ra&	H\dthree O\jon/H\dtwo DO\jon~	+	H\dtwo	&	1.25\\
H\dthree\jon~				+	HDO		&\ra&	H\dthree O\jon~			+	HD		&	1.25	&	H\dthree\jon~	+	DNC				&\ra&	H\dtwo CN\jon~			+	HD		&	2.00	\\
H\dtwo D\jon~				+	HNC		&\ra&	HDCN\jon~				+	H\dtwo	&	2.00	&	OH			+	D				&\ra&	OD					+	H			&	2.00		\\
H\dthree O\jon~			+	e\ijon	&\ra&	OH			+	H		+	H		&	1.25	&	H\dthree O\jon/H\dtwo DO\jon~	+	e\ijon 	&\ra&	H\dtwo O/HDO				+	H		&	1.25	\\
HD\dtwo O\jon~			+	e\ijon	&\ra&	D\dtwo O					+	H		&	1.25	&	HDO\jon/D\dtwo O\jon~		+	H\dtwo	&\ra&	H\dthree O\jon/H\dtwo DO\jon~	+	D		&	1.25		\\
\hline
H\dtwo CN\jon/HDCN\jon~ 	+	e\ijon	&\ra&	CN				+	H/D	+	H		&	2.00	&	H\dtwo CN\jon~+	e\ijon			&\ra&	HCN	/HNC			+	H			&	2.00		\\
HDCN\jon	~				+	e\ijon	&\ra&	HCN	/HNC				+	D		&	2.00	&	HDCN\jon~	+	e\ijon			&\ra&	DCN/DNC			+	H			&	2.00		\\
H						+	OD		&\ra&	HDO									&	2.00	&	D			+	OH				&\ra&	HDO									&	2.00		\\
\hline
HCO\jon/DCO\jon~			+	e\ijon	&\ra&	CO						+	H/D		&	1.25	&	HCO\jon~		+	D				&\ra&	DCO\jon~				+	H			&	2.00		\\
HOC\jon~					+	H\dtwo	&\ra&	HCO\jon~					+	H\dtwo	&	2.00	&	DOC\jon~		+	H\dtwo			&\ra&	HCO\jon/DOC\jon~		+	HD			&	2.00		\\
HCO\jon~					+	HCN		&\ra&	H\dtwo CN\jon~			+	CO		&	2.00	&	DCO\jon~		+	HCN				&\ra&	HDCN\jon	~			+	CO			&	2.00		\\
HCO\jon/DCO\jon~			+	H\dtwo O	&\ra&	H\dthree O\jon/H\dtwo DO\jon~	+	CO		&	1.50	&	HCO\jon~		+	HDO				&\ra&	H\dtwo DO\jon~		+	CO			&	1.50		\\	
HCO\jon~					+	OD		&\ra&	HDO\jon~				+	CO			&	2.00	&	HCO\jon~		+	CH\dtwo DOH		&\ra& CH\dtwo DOH\dtwo\jon~ 	+	CO			&	2.00		\\
HCO\jon~					+	CH\dthree OH	&\ra&	CH\dthree OH\dtwo\jon~ + CO			&	2.00	&	C\jon~		+	H\dtwo O			&\ra&	HCO\jon/HOC\jon~		+	H			&	2.00		\\
%
%
\hline
\end{tabular}
\end{table*}

Table~\ref{tab:topProbReact} {lists the subset of reactions} with cumulative correlation coefficients $c > 0.1$, which are the most problematic for the chemical evolution of the following species (and their isotopologues and isomers): H\dthree\jon, HCO\jon, HCN, H\dtwo O, CH\dthree OH, H\dthree O\jon, CH\dthree\jon, C\dtwo H\dtwo\jon~and CO. Because there are several additional key reactions with $0.05 < c < 0.1$, we list a more extensive table in the online material (Table~\ref{tab:probreact} of Appendix~\ref{sec:AppendixA}), including also reactions with $c > 0.05$ for the same set of species. 

As can clearly be seen, ion-neutral processes dominate Table~\ref{tab:topProbReact}, accompanied by a few dissociative recombination and neutral-neutral reactions as well as cosmic ray ionization of the two critical species: H\dtwo~and He. The last process may require more detailed description in astrochemical models, such as recently presented in \citet{2012A&A...537A...7R} and \citet{2012ApJ...756..157G}, so we assigned a relatively large uncertainty of a factor of 2 for this group of processes. Approximately half of the bimolecular reactions are connected to the chemical evolution of water and light hydrocarbons in the gas. Abundances of species mostly produced on grains, such as methanol, will not be strongly affected by the uncertainties. However, a small fraction of these species is still present in the gas in the center of dense cores, for which uncertainties in the gas-phase chemistry may become important. Many of the deuterated reactions in the table are produced by our cloning procedure, so their error coefficients are only an approximation, and can, in fact, be larger than estimated. Also, isomerization reactions for HOC\jon~and DOC\jon~with reaction rate uncertainties of a factor of two possess strong correlation coefficients ($0.3 - 0.4$). 

We find it clear that fractionation channels of the H\dthree\jon~and CH\dthree\jon~isotopologues require further study, as do reactions involving the isotopologues of H\dthree\jon~ reacting with CO, water, OH and their isotopologues, forming the initial steps towards more complex molecules. Reactions with H\dthree\jon~and H\dtwo D\jon~are both well represented in the list and initiate the ion-molecule chemistry, while HD\dtwo\jon~and D\dthree\jon~are often not abundant enough to have a significant effect in our models. H\dthree\jon~and H\dtwo D\jon~react with CO to form the isotopologues of HCO\jon, with OH and OD to form ionized water (H\dtwo O\jon, HDO\jon) as well as the water isotopologues, which strongly affect water abundances and D/H ratios. We also see many other interconnecting reactions among our set of key species. Several dissociative recombination reactions, which proceed very rapidly, show strong correlations. While their reaction rates can be accurately determined \citep{2006PhR...430..277F} the products and branching ratios of these reactions are not precisely known \citep[see e.g.][]{2010A&A...522A..90H, 2006FaDi..133..177G}.

\begin{figure}[!htb]
\centering
\includegraphics[width=0.45\textwidth]{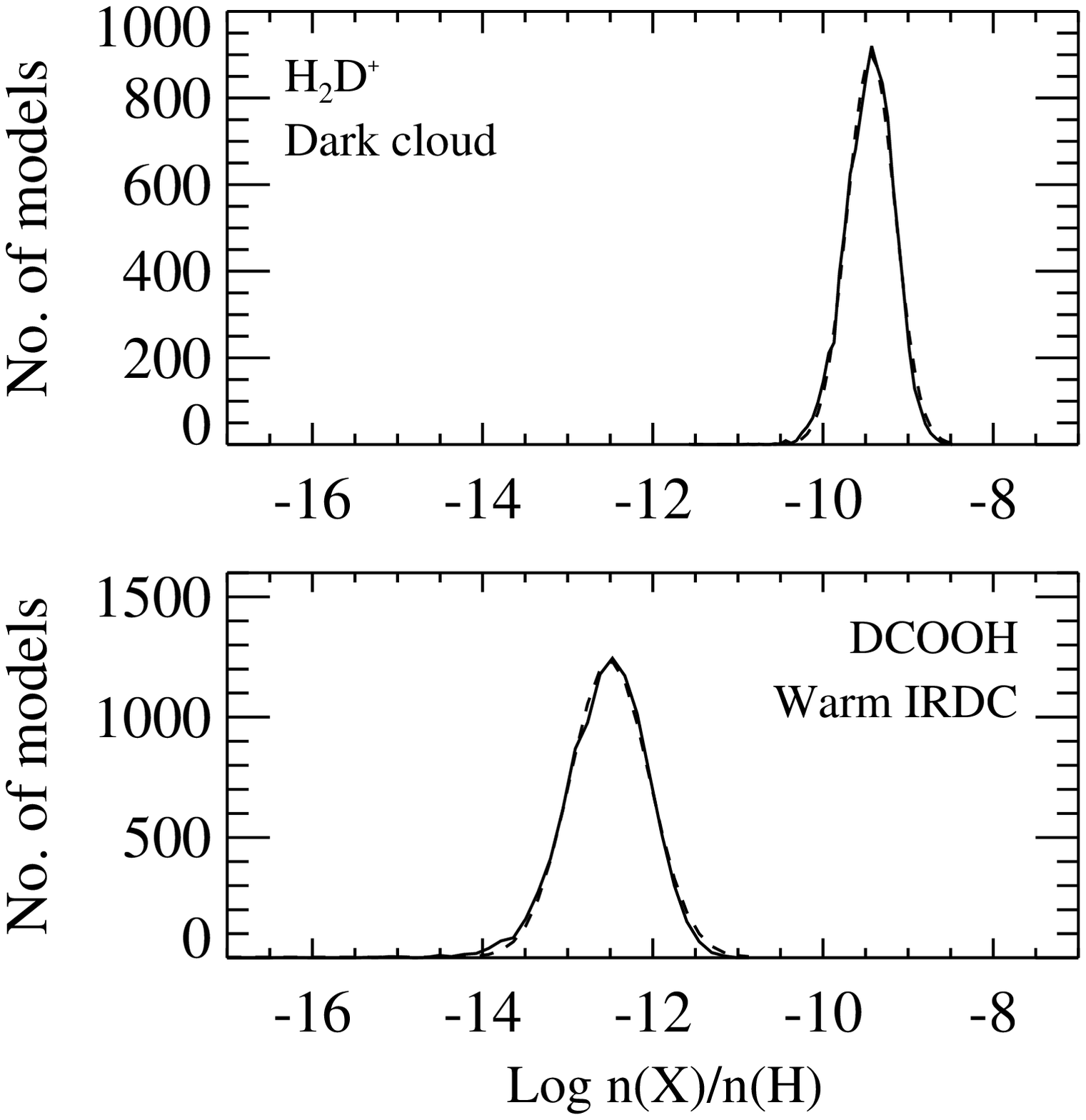}
\caption{Distributions of abundances from 10$\,$000 chemical runs for H\dtwo D\jon~in dark clouds (top) and DCOOH in warm IRDCs (bottom). Plots also show fitted Gaussian distributions (dashed lines). }
\label{fig:histAbu}
\end{figure}

\begin{figure}[!htb]
\centering
\includegraphics[width=0.45\textwidth]{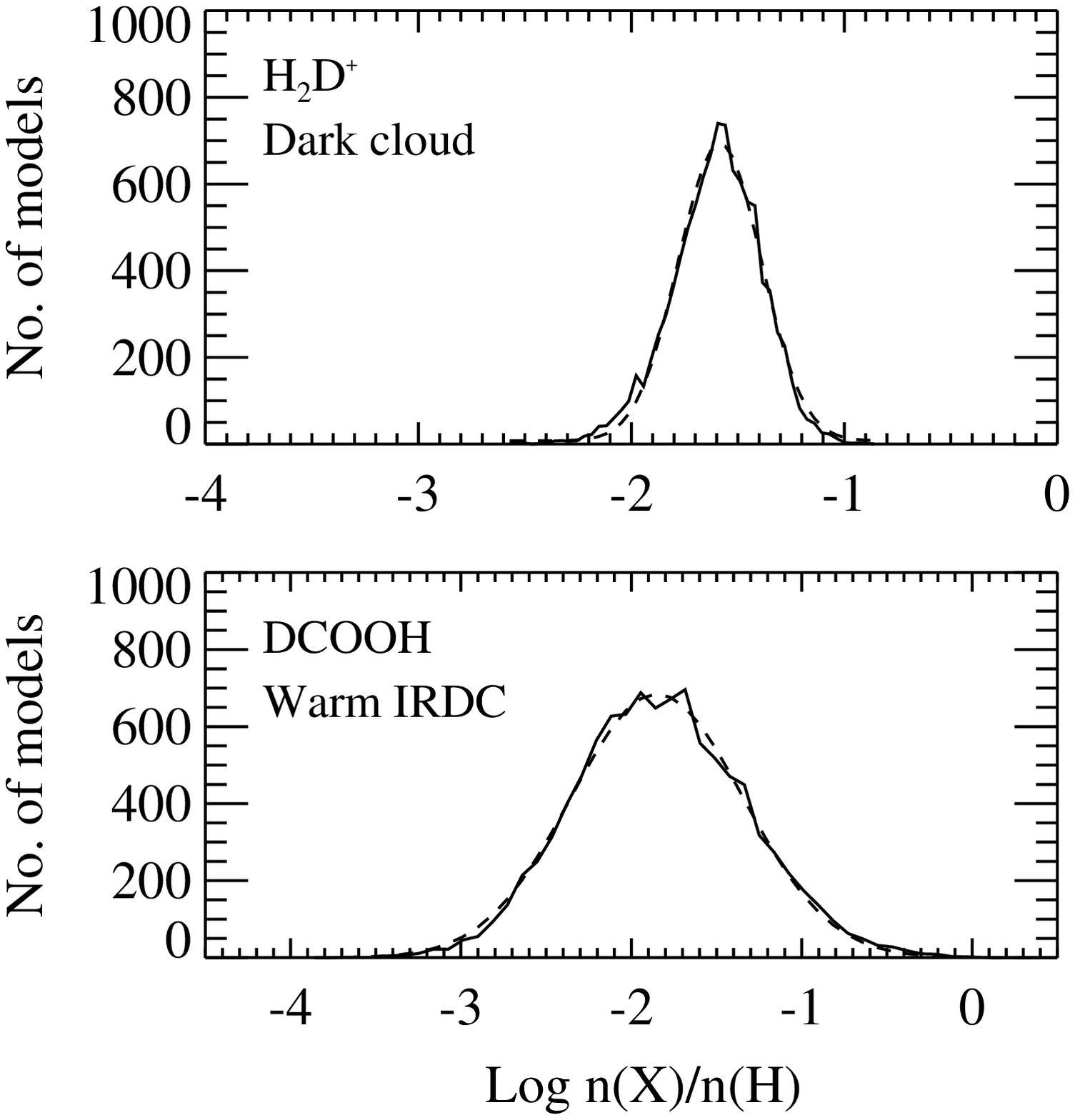} 
\caption{Distributions of D/H ratios from 10$\,$000 chemical runs for H\dtwo D\jon~in dark clouds (top) and DCOOH in warm IRDCs (bottom). Plots also show fitted Gaussian distributions (dashed lines). }
\label{fig:histDH}
\end{figure}

\begin{figure}[!htb]
\centering
\includegraphics[width=0.4\textwidth, angle=90]{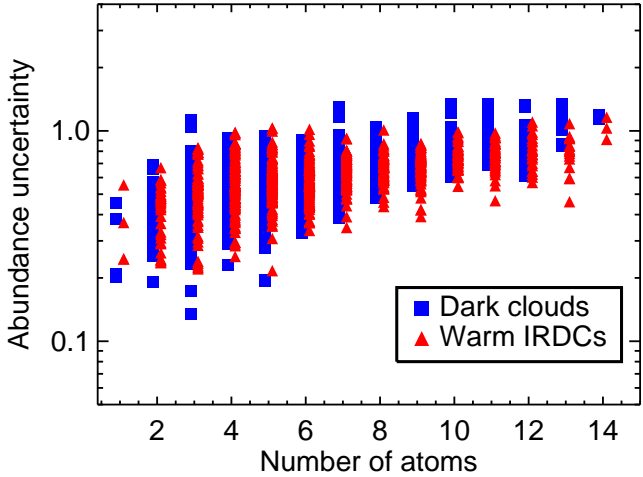}
\caption{The $1 \sigma$ abundance uncertainties {in orders of magnitude} for up to triply-deuterated species as a function of the number of atoms in a molecule. The dark cloud model results are denoted by squares and the warm infrared dark cloud model by triangles. A colored version of the plot is available in the online version. }
\label{fig:levelDist}
\end{figure}

\begin{figure}[!htb]
\centering
\includegraphics[width=0.4\textwidth, angle=90]{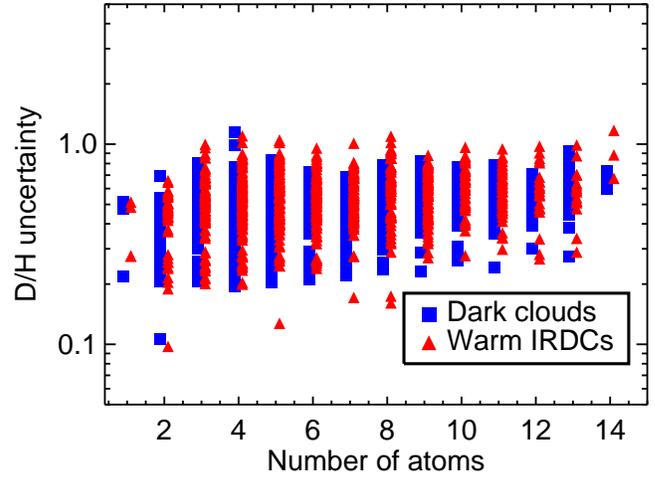}
\caption{The $1 \sigma$ uncertainties of the calculated D/H ratios {in orders of magnitude} for up to triply-deuterated species as a function of the number of atoms in a molecule. The dark cloud model results are denoted by squares and the warm infrared dark cloud model by triangles. A colored version of the plot is available in the online version. }
\label{fig:DHdisp}
\end{figure}

\begin{figure}[!htb]
\centering
\includegraphics[width=0.38\textwidth, angle=90]{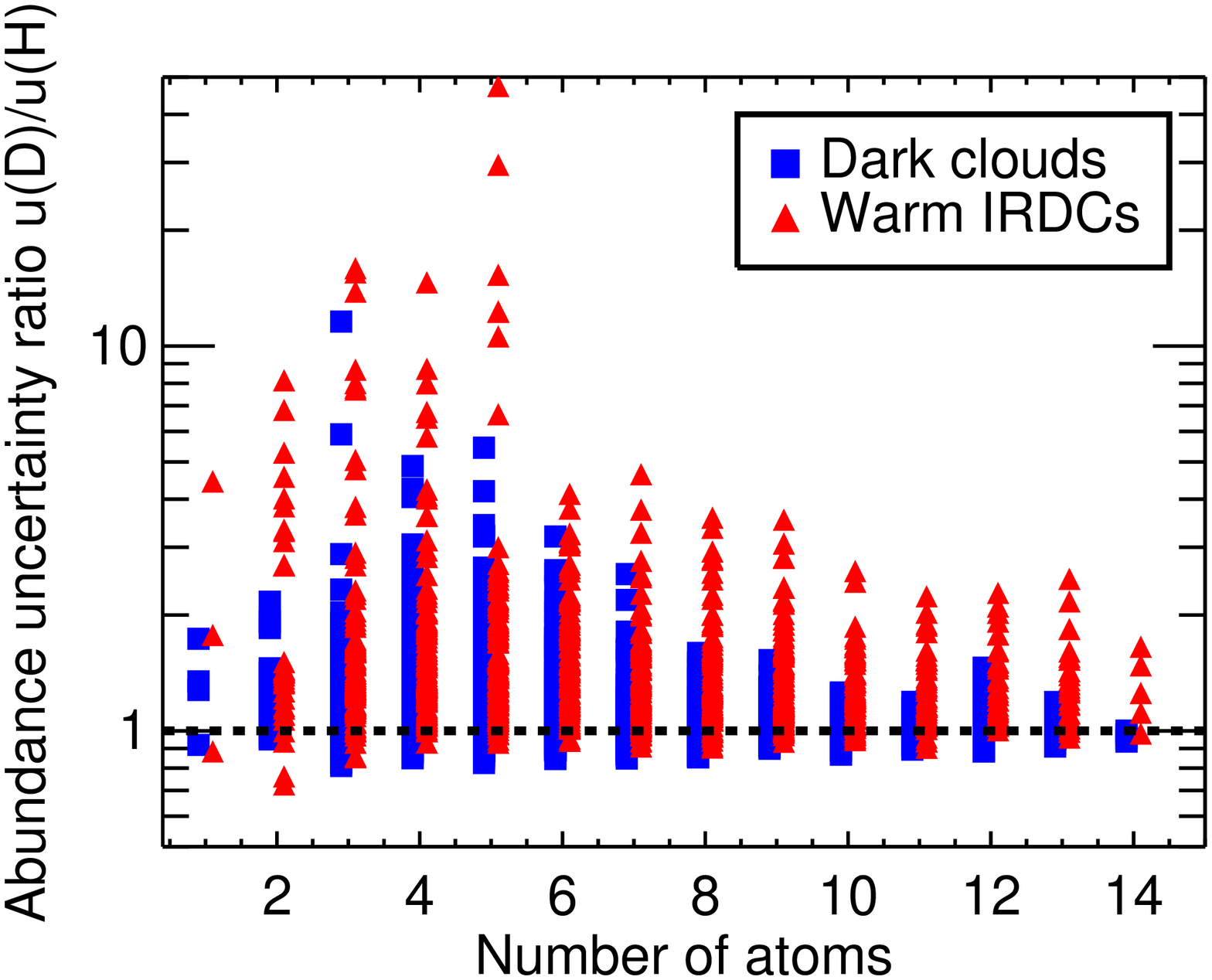}
\caption{Ratios of abundance uncertainties between un-deuterated and deuterated species with up to three D-atoms, as a function of the number of atoms in a molecule. The dark cloud model results are denoted by squares and the warm infrared dark cloud model results by triangles. The dashed line is added to identify where ratios are close to unity. A colored version of the plot is available in the online version.}
\label{fig:diffDist}
\end{figure}

\subsection{Uncertainties}
\label{sec:unc}

After calculating time-dependent abundances for all the realizations of the chemical network with varied reaction rate coefficients, and for each considered physical model, we fit Gaussians to the resulting abundance distributions at 1~Myr for all species. We carefully checked that such an approximation could be applied to the abundance distributions and found it to be the case for almost all species. In Figure~\ref{fig:histAbu}, we show examples of abundance distributions, with their Gaussian fits, for H\dtwo D\jon~and DCOOH in dark cloud and warm infrared dark cloud environments, respectively, both showing good fits. From the Gaussian fits, we then determined the full width half-maximum value (FWHM, $2.35\sigma$) of these distributions and used the $1 \sigma$ values to quantify the spread in abundances, which we henceforth refer to as the abundance uncertainties. We also applied the same procedure for the calculated D/H ratios, and show examples of the resulting D/H ratio distributions in Figure~\ref{fig:histDH} for water and formic acid in the same two regions. We find good fits to all these distributions with estimated 1 $\sigma$ values of factors of 1.9 and 2.9 for the abundance distribution of water and formic acid, respectively, while the values for D/H ratios are a factor of 1.4 and 3.2, respectively.

In Figure~\ref{fig:levelDist}, we plot the 1$\sigma$ abundance uncertainties for deuterated species with up to three D-atoms at 1~Myr as a function of the number of atoms for both environments. There are two major trends visible in this plot. First, the abundance uncertainties are in general lower in the case of the IRDCs models compared with cold dark clouds. 

At such a low temperature ($10-20$~K) many reactions with barriers cannot proceed, lowering the overall chemical complexity and thus the cumulative rate uncertainties. {One would expect uncertainties to be the lowest for dark clouds, but we suggest that as D/H ratios and abundances of deuterated species are also higher in colder environments, more reactions can occur to increase uncertainties.} Second, there is a strong trend of increasing abundance uncertainties with the number of atoms in species. This is obvious as the more atoms a species has, the more reaction pathways lead to its production and destruction from initial composition, and thus the higher is the accumulating effect of their uncertainties on modeled abundances \citep[see also the discussion in][]{2008ApJ...672..629V}. 

In general, using the $1\sigma$ confidence level, the abundances and column densities of species made of $\la 3$ atoms (e.g., CO, HCO$^+$, DCO$^+$) are uncertain by factors $1.5-5$, those for species made of $4-7$ atoms are uncertain by a factor of $1.5-7$, and those for more complex species made of $>7$ atoms are uncertain by a factor of $2 - 10$. The uncertainties for D- and H-bearing species are very similar in dark cloud environments. In warm IRDCs the typical uncertainties of larger H-bearing species ($>4$ atoms) are approximately a factor of two lower compared to D-bearing species, as de-fractionation begins at these elevated temperatures (25 K). Our estimates for the abundance uncertainties for deuterated species are comparable to the abundance uncertainties of un-deuterated species in protoplanetary disks \citep{2008ApJ...672..629V}, as well as diffuse and dark dense clouds \citep[e.g.,][]{2004AstL...30..566V, 2010A&A...517A..21W}. Chemically simple species containing Mg, Na, and Si tend to have high uncertainties reaching up to 3 orders of magnitude in abundances because their chemical pathways remain poorly investigated. Also large molecular species, be they rather abundant hydrocarbons (C$_{n}$H$_{m}$ with n, m $\gtrsim$ 4), with fractional abundances up to $10^{-9}-10^{-7}$, or complex and less abundant organic species (e.g. methyl formate, dimethyl ether), have large error bars in the computed abundances. For this latter group, the uncertainties can reach more than one order of magnitude. 

In addition to the abundance uncertainties, we also determine uncertainties in the resulting D/H ratios, as shown in Figure~\ref{fig:DHdisp}. Overall, the uncertainties in D/H ratios are lower when compared with the uncertainties of the corresponding H- and D-bearing isotopologues. This is because abundances of the individual isotopologues are often affected by the rate uncertainties in the same way, given that a majority of the deuterium fractionation processes are cloned, thus inheriting the rate of the `ancestor' reaction. We find the same trends as for the abundance uncertainties, but only a hint of increasing uncertainties with number of atoms. Uncertainties for D/H ratios are generally about half of one order of magnitude, but may vary between a factor of 2 and 10. As in the case of the abundance uncertainties, we find the largest D/H uncertainties for large hydrocarbons (C$_{n}$H$_{m}$, with $n, m \gtrsim$ 4), complex organics and species containing Mg, Na, and Si. 

The question remains how the uncertainties compare between deuterated and un-deuterated species. To illustrate the overall relative uncertainties between deuterated species and their un-deuterated analogues, we plot in Figure~\ref{fig:diffDist} the ratios of abundance uncertainties of up to triply-deuterated species and their un-deuterated analogues as a function of the number of atoms. Note that these relative uncertainties are not the same as the uncertainties in D/H ratios; the former can be labeled as u(D)/u(H), while the latter can be labeled as u(D/H), where u stands for uncertainty. Hence, the u(D)/u(H) ratio allows us to compare the relative errors between deuterated and un-deuterated species.

A majority of deuterated species show larger abundance uncertainties with respect to their un-deuterated analogues. There are two major reasons for this behavior. First, the majority of reactions with deuterated species originate from our cloning procedure, and thus have larger assumed uncertainties. Second, to produce a deuterium isotopologue of a molecule additional chemical pathways (e.g., isotope exchange processes) are required, increasing the accumulation of rate uncertainties. For many hydrocarbons (C$_{n}$H$_{m}$, C$_{n}$H$_{m}$\jon~with $m,n = 2,3$) the abundance uncertainties of their deuterated isotopologues are comparable to those of the main isotopologues (with ratios of $\sim 0.7-2$). These hydrocarbons form through ions of hydrocarbons reacting with H\dtwo~or smaller neutral hydrocarbons, such as CH$_4$ and C\dtwo H\dthree, and their reactions originate purely from the cloning procedure.

For a limited number of species, the abundance uncertainties of their deuterated isotopologues are even smaller than for the main isotopologues (with ratios of $\sim 0.7-0.9$; see Figure~\ref{fig:diffDist}), e.g., C\dtwo D\jon~and D\ijon. These are simple radicals and ions produced by a limited set of reactions, with relatively well-known rate coefficients and thus small uncertainties. The relevant deuterium fractionation chemistry is also limited and has comparably low uncertainties ($\sim 0.7 - 1.1$). Note that the spread in abundance uncertainties ratios appears to decrease with increasing number of atoms in species, getting closer to unity. This effect in Fig.~\ref{fig:diffDist} occurs because we plot only species with relative abundances exceeding $10^{-25}$ (with respect to hydrogen), whose numbers decrease substantially with size. If we also add species with such low abundances, this feature disappears and the trend in the uncertainties between small and large molecules is similar. 

{For} HD and D\dtwo, we find that the ratio of their abundance uncertainties to that of H$_2$ can reach very large values, $\sim1\, 000$ and $\sim10\, 000$, respectively. This effect occurs because abundances of H\dtwo~are very well constrained (uncertainties are $\sim 10^{-5}$), while HD and D\dtwo~have typical values of abundance uncertainties up to one order of magnitude.

\begin{figure}[!htb]
\centering
\includegraphics[width=0.35\textwidth, angle=90]{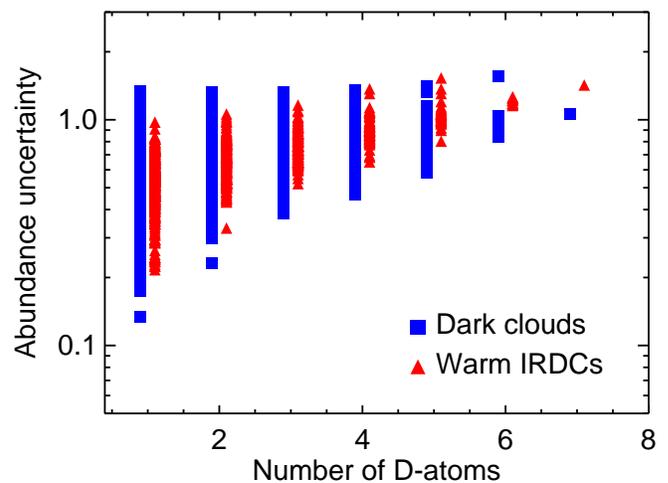}
\caption{Abundance uncertainties {in orders of magnitude} for deuterated species as a function of number of D-atoms. The dark cloud model results are denoted by squares and the warm infrared dark cloud model results by triangles. A colored version of the plot is available in the online version. }
\label{fig:DlvlDisp}
\end{figure}

{Finally, in Figure~\ref{fig:DlvlDisp} we plot the abundance uncertainties of deuterated species as a function of number of D-atoms, once again restricted to species with relative abundances $> 10^{-25}$. We see an increase in uncertainty with number of D-atoms, as expected from the increasing number of reactions involved for subsequently adding more D-atoms. We also notice a wider spread in uncertainties for species with lower levels of deuteration because smaller species, such as OD, HDO, DCO\jon, often have lower uncertainties because there are not as many steps involved in their formation compared to larger species. Larger multiply-deuterated species cannot have these low uncertainties as there are too many steps involved in their formation. Once again we also find that dark clouds have a larger spread in uncertainties than warm IRDCs. Again we argue that this difference occurs because dark cloud environments overall have higher abundances of deuterated species which are processed through more reactions.}

\subsection{General trends in D/H distributions}
\label{sec:gen_trends}

\begin{figure*}[!htb]
\centering
\includegraphics[width=0.95\textwidth]{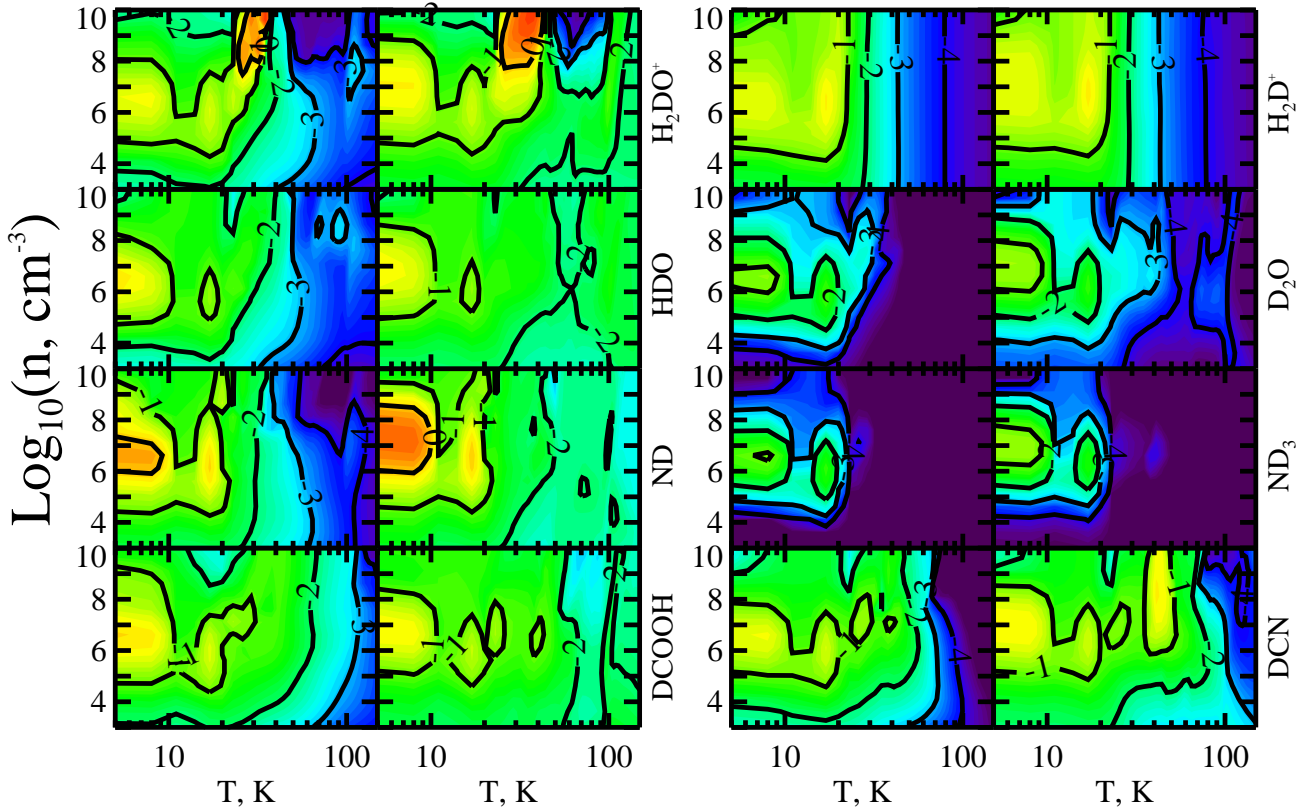}
\caption{Plots of the D/H distributions at 1~Myr for our models. Each panel is a contour plot displaying the distribution of specific logarithmic D/H ratios. In the online version, fractionation levels with elemental (low), intermediate, and high D/H fractions are indicated with a continuous scale starting with dark blue ($\leq 10^{-5}$), green (10$^{-2}$), yellow (10$^{-1}$), and red ($\geq 1$) colors. If we label the columns as 1, 2, 3, and 4 from left-to-right, columns 1 and 3 show results from the  ``Primordial'' model, while columns 2 and 4 show results from the evolutionary model. Columns 1 and 2 show ratios with significant differences from the norm in their strong dependence on initial abundance, where significant differences of more than one order of magnitude are also visible at high temperatures $>$ 100 K. Columns 3 and 4 show species with an unusually strong dependence on temperature (with elemental D/H ratio of $\sim 1.5\,10^{-5}$ achieved either at low, $\sim 20-40$~K or high, $\sim 80$~K, temperatures).}
\label{fig:modelCompare}
\end{figure*}

\begin{table*}
\centering
\caption{Species showing strong dependences on initial abundances or temperature \label{tab:deps}}
\begin{tabular}{c|c|c}
\hline
\tablewidth{0.9\textwidth}
Initial abundances										&				\multicolumn{2}{c}{~~~~~~~~~~Temperatures}										\\
													&				$\sim 30$~K							&		$\sim 80$~K					\\
\hline
C\dtwo HD, C\dthree HD, C$_{6}$D, CD, CD\dtwo, CD\dthree OH			&	C\dtwo D\dtwo, C\dtwo HD\jon, CD$_4$					&	C\dtwo HD\jon, CH\dtwo D\jon, CHD\dtwo\jon	\\
CH\dtwo D\dtwo, CH\dtwo DCN, CH\dtwo DOH, CHD\dtwo OH	&	CHD, CHD\dthree, N\dtwo D\jon 						&	CH\dthree D		\\
CHDCO, D\dtwo CO, D\dtwo CS, DC\dthree N, DC$_5$N		&	D\dtwo O, D\dtwo S, D\dthree\jon, D\dthree O\jon, DCN,		& 							\\
DCOOCH\dthree, DCOOH, DCS\jon, DNCO, H\dtwo DO\jon				&	DCO\jon, DNC, DOC\jon, H\dtwo D\jon					&							\\	
HDCO, HDO, HDS, ND, NH\dtwo D, NHD\dtwo, OD				&	HD\dtwo\jon, HD\dtwo O\jon, N\dtwo D\jon, ND\dthree		&							\\
\hline							
\end{tabular}
\end{table*}
In this section, we discuss general trends in the modeled D/H ratios in our 2D-parameter space (see Section~\ref{sec:phys_model}). The computed D/H fractionation ratios at the final time of 1~Myr are shown in Figure~\ref{fig:modelCompare} for the { ``Primordial''  (left panels of each separate block) and ''Evolution'' model (right panels of each separate block), which were introduced in Section~\ref{sec:initabund},} as functions of density and temperature for the following key gaseous molecules: H$_{2}$DO$^{+}$, H$_{2}$D$^{+}$, HDO, D$_{2}$O, ND, ND$_{3}$, DCOOH and DCN. For most species, the D/H ratios can reach high values of $\ga 10^{-3}$ at $T\la 30-80$~K. At higher temperatures, $\ga 100$~K, the computed D/H ratios begin to approach the elemental ratio of $\approx 1.5\times 10^{-5}$. The higher D/H ratios of $\ga 0.1$ have been observed for many species in the ISM, such as CH$_2$DOH, D$_2$CO \citep{2002P&SS...50.1267C}, D$_2$O \citep{2007ApJ...659L.137B}, H\dtwo D\jon~\citep{2003A&A...403L..37C}, HDO \citep{2011A&A...527A..19L}, NH$_2$D \citep{2003A&A...403L..25H, 2005A&A...438..585R}, NHD$_{2}$ \citep{2005A&A...438..585R}. For comparison of our model results with observations, see Section~\ref{sec:observe}.

We isolate species that are either mostly sensitive to kinetic temperature or initial abundances, {the former referring to the standard temperature dependence for deuterium fractionation, and} {the latter referring to the difference between the initial abundances in the ``Primordial'' and evolutionary models.} We note that usually species for which D/H ratios depend {more strongly} on the initial abundances also show (a weaker) temperature dependence{, because the release of CO and other radicals decreases the abundances of the H\dthree\jon~isotopologues which lowers the efficiency of transferring D atoms to other species}. These two groups are listed in Table~\ref{tab:deps}. The temperature-dependent species can be further divided into two subgroups by the temperature where D/H ratios decrease most sharply, and hence where deuterium fractionation becomes less pronounced: (1) at low temperatures, $T \sim 20-40$~K (related to the fractionation via H$_3^+$ isotopologues) and (2) at higher temperatures $T \sim 80$~K (related to the fractionation via CH$_2$D$^+$ and C$_2$HD$^+$). 

The left two columns of Figure~\ref{fig:modelCompare} show species that demonstrate a strong dependence of D/H ratios on the initial abundances while the right two columns show species with a strong temperature gradient for D/H ratios and no distinct dependence on initial abundances. {We define here a strong dependence on the initial abundance as an overall variation by a factor of 5 in D/H ratios at temperatures $> 100$ K for the final time step. For the case of HDO in Figure~\ref{fig:modelCompare}, D/H ratios are approximately 2 orders of magnitude higher in the ``Evolution'' model compared to the  ``Primordial''  model at temperatures $> 100$ K and hence its D/H ratios are considered to depend strongly on the initial abundance.} Frozen molecules show D/H distributions similar to their gas-phase analogues as we do not specifically consider selective substitution of H by D in surface species (yet we do consider production of deuterated molecules by surface chemistry). Therefore, we do not discuss the distribution of D/H for ices separately. 

\begin{figure}[!htb]
\centering
\includegraphics[width=0.47\textwidth]{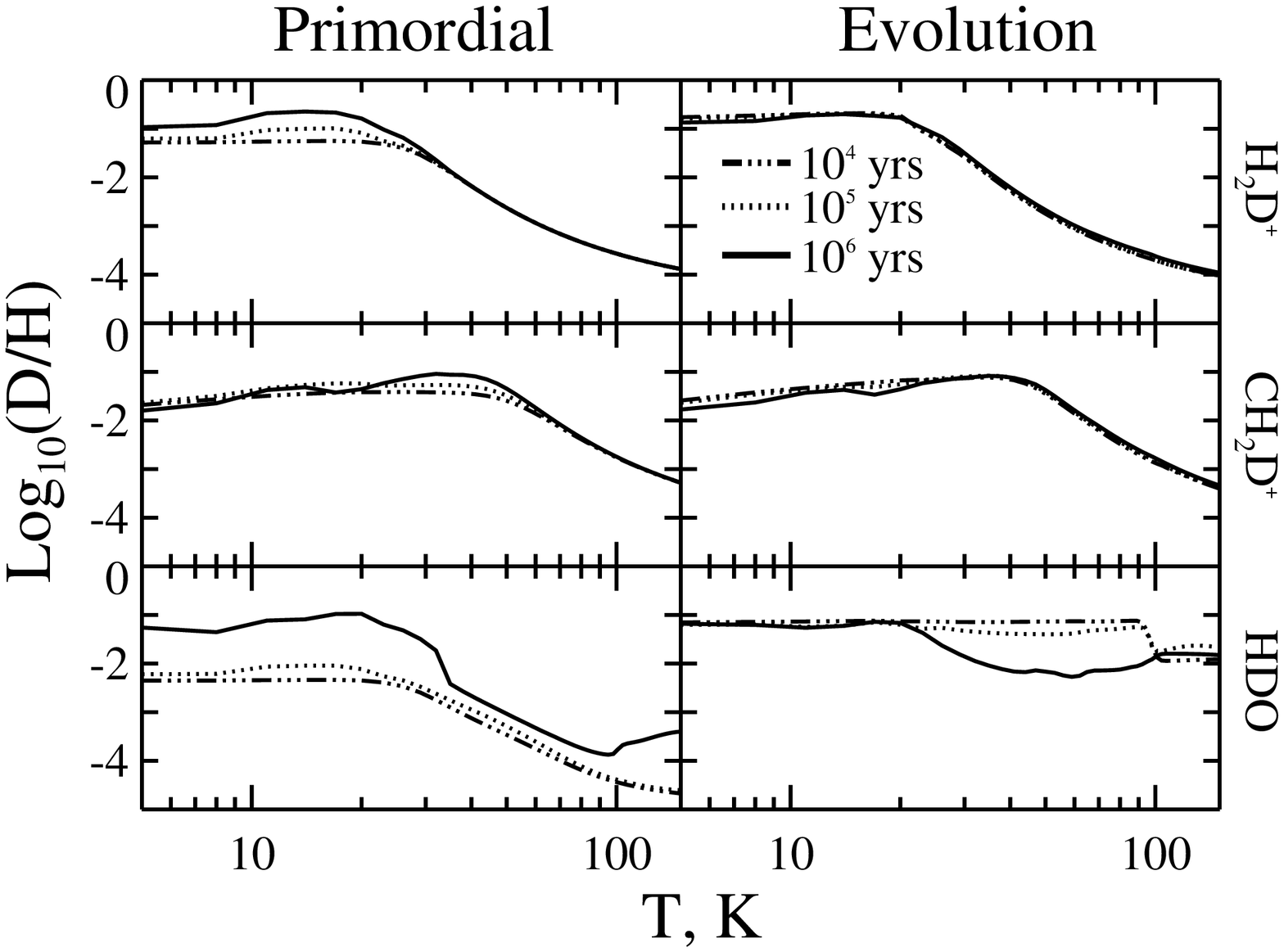}\\
\caption{D/H ratios for H\dtwo D\jon, CH\dtwo D\jon~and HDO as a function of temperature at a density $n_{\rm H} = 10^4$ cm$^{-3}$ and for three specific times: $10^4, 10^5, 10^6$ years. \label{fig:IAcompares}}
\end{figure}

The placement of species in different groups can be explained by the relative pace of their chemical evolution. If steady-state abundances are reached by 1 Myr in the ``Primordial'' stage, then they remain at steady state throughout the evolutionary model unless the physical conditions are changed. In Figure~\ref{fig:IAcompares} we show the evolution of D/H ratios at three separate times of 10$^4, 10^5, 10^6$ years, and at a density $n_{\rm H} = 10^4$ cm$^{-3}$ as a function of temperature for H\dtwo D\jon, CH\dtwo D\jon~and HDO. We discuss the specific case for the evolution of HDO later and begin with the two ions (two top panels) which are highly reactive, with fast chemical timescales associated with ion-neutral reactions and dissociative recombinations. Thus, the chemical ``memory'' of the pristine state of the fast-evolving deuterated species is completely lost at the final considered time of 1~Myr, with no apparent differences between abundances computed with the two distinct initial abundance sets.

{We find that the majority of neutral species, which includes several H-, C-, O-, N- and S-bearing species, show a dependence on the initial abundance as their evolution is significantly slower compared to ions. These are predominantly neutral species unless they are strongly associated with one of the major ions (such as H\dthree\jon, CH\dthree\jon, HCO\jon~and their isotopologues). This is illustrated in Figure~\ref{fig:IAcompares} for HDO (bottom panel), where we see differences in D/H ratios of $\sim$ 3 orders of magnitude between the two models at higher temperatures approaching 100 K. In the  ``Primordial''  model we start with atomic abundances and the water D/H ratios are lower for ices than for gas vapor because. This is because the formation of key ingredients for HDO ice through OD + H or OH + D is slower gas-phase fractionation via ion-molecule processes. In contrast, the ``Evolution'' model starts with high abundances and D/H ratios of water, mostly frozen onto grains. At low temperatures $<$ 30 K water quickly reach the steady-state abundances while at higher temperatures, because gas-phase fractionation is not efficient, the D/H ratios of water vapor decreases with increasing temperature. Desorption play an important role at temperatures above $20-30$~K where, even in highly obscured regions, CRP-driven desorption occurs. At temperatures $\gtrsim 100$ K we see a drop in D/H ratios down to similar values of water ice because desorption replenishes the water vapor constantly and the water vapor also inherits the same D/H ratios as the ice ($\sim 10^{-2}$). }

H\dtwo DO\jon~was the only ion we found showing a significant difference in the distribution between the two sets of initial abundances, which is related to the protonation of HDO by H\dthree\jon~and which thus behaves chemically similar to HDO (see above). Complex neutral species, which are not a direct outcome of fast dissociative recombination reactions, evolve more slowly via neutral-neutral and surface processes, and thus show a dependence on the adopted initial abundances. The high initial abundances of deuterated species accumulated at 10~K in the ``Evolution'' model enable rich deuterium chemistry during the entire time span of 1~Myr even at warm temperatures of $T\la 50-100$~K. As a result, some species show almost uniformly high D/H ratios of $\gtrsim 10^{-3}$ at $T =10-150$~K (see Figure~\ref{fig:modelCompare}, left column). On the other hand, the same species in the  ``Primordial''  model (with only HD available initially) show a significant decrease in the final D/H ratios toward warmer temperatures ($\gtrsim 40-80$~K). The effect is strong for saturated species that are at least partly formed on dust surfaces, and thus are sensitive to the choice of the initial abundances (and temperature), such as water or ammonia isotopologues. Also, {most} multiply-deuterated species, whose formation involves several isotope exchange reactions, show strong dependence on the initial abundances. 

Ions constitute the majority of species for which calculated D/H ratios are only dependent on temperature (see Figure~\ref{fig:modelCompare}, right two columns). Species such as the H\dthree\jon~isotopologues that are sensitive to freeze-out (or specifically, the freeze-out of CO), ``daughter'' molecules such as DCO$^+$, as well as (ionized) light hydrocarbons related to CH$_2$D$^+$ and C$_2$HD$^+$, all show a strong temperature dependence. From the exact gas kinetic temperature, labeled the critical temperature, at which the D/H ratios start to approach the elemental value, we can separate these species into two subgroups. The deuterated species formed via low-temperature fractionation channels involving isotopologues of H$_3^+$ have a critical temperature of $20 - 30$~K, whereas other deuterated species synthesized via high-temperature fractionation channels involving CH$_2$D$^+$ and C$_2$HD$^+$ show a higher critical temperature of $\sim 80$~K \citep{2009A&A...508..737P}. Most multi-deuterated species belong exclusively to the low-temperature group as do the majority of ions, which inherit the temperature dependence from the H\dthree\jon~isotopologues via proton/deuteron transfer reactions. The best example of such species is H\dtwo D\jon, shown in Figure~\ref{fig:modelCompare}. It shows a D/H turnover point at 20~K, after which the fractionation ratios decrease smoothly and reach levels of $\sim 10^{-5}$ at temperatures of $\mathord{\sim}$100~K. 

There are only a few species that show a dependence on temperature where deuterium fractionation is reduced at $T \sim 80$ K. Those are directly synthesized from deuterated light hydrocarbon ions such as CH\dtwo D\jon, which retain high D/H ratios at elevated temperatures, e.g. DCN and CH$_2$D$_2$. HDCS is unique in showing no clear difference between the two models at least within our defined temperature range, as its chemical evolution depends almost equally on the two distinct deuterium fractionation routes (via H$_3^+$ and light hydrocarbon isotopologues) via the formation of the ion H\dtwo DCS\jon,

Also, we find that the overall sensitivity of the calculated D/H ratios to density is weak, except for species sensitive to the freeze-out of CO and other radicals, i.e. H$_2$D$^+$, HD$_2^+$, and their direct dissociative recombination products. These species experience a rapid decrease in fractionation ratios toward lower densities of $< 10^{4}$ cm$^{-3}$, at which CO cannot severely deplete onto the dust grains even after 1~Myr of evolution. However, if we would follow the evolution of our chemical model beyond 1~Myr, at some moment the CO depletion can become severe enough for low-density regions to allow high D/H ratios for the H$_3^+$ isotopologues (if the adopted non-thermal desorption rate is not too high). 

To understand in more detail the differences between the evolution of deuterated molecules, we have analyzed the chemistry of several key isotopologues. We considered four distinct astrophysical environments: densities of 10$^4$ and 10$^8$~cm$^{-3}$, and temperatures of 10 and 80~K for both the  ``Primordial''  and ``Evolution'' initial abundance sets. We list key formation and destruction pathways for the assorted species and their main reactants as online material (see Appendix~\ref{sec:Dominant}).

\section{Discussion}
\label{sec:discussion}
In recent years, it has been realized that the ortho/para ratio of H$_{2}$ and other species in a source can strongly affect the degree of deuterium fractionation. For example, the backward endothermic reaction between o-H$_{2}$ and H$_{2}$D$^{+}$ can proceed far more rapidly at low temperatures such as 10 K than the corresponding reaction involving p-H$_{2}$, and so reduce the degree of deuterium fractionation if there is a sufficient amount of o-H$_{2}$ \citep[e.g.,][]{1990MNRAS.247..500F, 1992A&A...258..479P, 2006A&A...449..621F, 2010A&A...509A..98S, 2013A&A...551A..38P}. The fraction of H$_{2}$ in its ortho levels is small but not well known, however, and can only be obtained by careful analysis of reduced fractionation levels in specific sources. Given the complexity and huge size of our chemical network, with many reactions containing ortho and para reactants, we chose not to include the different states of deuterated and non-deuterated species for the moment, and leave this to a separate study. Thus, our network is rather accurate for modeling dense, cold ISM phases, where H$_2$ mainly exists in its para state \citep[e.g.,][]{2009A&A...494..623P} and higher temperature phases where the degree of fractionation is in any case low.

Another important effect concerning deuterium fractionation that is not fully treated by our approach is the evolution of physical conditions, such as a steady warm-up phase of a protostellar envelope leading to the formation of a hot core/corino. Our simple evolutionary model, where we simulate a TMC-1 environment for 1 Myr and use these final abundances as the initial abundances for the modeling is only a ``poor man's'' approach for such a study \citep[see, e.g.,][]{2011ApJ...741L..34C,2012ApJ...760...40A}. Certainly this abrupt change from TMC-1 temperature and density to other temperatures and densities does not catch the gradual evolution of the physical conditions from one phase to the next. As we showed in Section~\ref{sec:results}, D/H ratios for a large number of the deuterated species are mainly dependent on temperature, while density has an effect on the evolution of a limited number of species. The gradual increase in temperature throughout the collapse of a dark cloud and the formation of a (proto)star will undoubtedly affect the evolution of some deuterated species in ways uncatchable by our approach. Examples include the release of key ice species into the gas-phase or the steady increase in mobility of surface radicals, which produce complex organic molecules \citep{2008ApJ...682..283G}. However, the effects on predicted D/H ratios are likely to be only minor as the final stages of collapse are rapid, so that deuterium fractionation of the gas-phase species stays until the material winds up in the actual protostar or in the protoplanetary disk surrounding it \citep{2008ApJ...674..984A, 2012ApJ...760...40A}. 

\subsection{Observations}
\label{sec:observe}

\begin{table}
\caption{Observable deuterated species with ALMA in dark clouds. \label{tab:mostDeuteratedALMAdc}}
\tiny
\begin{tabular}{lccccc}	
\hline
\tablewidth{0.30\textwidth}
Species			&	 n(x)/n(H\dtwo) [$\times 10^{-11}$]	& 	Line							& 	Frequency [GHz]	& 	Line flux [mK]	&	ALMA band	\\
\hline
D\dtwo CO		&	1.05							&	3 0 3 	$\rightarrow$~2 0 2		&	174.413			&	33.48		&	5	\\
				&								&	4 0 4 	$\rightarrow$~3 0 3		&	231.410			&	28.48		&	6	\\
				&								&	5 0 5		$\rightarrow$~4 0 4		&	287.486			&	14.99		&	7	\\
D\dtwo O			&	53.6							&	1 1 0 	$\rightarrow$~1 0 1		&	316.800			&	1899.80		&	7	\\
				&								&	2 0 2 	$\rightarrow$~1 1 1 		&	468.247			&	853.66		&	8	\\
				&								&	2 1 1 	$\rightarrow$~2 0 2		&	403.562			&	198.18		&	8	\\
D\dtwo S			&	0.0634						&	2 2 0 	$\rightarrow$~1 1 1 		&	669.787			&	1.40			&	9	\\
{DCN}$^{*}$	&	5.21							&	1		$\rightarrow$~0		&	115.271			&	6648.00		&	3	\\
				&								&	2		$\rightarrow$~1		&	230.538			&	5272.00		&	6	\\
				&								&	3		$\rightarrow$~2		&	345.796			&	3860.00		&	7	\\
{DCO\jon}	&	16.04						&	2		$\rightarrow$~1		&	144.077			&	2196.40		&	4	\\
				&								&	4		$\rightarrow$~3		&	288.144			&	2054	.53		&	7	\\
				&								&	5		$\rightarrow$~4		&	360.170			&	801.50		&	7	\\
{DNC}		&	11.01						&	2		$\rightarrow$~1		&	152.610			&	1047.68		&	4	\\
				&								&	4		$\rightarrow$~3		&	305.207			&	866.60		&	7	\\
				&								&	6		$\rightarrow$~5		&	457.776			&	66.33		&	8	\\
{H\dtwo DO\jon}&	29.93						&	1 0 1 1 	$\rightarrow$~2 1 1 0	&	250.914			&	9.84			&	6	\\
{HDCO}		&	24.69						&	4 0 4		$\rightarrow$~3 0 3		&	256.586			&	616.96		&	6	\\
				&								&	2 0 2 	$\rightarrow$~1 0 1		&	128.813			&	513.23		&	4	\\
				&								&	3 1 2 	$\rightarrow$~2 1 1		&	201.341			&	313.03		&	5	\\
HDCS			&	0.7897						&	5 0 5		$\rightarrow$~4 0 4		&	154.885			&	4.12			&	4	\\
				&								&	3 0 3		$\rightarrow$~2 0 2		&	92.982			&	3.12			&	3	\\
				&								&	6 0 6		$\rightarrow$~5 0 5		&	185.693			&	3.01			&	5	\\
HDS				&	1.630						&	2 0 2		$\rightarrow$~1 0 1		&	477.764			&	25.62		&	8	\\
				&								&	1 0 1 	$\rightarrow$~0 0 0 		&	244.556			&	18.22		&	6	\\
				&								&	1 1 0		$\rightarrow$~1 0 1		&	195.559			&	9.05			&	5	\\
ND				&	27.03						&	1 0 1 2	$\rightarrow$~0 1 2 3	&	491.934			&	294.70		&	8	\\		
				&								&	1 0 1 2	$\rightarrow$~0 1 1 2	&	491.969			&	124.38		&	8	\\
				&								&	1 0 1 1	$\rightarrow$~0 1 2 2	&	491.917			&	123.84		&	8	\\
ND\dthree			&	0.96							&	2 0 0 	$\rightarrow$~1 0 1		&	614.933			&	73.05		&	9	\\
				&								&	2 1 1 	$\rightarrow$~1 1 0		&	618.125			&	60.59		&	9	\\
				&								&	2 1 0		$\rightarrow$~1 1 1		&	614.968			&	59.74		&	9	\\
{NH\dtwo D}	&	296.70						&	1 1 0 1	$\rightarrow$~0 0 0 0 	&	494.455			&	10000.00		&	8	\\
				&								&	1 1 1 0	$\rightarrow$~1 0 1 1	&	85.926			&	870.34		&	3	\\
				&								&	1 0 1 1	$\rightarrow$~0 0 0 1	&	332.782			&	732.35		&	7	\\
{NHD\dtwo}	&	14.60						&	2 1 1 0	$\rightarrow$~1 0 1 1	&	699.224			&	874.07		&	9	\\
				&								&	2 1 1 1	$\rightarrow$~1 0 1 0	&	709.350			&	456.67		&	9	\\
				&								&	1 1 1 0	$\rightarrow$~0 0 0 0	&	335.514			&	40.68		&	7	\\						
			\hline\\[-7.0pt]
\multicolumn{6}{l}{The table is limited to species with relative abundances $> 10^{-12}$.}\\
\multicolumn{6}{l}{Species detected unambiguously or in a preliminary manner in the specific environments are marked in boldface.}\\
\multicolumn{6}{l}{Parameters for line flux calculations: 10~K, $10^{4}$ cm$^{-3}$, 10$^{22}$ cm$^{-2}$}\\
\multicolumn{6}{l}{$^{*}$ DCN should be observable, but is not listed in CDMS or RADEX, instead calculated RADEX values for HCN are listed. }
\end{tabular}
\end{table}

\begin{table}
\caption{Observable deuterated species with ALMA in infrared dark clouds. \label{tab:mostDeuteratedALMAirdc}}
\tiny
\begin{tabular}{lccccc}	
\hline
\tablewidth{0.30\textwidth}
Species			&	 n(x)/n(H\dtwo) [$\times 10^{-11}$]	& 	Line							& 	Frequency [GHz]	& 	Line flux [mK]	&	ALMA band	\\
\hline
C$_4$D			&	0.039						&	11 12 	$\rightarrow$~10 11		&	97.140			&	2.36			&	3	\\
				&								&	12 13	$\rightarrow$~11 12		&	105.971			&	2.35			&	3	\\
				&								&	10 11 	$\rightarrow$~9 10		&	88.308			&	2.25			&	3	\\
CH\dtwo D\jon		&	0.045						&	2 1 2		$\rightarrow$~1 1 1		&	490.012			&	44.05		&	8	\\
				&								&	1 0 1		$\rightarrow$~0 0 0		&	278.692			&	10.68		&	7	\\
				&								&	2 1 1		$\rightarrow$~2 1 2		&	201.754			&	1.62			&	5	\\
CH\dtwo DCN		&	0.0038						&	9 0 9		$\rightarrow$~8 0 8		&	156.281			&	7.47			&	4	\\
				&								&	10 0 10	$\rightarrow$~9 0 9		&	173.639			&	6.73			&	5	\\
				&								&	6 0 6		$\rightarrow$~5 0 5		&	104.198			&	6.00			&	3	\\
D\dtwo CO		&	0.011						&	5 0 5		$\rightarrow$~4 0 4		&	287.486			&	63.48		&	7	\\
				&								&	6 0 6		$\rightarrow$~5 0 5		&	342.522			&	48.67		&	7	\\
				&								&	3 0 3		$\rightarrow$~2 0 2		&	174.413			&	45.25		&	5	\\
D\dtwo O			&	3.10							&	1 1 0		$\rightarrow$~0 0 0		&	317.800			&	30000.00		&	7	\\
				&								&	2 0 2		$\rightarrow$~1 0 1		&	468.247			&	30000.00		&	8	\\
				&								&	2 1 1 	$\rightarrow$~1 0 0		&	403.562			&	24681.37		&	8	\\
{DCN}$^{*}$	&	0.38							&	1		$\rightarrow$~0		&	88.634			&	21800.00		&	3	\\
				&								&	2		$\rightarrow$~1		&	177.258			&	20590.00		&	5	\\
				&								&	3		$\rightarrow$~2		&	265.886			&	19040.00		&	6	\\
{DCO\jon	}	&	47.09						&	5		$\rightarrow$~4		&	360.170			&	30000.00		&	7	\\
				&								&	3		$\rightarrow$~2		&	216.113			&	30000.00		&	6	\\
				&								&	6		$\rightarrow$~5		&	432.189			&	30000.00		&	8	\\
{DNC}		&	0.81							&	3		$\rightarrow$~2		&	228.910			&	24216.75		&	6	\\
				&								&	6		$\rightarrow$~5		&	457.776			&	14474.58		&	8	\\
				&								&	2		$\rightarrow$~1		&	152.610			&	11600.74		&	4	\\
DOC\jon			&	0.00062						&	3		$\rightarrow$~2		&	229.149			&	15.36		&	6	\\
				&								&	5		$\rightarrow$~4		&	318.885			&	15.26		&	7	\\
				&								&	6		$\rightarrow$~5		&	458.237			&	9.15			&	8	\\
H\dtwo DO\jon		&	3.39							&	1 0 1 1	$\rightarrow$~2 1 1 0	&	250.914			&	2582.00		&	6	\\
				&								&	3 3 1 0	$\rightarrow$~2 2 1 1	&	649.653			&	368.10		&	9	\\
				&								&	3 3 0 0	$\rightarrow$~2 2 0 1	&	632.902			&	342.59		&	9	\\
HDCO			&	0.10							&	5 0 5		$\rightarrow$~4 0 4		&	319.770			&	629.27		&	7	\\
				&								&	3 0 3		$\rightarrow$~2 0 2		&	192.893			&	502.15		&	5	\\
				&								&	4 1 3		$\rightarrow$~3 1 2		&	268.292			&	426.06		&	6	\\
HDCS			&	0.0065						&	7 0 7		$\rightarrow$~6 0 6		&	216.662			&	6.88			&	6	\\
				&								&	5 0 5		$\rightarrow$~4 0 4		&	154.885			&	6.46			&	4	\\
				&								&	8 0 8		$\rightarrow$~7 0 7		&	247.488			&	5.69			&	6	\\
HDS				&	0.018						&	1 1 1		$\rightarrow$~0 0 0		&	389.041			&	70.17		&	8	\\
				&								&	2 0 2		$\rightarrow$~1 0 1		&	477.764			&	62.06		&	8	\\
				&								&	3 0 3		$\rightarrow$~2 0 2		&	691.498			&	45.74		&	9	\\
ND				&	0.16							&	1 0 1 2	$\rightarrow$~0 1 2 3	&	491.934			&	357.79		&	8	\\
				&								&	1 0 1 2	$\rightarrow$~0 1 1 2	&	491.969			&	150.99		&	8	\\
				&								&	1 0 1 1	$\rightarrow$~0 1 2 2	&	491.917			&	150.36		&	8	\\
ND\dthree			&	0.0080						&	2 1 1		$\rightarrow$~1 1 0		&	618.125			&	125.84		&	9	\\
				&								&	2 1 0		$\rightarrow$~1 1 1		&	614.968			&	124.41		&	9	\\
				&								&	1 0 1		$\rightarrow$~0 0 0		&	309.909			&	68.34		&	7	\\
{NH\dtwo D}	&	27.15						&	1 1 0 1	$\rightarrow$~0 0 0 0	&	494.455			&	30000.00		&	8	\\
				&								&	2 0 2 1	$\rightarrow$~1 0 1 1	&	649.916			&	17102.87		&	9	\\
				&								&	1 1 1 0	$\rightarrow$~1 0 1 1	&	85.926			&	16766.37		&	3	\\
NHD\dtwo			&	0.45							&	1 1 0 1	$\rightarrow$~0 0 0 0	&	388.652			&	4199.24		&	8	\\
				&								&	2 1 1 1	$\rightarrow$~1 0 1 0	&	709.350			&	3793.42		&	9	\\
				&								&	2 0 2 0	$\rightarrow$~1 1 0 1	&	410.491			&	296.35		&	8	\\
			\hline\\[-7.0pt]
\multicolumn{6}{l}{The table is limited to with relative abundances $> 10^{-12}$.}\\
\multicolumn{6}{l}{Species detected unambiguously or in a preliminary manner in the specific environments are marked in boldface.}\\
\multicolumn{6}{l}{Parameters for line flux calculations: 25~K, $10^{5}$ cm$^{-3}$, 10$^{24}$ cm$^{-2}$}\\
\multicolumn{6}{l}{$^{*}$ DCN should be observable, but is not listed in CDMS or RADEX, instead calculated RADEX values for HCN are listed. }

\end{tabular}
\end{table}

\begin{table}
\caption{Observable deuterated species with ALMA in high-mass protostellar objects. \label{tab:mostDeuteratedALMAhmpo}}
\tiny
\begin{tabular}{lccccc}	
\hline
\tablewidth{0.30\textwidth}
Species			&	 n(x)/n(H\dtwo) [$\times 10^{-11}$]	& 	Line							& 	Frequency [GHz]	& 	Line flux [mK]	&	ALMA band\\
\hline
C$_4$D			&	0.95							&	23 24 	$\rightarrow$~22 23		&	203.104			&	28.13		&	5	\\
				&								&	22 23	$\rightarrow$~21 22		&	194.275			&	27.98		&	5	\\
				&								&	24 25	$\rightarrow$~23 24		&	211.932			&	27.96		&	6	\\
C$_6$D			&	0.018						&	42 -1 43	$\rightarrow$~41 1 42	&	112.462			&	1.97			&	3	\\
				&								&	41 -1 42	$\rightarrow$~40 1 41	&	109.856			&	1.97			&	3	\\
				&								&	41 1 42	$\rightarrow$~40 -1 41	&	109.816			&	1.97			&	3	\\
D\dtwo CO		&	0.0017						&	8 0 8		$\rightarrow$~7 0 7		&	499.596			&	2.44			&	8	\\
				&								&	11 0 11	$\rightarrow$~10 0 10	&	605.672			&	2.21			&	9	\\
				&								&	7 0 7		$\rightarrow$~6 0 6		&	396.517			&	2.18			&	8	\\
D\dtwo CS		&	0.000658						&	12 0 12	$\rightarrow$~11 0 11	&	337.599			&	1.79			&	7	\\
				&								&	14 0 14	$\rightarrow$~13 0 13	&	392.631			&	1.78			&	8	\\
				&								&	11 0 11	$\rightarrow$~10 0 10	&	309.916			&	1.71			&	7	\\
D\dtwo O			&	1.87							&	2 1 1		$\rightarrow$~2 0 2		&	403.562			&	5768.78		&	8	\\
				&								&	4 2 2		$\rightarrow$~3 1 1		&	692.244			&	3398.71		&	9	\\
				&								&	2 0 2		$\rightarrow$~1 0 1		&	468.247			&	3334.45		&	8	\\
{DCN}$^{*}$	&	0.21							&	1		$\rightarrow$~0		&	88.634			&	71760.00		&	3	\\
				&								&	2		$\rightarrow$~1		&	177.258			&	70430.00		&	5	\\
				&								&	3		$\rightarrow$~2		&	265.886			&	68680.00		&	6	\\
DCO\jon			&	23.91						&	9		$\rightarrow$~8		&	648.193			&	75000.00		&	9	\\
				&								&	6		$\rightarrow$~5		&	432.189			&	75000.00		&	8	\\
				&								&	5		$\rightarrow$~4		&	360.170			&	75000.00		&	7	\\
DCS\jon			&	0.0023						&	12		$\rightarrow$~11		&	432.338			&	5.49			&	8	\\
				&								&	10		$\rightarrow$~9		&	360.307			&	5.27			&	7	\\
				&								&	13		$\rightarrow$~12		&	468.347			&	5.23			&	8	\\
{DNC}		&	0.41							&	9		$\rightarrow$~8		&	686.553			&	7042.88		&	9	\\
				&								&	6		$\rightarrow$~5		&	457.776			&	6233.04		&	8	\\
				&								&	4		$\rightarrow$~3		&	305.206			&	3003.38		&	7	\\
H\dtwo DO\jon		&	2.27							&	3 3 1 0	$\rightarrow$~2 2 1 1	&	649.653			&	2352.22		&	9	\\
				&								&	3 3 0 0	$\rightarrow$~2 2 0 1	&	632.902			&	2258.85		&	9	\\
				&								&	4 1 3 0	$\rightarrow$~3 0 3 1	&	716.959			&	1629.36		&	9	\\
HDCO			&	0.093						&	10 0 10	$\rightarrow$~9 0 9		&	625.688			&	136.82		&	9	\\
				&								&	7 0 7		$\rightarrow$~6 0 6		&	444.229			&	134.73		&	8	\\
				&								&	8 1 8		$\rightarrow$~7 1 7		&	491.937			&	129.31		&	8	\\
HDCS			&	0.56							&	13 0 13	$\rightarrow$~12 0 12	&	400.766			&	151.96		&	8	\\
				&								&	11 0 11	$\rightarrow$~10 0 10	&	339.646			&	148.23		&	7	\\
				&								&	14 0 14	$\rightarrow$~13 0 13	&	431.218			&	145.25		&	8	\\
ND				&	3.70							&	1 0 1 2	$\rightarrow$~0 1 2 3	&	491.934			&	911.92		&	8	\\				
				&								&	1 0 1 2	$\rightarrow$~0 1 1 2	&	491.969			&	384.82		&	8	\\
				&								&	1 0 1 1	$\rightarrow$~0 1 2 2	&	491.917			&	383.23		&	8	\\
ND\dthree			&	0.0031						&	2 1 1		$\rightarrow$~1 0 0		&	618.125			&	4.24			&	9	\\
				&								&	2 1 0		$\rightarrow$~1 0 0		&	614.968			&	4.20			&	9	\\
				&								&	1 0 1		$\rightarrow$~0 0 0		&	309.909			&	1.14			&	7	\\
NH\dtwo D		&	25.07						&	1 1 0 1	$\rightarrow$~0 0 0 0	&	494.455			&	8876.12		&	8	\\
				&								&	3 3 0 0	$\rightarrow$~3 2 2 1	&	618.142			&	5010.99		&	9	\\
				&								&	2 2 0 1	$\rightarrow$~2 1 2 0	&	475.890			&	4998.31		&	8	\\
NHD\dtwo			&	0.37							&	2 1 1 1	$\rightarrow$~1 0 1 0	&	709.350			&	301.28		&	9	\\
				&								&	1 1 0 1	$\rightarrow$~0 0 0 0	&	388.652			&	158.40		&	8	\\
				&								&	3 1 2 1	$\rightarrow$~2 2 0 0	&	672.370			&	40.29		&	9	\\
			\hline\\[-7.0pt]
\multicolumn{6}{l}{The table is limited to species with relative abundances $> 10^{-12}$.}\\
\multicolumn{6}{l}{Species detected unambiguously or in a preliminary manner in the specific environments are marked in boldface.}\\
\multicolumn{6}{l}{Parameters for line flux calculations: 75~K, $10^{5}$ cm$^{-3}$, 10$^{24}$ cm$^{-2}$} \\
\multicolumn{6}{l}{$^{*}$ DCN should be observable, but is not listed in CDMS or RADEX, instead calculated RADEX values for HCN are listed. }
\end{tabular}
\end{table}

\begin{table*}
\centering
\caption{Fractional abundances and D/H ratios: deuterated ices. \label{tab:mostDeuteratedIces}}
\begin{tabular}{lcc|lcc|lcc}	
\hline
\tablewidth{0.70\textwidth}
				&	Dark Clouds		& 					&				& 	Warm IRDCs			&						&					&	HMPOs 										\\
Species			& 	Abundance		& 	D/H ratio			& 	Species		&	 Abundances			& 	D/H ratios				& 	Species			& 	Abundances			& 	D/H ratios				\\
\hline
HDO				& 3.90 $\times 10^{-7}$	& 4.26 $\times 10^{-3}$	&	HDO			& 	6.77 $\times 10^{-8}$ 	&	2.94 $\times 10^{-3}$	&	HDO				&	8.68 $\times 10^{-7}$	&	1.27 $\times 10^{-2}$	\\
CH\dthree D		& 1.41 $\times 10^{-7}$	& 8.70 $\times 10^{-3}$	&	HDO\dtwo		&	4.06 $\times 10^{-9}$ 	&	1.78 $\times 10^{-3}$	&	NHDOH			&	1.73 $\times 10^{-7}$	&	9.25 $\times 10^{-2}$	\\
NH\dtwo D		& 6.75 $\times 10^{-8}$	& 6.14 $\times 10^{-3}$	& 	NH\dtwo D	&	2.48	$\times 10^{-9}$ 	&	7.40 $\times 10^{-3}$	&	NH\dtwo D		&	1.68 $\times 10^{-7}$	&	2.21 $\times 10^{-2}$	\\
C\dthree HD		& 5.38 $\times 10^{-9}$	& 9.61 $\times 10^{-3}$	&	C\dthree HD	&	6.62 $\times 10^{-10}$	&	7.75 $\times 10^{-4}$	&	D\dtwo O			&	1.11 $\times 10^{-8}$	&	1.63 $\times 10^{-4}$	\\
CH\dtwo D\dtwo	& 1.89 $\times 10^{-9}$	& 1.17 $\times 10^{-4}$	&	D\dtwo O		&	4.68 $\times 10^{-10}$	&	1.15 $\times 10^{-5}$	&	HDO\dtwo			&	5.01 $\times 10^{-9}$	&	7.95 $\times 10^{-2}$	\\
DNO				& 1.72 $\times 10^{-9}$	& 1.06 $\times 10^{-1}$	&	CHDCO		&	2.77 $\times 10^{-10}$	&	1.83 	$\times 10^{-3}$	&	ND\dtwo OH		&	3.14 $\times 10^{-9}$	&	1.68 $\times 10^{-3}$	\\
D\dtwo O			& 1.05 $\times 10^{-9}$	& 1.15 $\times 10^{-5}$	& 	DCN			&	2.30 $\times 10^{-10}$	&	1.04 $\times 10^{-3}$	&	NHD\dtwo			&	2.65 $\times 10^{-9}$	&	3.29 $\times 10^{-4}$	\\
HDS				& 6.72 $\times 10^{-10}$	& 1.15 $\times 10^{-2}$	&	DCOOH		&	1.72 $\times 10^{-10}$	&	4.31	$\times 10^{-2}$	&	C$_{5}$HD		&	2.45 $\times 10^{-9}$	&	1.25 $\times 10^{-2}$	\\
NHD\dtwo			& 2.48 $\times 10^{-10}$	& 2.25 $\times 10^{-5}$	& 	DNC			&	1.57 $\times 10^{-10}$	&	1.11 $\times 10^{-2}$	&	C$_{6}$HD		&	9.96 $\times 10^{-10}$	&	1.72 $\times 10^{-2}$	\\
DCN				& 1.73 $\times 10^{-10}$	& 5.90 $\times 10^{-3}$	&	CH\dthree D	&	1.53 $\times 10^{-10}$	&	1.09 $\times 10^{-1}$ 	&	D\dtwo O\dtwo		&	8.15 $\times 10^{-10}$	&	1.29 $\times 10^{-2}$	\\
			\hline\\[-7.0pt]
\multicolumn{9}{l}{The table is limited for each environment to 10 species or species with relative abundances $> 10^{-12}$.}\\
\multicolumn{9}{l}{Species detected unambiguously or in a preliminary manner in the specific environments are marked in {boldface}.}\\
\multicolumn{9}{l}{Dark clouds 10~K, $10^{4}$ cm$^{-3}$ $-$ Warm IRDCs 25~K, 10$^{5}$ cm$^{-3}$ $-$ HMPOs 75~K, 10$^{5}$ cm$^{-3}$}
\end{tabular}
\end{table*}

{We have listed in Tables~\ref{tab:mostDeuteratedALMAdc}-\ref{tab:mostDeuteratedALMAhmpo} we list the the most abundant, (potentially) observable deuterated species with ALMA in dark clouds, infrared dark clouds (IRDCs), and high-mass protostellar objects (HMPOs), respectively. In order to predict the observability of molecules in these different environments we use the Cologne Database of Molecular Spectroscopy \citep[CDMS; ][]{2001A&A...370L..49M, 2005JMoSt.742..215M} to calculate line fluxes under local thermal equilibrium conditions (LTE) for various deuterated molecules with transitions observable by ALMA. For these estimates we assume the following parameters for the different environments: for dark clouds temperature 10~K, number density $10^{4}$ cm$^{-3}$ and H$_{2}$ column density 10$^{22}$ cm$^{-2}$ \citep{2013A&A...551A..98L}, for warm IRDCs 25~K, 10$^{5}$ cm$^{-3}$ and H$_{2}$ column density 10$^{24}$ cm$^{-2}$ \citep[e.g.][]{2008ApJ...689.1141R}, and for HMPOs 75~K, 10$^{5}$ cm$^{-3}$ and column density 10$^{24}$ cm$^{-2}$ \citep[e.g.][]{2007A&A...466.1065B, 2008A&A...490..213R}. There are known differences in H$_{2}$ column densities between interferometry and single-dish observations \citep{2009A&A...499..149V}. While we concentrate here on the higher end of column densities for our calculations, the calculated line fluxes can be adopted to the lower column densities by simply dividing them by 10. }

{For dark clouds and warm IRDCs we implement the atomic initial abundances, while for HMPOs we implemented instead the ``Evolution'' model (with initially high D/H ratios). Line intensities for a few molecules (e.g. DCO\jon) can also be estimated using the non-LTE molecular radiative transfer tool RADEX \citep{2007A&A...468..627V}. Because RADEX and our calculations using CDMS gave different results we scaled our CDMS calculations to match RADEX values for linear molecules (e.g. DCO\jon). We look for transitions in bands 3 - 9 for ALMA (84 - 720 GHz with gaps between bands, see \url{science.nrao.edu/facilities/alma/observing}) and consider a line sensitivity limit of 1 mK for all bands. In Tables~\ref{tab:mostDeuteratedALMAdc}-\ref{tab:mostDeuteratedALMAhmpo} are only the three strongest transitions listed, while a full list is available upon request from the authors.}

{Amongst these listed species are tracers (through D/H ratios) of initial abundances (e.g. D\dtwo CO, HDO, NH\dtwo D, CH\dtwo D\jon) and temperature (D\dtwo O, DCO\jon, DCN, D\dtwo S) as we have listed in Table~\ref{tab:deps}. 
A few ions are observable with ALMA, such as DCO\jon , H\dtwo DO\jon , CH\dtwo D\jon. We find that the tracer of ionization in the cold ISM regions with high depletion, H\dtwo D\jon~ and HD\dtwo\jon, will be hard to detect with ALMA, as found by \citet{2011A&A...533A.143C} for protoplanetary disks. We note that all isotopologues of ammonia are easily observable in the all three ISM environments. The metastable doublet lines of NH$_3$ are used to constrain gas temperature \citep[e.g.][]{1983ARA&A..21..239H, 2009MNRAS.399..425M}. We think that using the relative abundances of the minor NH$_3$ isotopologues one could also discern the previous temperature history of the environment. We determine that several sulphur-bearing species, such as HDCS, D\dtwo CS, HDS, should be observable by ALMA, and as sulphur chemistry is not well understood yet, observations of these species could serve as proxies for future studies. Finally we note that the HDO lines are expected to be observable with ALMA, but it is not included into CDMS, and we could not calculate its line intensities assuming LTE. On the other hand, water has a complex level structure, with some of the lines that are masing and many that become highly optically thick, and the escape probability non-LTE method of RADEX is not capable of modeling its line intensities reliably. For estimation of the water line fluxes one has to perform a full line radiative transfer modeling for each individual object. }

{For deuterated ices we only list the most abundant species in Table~\ref{tab:mostDeuteratedIces} and do not try to predict their observability. Amongst the deuterated ices we find several polyynes (C$_{n}$H\dtwo, with $n > 4$) are abundant, especially at temperatures $> 10$ K. These species have been observed in the Solar system such as in Titan's atmosphere \citep[e.g.][]{2009Icar..202..620T}, but should also be abundant in ices in interstellar space. Several organics such as deuterated formic acid and hydroxylamine (both singly- and doubly-deuterated) are abundant, even among the most abundant at higher temperatures ($\gtrsim 25$ K). Also both ammonia and water (singly- and doubly-deuterated) show high abundances. } 

Uncertainties for the predicted abundances are factors of $\sim 1.5-5$ for species made of $\lesssim 3-4$ atoms and $\gtrsim 1.5-10$ for more complex ones. Apart from the estimated abundance and D/H ratio uncertainties due to the errors in the rate coefficients, we also have error bars associated with the exact estimation of physical properties of the observed environment (uncertainties in dust emissivity, temperature, poorly known dust-to-gas mass ratio, etc.) Also, many of the published observed D/H values are based on measurements made from several sources, for which derived physical properties can differ significantly. 

We now discuss how our model results compare with assorted observations. \citet{1995ApJ...447..760V} observed several deuterated species towards the protobinary source IRAS 16293, namely, DCO\jon, DCN, C\dtwo D, HDS, HDCO, NH\dtwo D, detected in the different regions around the protostar. The first region is the warm and dense inner core ($T \gtrsim 80$~K, n$_{\rm{H}} \sim 10^{7}$ cm$^{-3}$) found to be rich with organic molecules, the second is the circumbinary envelope with $T \sim 40$~K and $n_{\rm{H}} \sim 10^{6}-10^{7}$ cm$^{-3}$, where molecules such as DCO\jon~and HDCO where found, and the third is the colder, low-density outer part of the envelope with $T\sim 10-20$~K and n$_{H}\sim 10^{4}-10^{5}$ cm$^{-3}$ with radicals such as CN, C\dtwo H, C\dthree H\dtwo. We find good agreement between our calculated values and those derived for all observed species. As an example, HDCO shows a D/H ratio of 0.13 in our ``Evolution'' model compared with the observed value of 0.14, and for DCN the D/H ratio is 0.027 compared with the observed ratio of 0.013. The worst agreement we find is for C\dtwo D and DCO\jon~with a difference of a factor of $\sim 5$ in the D/H ratios, which is still acceptable agreement. Uncertainties for these species range between factors of $2 - 5$, with largest uncertainties for C\dtwo D and DCN. Considering these uncertainties, our predicted D/H values are in agreement with the IRAS-16293 observations. 

\citet{2003A&A...403L..37C} detected ortho-H\dtwo D\jon~towards the prestellar core L1544, with derived abundances of 7.2~$\times 10^{-10}$ and 3.2~$\times 10^{-10}$ at the peak and off-peak positions, respectively. For our ÒPrimordialÓ model with a core density $n_{\rm {H_2}} = 10^{6}$ cm$^{-3}$, {a temperature 7 K,} and an appropriate equilibrium o/p ratio of 3:1 taken from \citet{2004A&A...418.1035W}, we find reasonable agreement with calculated abundances of 0.7 $-$ 4.8 $\times$ 10$^{-10}$ in the core (peak position) and 2.3 $-$ 4.2 $\times$ 10$^{-10}$ at the off-peak position, {with a lower density,} assuming a density of {10$^{4} - 10^{5}$} cm$^{-3}$. {We however find a higher abundance at the off-peak position; there are several possible reasons for the discrepancy such as incorrect treatment of the ortho-para species in the network or the lack of a detailed physical model. A detailed study is not within the scope of this paper but we note that abundances are within a factor of 2-3 of observed abundances. }

\citet{2004ApJ...608..341S} observed H\dtwo D\jon, {DCO\jon, HCO\jon, HDO and H\dtwo O} toward the protobinary source IRAS 16293 {(A and B)} as well as the cold prestellar object IRAS 16293 E. They measured {the H\dtwo D\jon} abundance {to be} 2~$\times 10^{-9}$ in the cold, outer envelope with n$_{{\rm H}_2}$ = 10$^{4} - 10^{5}$ cm$^{-3}$ and $T < 20$ K, where our ÒEvolutionÓ model predicts a similar abundance of $\sim 10^{-9}$. In the inner envelope the temperature is higher, depleting the H\dthree\jon~isotopologues as CO returns to the gas phase, and the abundance decreases to $\sim 10^{-12}$ cm$^{-3}$. The temperature in the inner envelope is not well constrained, but with a central density of $\sim 10^{6}$ cm$^{-3}$ we find the best agreement at temperatures $\approx 30$ K, with an abundance of 1.2~$\times 10^{-12}$, and the agreement worsens if the temperature is increased, as more CO returns into the gas phase and the overall deuterium fractionation ceases. {DCO\jon~and HCO\jon~were observed with abundances $2 \times 10^{-11}$ and $< 1\times 10^{-9}$ respectively, which agree with our model values from 17 $-$ 26 K with DCO\jon~abundances 8.4 $\times 10^{-11} - 8.2 \times 10^{-11}$ and HCO\jon abundances~4.7 $\times 10^{-10} - 1.1 \times 10^{-9}$, leading to a D/H ratio of 0.078 $-$ 0.18. HDO and H\dtwo O were observed with abundances 3$ \times 10^{-10}$ and $3 \times 10^{-7} - 4 \times 10^{-9}$ respectively, which agree reasonably well with our modeled abundances of 5.9 $\times 10^{-12} - 6.3 \times 10^{-11}$ and 9.9 $\times 10^{-10} - 1.7 \times 10^{-8}$, respectively. Lastly, HDO, DCO\jon~ and HCO\jon~were also observed in the prestellar core object IRAS 16293 E. From estimated temperatures 16 - 25 K and densities $1.1 - 1.6 \times 10^6$ cm$^{-3}$ we estimate DCO\jon~abundances to be 7.0 $\times 10^{-11} - 2.9 \times 10^{-9}$ and HCO\jon abundances ~8.9 $\times 10^{-11} - 1.10 \times 10^{-9}$, leading to a D/H ratio 0.26 $-$ 0.79. This agrees well with the observed abundances of $5.0 \times 10^{-11}$ and $1.0 \times 10^{-10}$ for DCO\jon~and HCO\jon~with a D/H ratio 0.5. We find the same agreement for HDO with an observed abundance $2 \times 10^{-10}$ comparable to our modeled abundance range of 3.0 - 7.5 $\times 10^{-10}$. }

\citet{2012A&A...539A.132C} observed multiple lines of HDO and H\dtwo $^{18}$O towards IRAS 16293A with an estimated D/H $\sim 0.034$ in the hot corino region and D/H$\sim 0.005$ in the outer envelope, utilizing a standard isotopic ratio of H\dtwo $^{18}$O/H\dtwo $^{16}$O = 500. In order to reproduce observed line emission, they added an outer absorbing layer with an H\dtwo O column density of 1.23 $\times 10^{13}$ cm$^{-2}$. Depending on the exact choice of density and temperatures, our models give for the cold envelope ($n_{\rm H} \sim 10^{5}$ cm$^{-3}$, T $\lesssim 30$ K) D/H ratios of $\sim 0.01 - 0.1$, for both sets of initial abundances, in agreement with observations. For hot cores ($n_{\rm H} \sim 10^{8}$ cm$^{-3}$, T $\sim 150$ K) the  ``Primordial''  model estimates the D/H ratio to be $0.0001 - 0.001$, while the ``Evolution'' model predicts that the D/H ratio is $\sim 0.0001 - 0.01$. Although our evolutionary model is a poor solution and does not account for the gradual warm-up of the environment, our predicted D/H ratios from the ``Evolution'' model are in agreement with those estimated by Coutens et al., albeit only with upper limits of our estimates. 

The radical OD was observed for the first time outside of the Solar system by \citet{2012A&A...542L...5P} along the line of sight towards the low-mass protostar IRAS 16293A. They also observed HDO and found a high OD/HDO ratio of $\sim 10-100$. Parise et al. compared their observations to the modeled values of OH/H\dtwo O and found their calculated values to be too low. The agreement was slightly better when they implemented a simple evolutionary model with increasing temperature with time, but the result was still lower than observations, with the highest modeled values reaching 5.7. Studying the chemical evolution in our ÓPrimordial modelÓ for temperatures $T < 30$ K and densities $n_{\rm H} = 10^{4} - 10^{6}$ cm$^{-3}$, we find that the large OD/HDO ratio is mainly due to the efficient reaction OH + D $\longrightarrow$ OD + H, as originally suggested by \citet{1989ApJ...340..906M}. Via this reaction, the OD/HDO ratios can reach values approximately one order of magnitude larger than OH/H\dtwo O. Furthermore, toward temperatures $\sim 30$ K the OD/HDO ratios can even be as high as two orders of magnitude. Thus, our model is in agreement with the observed OD/HDO ratios from \citet{2012A&A...542L...5P}, without the need for a warm-up phase.

\subsection{Earlier Models}

\citet{2000A&A...361..388R} {and \citet{2000A&A...364..780R}} have investigated the chemical evolution with deuterium fractionation for temperatures 10$-$100 K and densities $3\times 10^3-3\times 10^8$~cm$^{-3}$ on a less resolved grid, consisting of only 100 points. They used a time-dependent chemical gas-phase model based on the UMIST'95 database \citep{1997A&AS..121..139M}. Their resulting network consists of $\sim$ 300 species linked by $>$5\,000 reactions, but only includes singly-deuterated species and limited surface chemistry for H\dtwo~and HD. We compared the results between our models for a number of species, including DCO\jon, HDCO, DCN, DNC and DC$_{5}$N, looking at the distribution of D/H fractionation ratios and time-dependent abundances at $10^5$~yr, and we found good agreement between our models. We also studied the {molecular abundances} under conditions typical of the TMC-1 environment in our  ``Primordial''  model {and under these conditions} we found that the quantitative agreement in the D/H ratios is better than an order of magnitude for all species, with the worst agreement for NH\dtwo D where the ratio between the two models is 0.14. The comparison for D/H values is shown in Table~\ref{tab:modelComp}. The intrinsic uncertainty in the abundance of DC$_{5}$N as predicted in our sensitivity analysis is very large, $\sim 1-1.5$ orders of magnitude, and is comparable to the difference between our and Roberts' \& Millar's model. We note however that our modeled D/H ratios show a better agreement with the observations of DC$_{5}$N and HC$_{5}$N in TMC-1 by \citet{1981ApJ...251L..33M} than calculated values by Roberts $\&$ Millar. {In \citet{2000A&A...364..780R} expanded their study to include doubly-deuterated species, allowed species to freeze onto grains and looked at a different selection of species.} We compared their predictions with our results for singly- and doubly- deuterated isotopologues of NH\dthree, H\dtwo O, H\dtwo CO, and found reasonably good agreement for all singly-deuterated species. In our model, NH\dtwo D shows enhanced D/H ratios ($\sim 10^{-3} - 10^{-1}$) up to temperatures of $30-40$~K, while the enhanced D/H ratios in the model of Roberts \& Millar only appear up to $20-30$ K. For the doubly-deuterated species D\dtwo O, NHD\dtwo~and D\dtwo CO we predict similar D/H ratios to Roberts \& Millar up to temperatures of $\sim 50$~K, with values $\sim$ 10$^{-3} -10^{-1}$, while at larger temperatures our models diverge. Our model predicts a strong decrease in the respective D/H ratios to $\sim 10^{-5}$, while the D/H ratios of Roberts \& Millar decrease more smoothly and do not reach the same value until at $\sim$~100 K. 

\begin{table}
\centering
\caption{Comparison of D/H ratios for a TMC1-like environment (T = 10 K, $n_{\rm H} = 10^{4}$ cm$^{-3}$) between our model and \citet{2000A&A...361..388R}. \label{tab:modelComp}}
\begin{tabular}{c|ccc}
\hline
\tablewidth{0.45\textwidth}
Species				&	Our model	&	Roberts $\&$ Millar	&	Ratio		\\
\hline
NH\dtwo D			&	1.4 $\times 10^{-2}$ &	8.4 $\times 10^{-2}$		&	0.17				\\
HDCO				&	1.7 $\times 10^{-2}$	&	4.2 $\times 10^{-2}$		&	0.40				\\
DCN					&	2.4 $\times 10^{-2}$	&	0.9 $\times 10^{-2}$		&	2.7				\\
DNC					&	2.6 $\times 10^{-2}$	&	1.5 $\times 10^{-2}$		&	1.7				\\
C\dtwo D				&	1.1 $\times 10^{-2}$	&	1.1 $\times 10^{-2}$		&	1.0				\\
C$_{4}$D				&	1.6 $\times 10^{-2}$	&	0.4 $\times 10^{-2}$		&	4.0				\\
DCO\jon				&	3.9 $\times 10^{-2}$	&	1.9 $\times 10^{-2}$		&	2.1				\\
N\dtwo D\jon			&	8.6 $\times 10^{-3}$	&	2.5 $\times 10^{-2}$		&	0.34				\\
C\dthree HD			&	1.3 $\times 10^{-2}$	&	0.6 $\times 10^{-2}$		&	2.2				\\
C\dthree H\dthree D		&	1.6 $\times 10^{-2}$	&	8.3 $\times 10^{-2}$		&	0.19				\\
DC\dthree N			&	9.6 $\times 10^{-3}$	&	0.7 $\times 10^{-2}$		&	1.4				\\
DC$_{5}$N			&	1.2 $\times 10^{-2}$	&	2.3 $\times 10^{-2}$		&	0.52				\\
HDCS				&	1.8 $\times 10^{-2}$	&	4.0 $\times 10^{-2}$		&	1.4				\\
\hline
\end{tabular} 
\end{table}

In the study of \citet{2004A&A...424..905R}, the chemical evolution in a sample of prestellar cores using two subsets of the Rate'99 and osu.2003 chemical networks was compared. With the networks limited to include species with six or fewer carbon atoms and no surface chemistry, Roberts et al. used the chemical models to successfully explain observations of the CO depletion, and its relevance to the D\dtwo CO and HDCO fractionation ratios. If we compare the calculated fractionation ratios between the steady state abundances of Roberts et al. (see their Table~5) and our models for a TMC-1 environment, the D/H ratios for the majority of key species such as H\dtwo D\jon, N\dtwo D\jon, DCO\jon~and HDO agree reasonably well. However, we found significant discrepancies for D\dtwo O, HD\dtwo\jon, D\dthree\jon~and NHD\dtwo. The reason why our D/H ratios for doubly-deuterated species differ from those of Roberts et al. appears to be twofold. First, they have not considered surface chemistry and assumed that all atomic D that freezes out is immediately returned to the gas as HD. In contrast, in our model, the accreted deuterium atoms are incorporated in surface species and do not easily return to the gas phase. Second, they use only steady-state abundances while we use time-dependent abundances at 1 Myr.

\citet{2004A&A...418.1035W} studied steady-state chemistry in a completely depleted, low-mass prestellar core, with an emphasize on explaining observations of ortho-H\dtwo D\jon~towards L1544 by \citet{2003A&A...403L..37C}. While our model does not yet include nuclear-spin state chemistry, we compared the calculated abundances of the H\dthree\jon~isotopologues at densities of $n_{\rm H}~=~10^{5}, 10^{6}, 10^{7}$ cm$^{-3}$, $T = 10$~K assuming a cosmic ray ionization rate $\zeta= 3\times10^{-17}$~s$^{-1}$ (Note that at such conditions para-H$_2$ will be the dominant form of molecular hydrogen and thus fractionation will proceed with a high efficiency as assumed in our model). For both the  ``Primordial''  and ``Evolution'' models we found good overall agreement for the D/H ratios of the H\dthree\jon~isotopologues and the electron abundances. The only difference occurs at high densities of $\sim 10^{7}$~cm$^{-3}$, where the HD\dtwo\jon~and D\dthree\jon~abundances are one and two orders of magnitude lower in our model, respectively. We could not find out whether this difference increases at higher densities. The likely reason for such a discrepancy is the lack of surface chemistry and the assumption of complete freeze-out in the Walmsley et al. model. Even at such high densities and 10~K, the depletion is not complete in our model, so that the H$_3^+$ isotopologues can still be destroyed by ion-molecule reactions in addition to dissociative recombination with electrons or negatively charged grains. Moreover, in our model, atomic D released upon dissociative recombination can stick to a grain and be incorporated in the surface molecules. At $n_{\rm H} \ga 10^7$~cm$^{-3}$ and after 1~Myr of evolution, a substantial fraction of the gas-phase reservoir of the elemental D can be chemically 'transferred' to ices, unable to directly come back to the gas phase. Consequently, it will increase surface fractionation and abundances of deuterated ices.

\section{Conclusions}
\label{sec:conclusion}
We present an extended, publicly available chemical network for deuterium fractionation, with the most up-to-date reaction rate coefficients from laboratory measurements and theoretical studies. The new deuterium chemistry model does not yet include nuclear-spin state processes and is better suitable for cold ISM environments, $T\sim 10-20$~K. In this paper, we have tested this network by performing a benchmarking study of deuterium chemistry under dense ISM conditions with two distinct initial abundance sets. The limits of accuracy of the network have been investigated with a sensitivity analysis. The most problematic reactions for the chemical evolution of H\dthree\jon, HCO\jon, HOC\jon, HCN, HNC, H\dtwo O, CH\dthree OH, H\dthree O\jon, CH\dthree\jon, C\dtwo H\dtwo\jon~and their isotopologues as well as CO are listed or presented as online material. Ion-neutral and dissociative recombination reactions dominate the list, accompanied by a smaller number of neutral-neutral reactions and the cosmic ray ionization of H\dtwo~and He. 

In general, using the $1\sigma$ confidence level, the abundances and column densities of species made of $\la 3$ atoms (e.g., CO, HCO$^+$, DCO$^+$) are uncertain by factors $1.5-5.0$, those for species made of $4-7$ atoms are uncertain by a factor of $1.5-7$, and those for more complex species made of $>7$ atoms are uncertain by a factor of $2 - 10$. For D/H ratios the uncertainties are, for the same different ranges of molecule sizes, a factor of $1.6-5$, $1.6-10$ and $2.5-10$, respectively. 

Despite certain limitations of our model, it successfully explains the observed D/H ratios in dark clouds (10~K and $10^{4}$ cm$^{-3}$), prestellar cores ($T\lesssim 10~$K, n$\sim 10^{4}$ cm$^{-3}$), and protostellar envelopes (cold, $T\sim 30$~K, n$_{\rm {H}}\sim 10^5$ cm$^{-3}$ and warm, $T\sim 150$~K, n$_{\rm {H}} \sim 10^{8}$ cm$^{-3}$), for many key species including water, methanol, ammonia and many hydrocarbons. Our results show good agreement with previous model studies by \citet{2000A&A...364..780R}, \citet{2000A&A...361..388R}, \citet{2004A&A...424..905R} and \citet{2004A&A...418.1035W}. We also list the dominant formation and destruction pathways for DCO\jon, DCN and isotopologues of H\dthree\jon~and water in Appendix~\ref{sec:Dominant}. Finally, in Tables~\ref{tab:mostDeuteratedALMAdc}-\ref{tab:mostDeuteratedALMAhmpo} we have listed the most abundant, potentially detectable deuterated species in cold cores, and warm IRDCs and HMPOs, which can be searched for with ALMA. 
 
\section*{Acknowledgements}
The research leading to these results has received funding from the European Community's Seventh Framework Programme [FP7/2007-2013] under grant agreement no. 238258. D.S. acknowledges support by the Deutsche Forschungsgemeinschaft through SPP 1385: ``The first ten million years of the solar system -- a planetary materials approach'' (SE 1962/1-1 and SE 1962/1-2). E. H. wishes to acknowledge the support of the National Science Foundation for his astrochemistry program. He also acknowledges support from the NASA Exobiology and Evolutionary Biology program through a subcontract from Rensselaer Polytechnic Institute.

\bibliography{DeuteriumHotNColdEnv_final}{}
\bibliographystyle{aa}

\clearpage
\appendix
%
\section{Updated and added reactions to chemical network }
\label{sec:AppendixA}
{In this appendix, we list problematic reactions and rate coefficients (Table~\ref{tab:probreact}) identified in our sensitivity analysis for isotopologues and isomers of water, H\dthree\jon, HCO\jon~and HCN, as well as added and updated non-deuterated (Table~\ref{tab:netwOrg}) and deuterated (Table~\ref{tab:netwDeut}) reactions to our network. The added and updated reactions have been collected from several literature references as well as newly announced values (as of 2012-12-26) reported in the Kinetic Database for Astronomy (KIDA; \url{http://kida.obs.u-bordeaux1.fr/}). }

\begin{center}
\begin{longtable}{lclcccc}
\caption[Added and updated non-deuterium reactions. ]{Added and updated non-deuterium reactions.} \label{tab:netwOrg}\\
\tablewidth{0.9\textwidth}
\scriptsize
Reaction			&	&						&	{$\alpha$}		&	{$\beta$}		&	{$\gamma$}	&	Ref	\\
\hline
\endhead
%
C\jon~				+ H\dtwo 			&\ra&	CH\dtwo\jon								&	2.00 {\rm x} 10$^{-16}$	&	-1.30	&	-23		&	1\\
C\jon~				+ HCOOH			&\ra&	HCO\jon~ 		+	OH	+	C			&	8.00{\rm x} 10$^{-10}$	&	-0.50	&	0		&	5\\
C\dtwo H\dtwo\jon~ 		+ H\dtwo 			&\ra&	C\dtwo H$_4$\jon							&	2.90 {\rm x} 10$^{-14}$	&	-1.50	&	0		&	1\\
CH\dthree\jon~ 		+ H\dtwo O 		&\ra&	CH\dthree OH\dtwo\jon						&	5.50 {\rm x} 10$^{-12}$	&	-1.70	&	0		&	3\\
CH\dthree\jon~ 		+ H\dtwo			&\ra&	CH$_5$\jon								&	3.78 {\rm x} 10$^{-16}$	&	-2.30	&	22		&	4\\
CH\dthree O\dtwo\jon~ 	+ CH\dthree OH 	&\ra&	H$_5$C\dtwo O\dtwo\jon~	+ H\dtwo O		&	2.00{\rm x} 10$^{-9}$	&	-0.50	&	2810		&	5\\
CH\dthree OH\dtwo\jon~	+ HCOOH			&\ra&	HCOOH\dtwo\jon~			+CH\dthree OH 	&	3.63{\rm x} 10$^{-9}$	&	-0.50	&	685		&	5\\
CH$_5$\jon~			+ HCOOH			&\ra&	HCOOH\dtwo\jon~			+ CH$_4$			&	3.00{\rm x} 10$^{-9}$	&	-0.50	&	0		&	5\\
H\dtwo CN\jon~		+ C\dtwo H\dtwo	&\ra& 	C\dthree H$_4$N\jon						&	3.30{\rm x} 10$^{-16}$	&	-2.00	&	0		&	4\\
H\dtwo CN\jon~		+ HCOOH			&\ra&	HCOOH\dtwo\jon~			+ HCN			&	1.40{\rm x} 10$^{-9}$	&	-0.50	&	0		&	5\\
H\dthree\jon~			+ HCOOH			&\ra&	HCO\jon~ 		+ H\dtwo O	+ H\dtwo		&	3.90{\rm x} 10$^{-9}$	&	-0.50	&	0		&	5\\
H\dthree\jon~ 			+ O				&\ra&	OH\jon~ 					+ H\dtwo			&	7.98 {\rm x} 10$^{-10}$	&	-0.156&	-1.41		&	1\\
H\dthree\jon~ 			+ O				&\ra&	H\dtwo O\jon~ 				+ H 				&	3.42 {\rm x} 10$^{-10}$	&	-0.156&	-1.41		&	1\\
H\dthree CO\jon~		+ HCOOH			&\ra&	HCOOH\dtwo\jon~			+ H\dtwo CO		&	2.00{\rm x} 10$^{-9}$	&	-0.50	&	0		&	5\\
H\dthree O\jon~		+ C\dtwo H$_4$	&\ra& 	C\dtwo H$_5$OH\dtwo\jon 					&	1.90 {\rm x} 10$^{-14}$	&	-2.80	&	0.25		&	4\\	
H\dthree CO\jon~		+ H\dtwo CO		&\ra&	H$_5$C\dtwo O\dtwo\jon						&	8.15{\rm x} 10$^{-15}$	&	-3.00	&	0		&	6\\
H\dthree S\jon~		+ HCOOH			&\ra&	HCOOH\dtwo\jon~			+	H\dtwo S		&	2.00{\rm x} 10$^{-9}$	&	-0.50	&	0		&	5\\
HCOOH\dtwo\jon~ + CH\dthree OH 			&\ra&	CH\dthree OH\dtwo\jon~		+ 	HCOOH		&	2.29{\rm x} 10$^{-9}$	&	-0.50	&	0		&	5\\
He\jon~				+ HCOOH			&\ra&	HCO\jon~ 		+	OH	+	He			&	9.00{\rm x} 10$^{-10}$	&	-0.50	&	0		&	5\\
N\dtwo H\jon~			+ HCOOH			&\ra&	HCOOH\dtwo\jon~	+	N\dtwo				&	1.70{\rm x} 10$^{-9}$	&	-0.50	&	0		&	5\\
%
C	 				+ O\dtwo			& \ra &	CO				+ O			&	1.28 {\rm x} 10$^{-9}$	&	-0.32	&	0	&	4\\
C					+ OH				& \ra & 	CO 				+ H			&	1.15 {\rm x} 10$^{-10}$	&	-0.34	&	0	&	4\\
C\dtwo~ 				+ O				& \ra &	CO				+ C			&	2.00 {\rm x} 10$^{-10}$	&	-0.12	&	0	&	1\\
C\dtwo~				+ OCS			& \ra &	CO				+ C\dtwo S	&	1.00 {\rm x} 10$^{-10}$	&	0.00	&	0	&	4\\	
C\dtwo H				+ N				& \ra &	C\dtwo N 			+ H			&	1.00 {\rm x} 10$^{-10}$	&	0.18	&	0	&	1\\
C\dtwo H				+ O				& \ra &	CO				+ CH			&	1.00 {\rm x} 10$^{-10}$	&	0.00	&	0	&	1\\
C\dthree H 			+ O				& \ra &	C\dtwo H			+ CO			&	1.00 {\rm x} 10$^{-10}$	&	0.00	&	0	&	1\\
C\dthree O 			+ O				& \ra & 	C\dthree~			+ O\dtwo		&	1.00 {\rm x} 10$^{-10}$	&	0.00	&	0	&	1\\
CH					+ OCS			& \ra &	H		+ CO	 	+ CS			&	4.00 {\rm x} 10$^{-10}$	&	0.00	&	0	&	4\\	
CH					+ SO				& \ra &	H 		+ OCS				&	1.10 {\rm x} 10$^{-10}$	&	0.00	&	0	&	4\\	
CH					+ SO				& \ra &	CO		+ HS					&	9.00 {\rm x} 10$^{-11}$	&	0.00	&	0	&	4\\	
CH\dtwo~				+ H				& \ra &	CH 		+ H\dtwo				&	2.20 {\rm x} 10$^{-10}$	&	0.00	&	0	&	1\\
CH$_4$				+ CH 			& \ra &	C\dtwo H$_4$ 		+ H			&	1.06 {\rm x} 10$^{-10}$	&	-1.04	&	0	&	4\\
CN 					+ N				& \ra &	C		+ N\dtwo				&	1.00 {\rm x} 10$^{-10}$	&	0.00	&	0	&	1\\
CN					+ O				& \ra &	CO		+ N					&	2.60 {\rm x} 10$^{-11}$	&	-0.12	&	0	&	1\\
CN					+ O\dtwo 			& \ra &	OCN + O						&	1.99{\rm x} 10$^{-11}$	&	-0.63	&	0	&	4\\
H\dtwo~ 				+ CH 			& \ra &	CH\dtwo~ 		+ H			&	1.20 {\rm x} 10$^{-9}$	&	0.00	&	0	&	4\\
HNO					+ O				& \ra &	NO				+ OH			&	3.77 {\rm x} 10$^{-11}$	&	-0.08	&	0	&	1\\
NH					+ O				& \ra &	NO				+ H			&	6.60 {\rm x} 10$^{-11}$	&	0.00	&	0	&	1\\
NH\dtwo~				+ O				& \ra &	HNO				+ H			&	6.39 {\rm x} 10$^{-11}$	&	-0.10	&	0	&	1\\
NH\dtwo~				+ O				& \ra &	NH				+ OH			&	7.10 {\rm x} 10$^{-12}$	&	-0.10	&	0	&	1\\
NH\dthree~			+ CN				& \ra &	HCN				+ NH\dtwo	&	2.77 {\rm x} 10$^{-11}$	&	-0.85	&	0	&	1\\
O					+ C\dtwo S 		& \ra &	CO				+ CS			&	1.00 {\rm x} 10$^{-10}$	&	0.00	&	0	&	4\\	
O					+ OCS			& \ra &	CO				+ CS			&	1.00 {\rm x} 10$^{-10}$	&	0.00	&	0	&	4\\	
O					+ OH	 			& \ra &	O\dtwo~ 			+ H			&	4.00{\rm x} 10$^{-11}$	&	0.00	&	0	&	4\\
S					+ C\dtwo O		& \ra &	CO				+ CS			&	1.00 {\rm x} 10$^{-10}$	&	0.00	&	0	&	4\\	
S					+ HCO			& \ra &	H				+ OCS		&	8.00 {\rm x} 10$^{-11}$	&	0.00	&	0	&	4\\	
S					+ HCO			& \ra &	CO				+ HS			&	4.00 {\rm x} 10$^{-11}$	&	0.00	&	0	&	4\\	
\hline
%
C\dthree H\jon~ 	+ e\ijon	& \ra &	C\dtwo~ 	+ CH			&	6.00 {\rm x} 10$^{-9}$	&	-0.50	&	0	&	2\\
C\dthree H\jon~ 	+ e\ijon 	& \ra &	C\dtwo H 	+ C 			&	9.90 {\rm x} 10$^{-8}$	&	-0.50	&	0	&	2\\
C\dthree H\jon~ 	+ e\ijon 	& \ra &	C\dthree~	+ H			&	1.95 {\rm x} 10$^{-7}$	&	-0.50	&	0	&	2\\
C\dthree H\dtwo\jon~ + e\ijon 	& \ra &	C\dtwo~ 	+ CH\dtwo	&	1.44 {\rm x} 10$^{-8}$	&	-0.50	&	0	&	2\\
C\dthree H\dtwo\jon~ + e\ijon 	& \ra &	C\dtwo H	+ CH 		&	1.44 {\rm x} 10$^{-8}$	&	-0.50	&	0	&	2\\
C\dthree H\dtwo\jon~ + e\ijon 	& \ra &	C\dtwo H\dtwo~ + C		&	8.64 {\rm x} 10$^{-8}$	&	-0.50	&	0	&	2\\
C\dthree H\dtwo\jon~ + e\ijon 	& \ra &	C\dthree H 	+ H		&	1.66 {\rm x} 10$^{-7}$	&	-0.50	&	0	&	2\\
C\dthree H\dtwo\jon~ + e\ijon 	& \ra &	C\dthree~ + H\dtwo 		&	8.28 {\rm x} 10$^{-8}$	&	-0.50	&	0	&	2\\
C$_4$H\jon~ 		+ e\ijon	& \ra &	C$_4$		+ H		&	1.74 {\rm x} 10$^{-7}$	&	-0.50	&	0	&	2\\
C$_4$H\jon~ 		+ e\ijon 	& \ra &	C\dthree H 	+ C 		&	7.80 {\rm x} 10$^{-8}$	&	-0.50	&	0	&	2\\
C$_4$H\jon~ 		+ e\ijon	& \ra &	C\dtwo H 	+ C\dtwo 		&	4.80 {\rm x} 10$^{-8}$	&	-0.50	&	0	&	2\\
C$_5$\jon~		+ e\ijon	& \ra &	C$_4$		+ C		&	3.90 {\rm x} 10$^{-8}$	&	-0.50	&	0	&	2\\
C$_5$\jon~		+ e\ijon	& \ra &	C$_3$ 	+ C\dtwo		&	2.61 {\rm x} 10$^{-7}$	&	-0.50	&	0	&	2\\
C$_6$\jon~		+ e\ijon	& \ra &	C$_5$ 	+ C			&	1.80 {\rm x} 10$^{-7}$	&	-0.30	&	0	&	2\\
C$_6$\jon~		+ e\ijon	& \ra &	C$_4$		+ C\dtwo	&	2.20 {\rm x} 10$^{-7}$	&	-0.30	&	0	&	2\\
C$_6$\jon~ 		+ e\ijon	& \ra &	C$_3$ 	+ C\dthree	&	1.60 {\rm x} 10$^{-6}$	&	-0.30	&	0	&	2\\
C$_7$\jon~		+ e\ijon	& \ra &	C$_6$		+ C		&	2.30 {\rm x} 10$^{-8}$	&	-0.30	&	0	&	2\\
C$_7$\jon~		+ e\ijon	& \ra &	C$_5$		+ C\dtwo	&	4.37 {\rm x} 10$^{-7}$	&	-0.30	&	0	&	2\\
C$_7$\jon~		+ e\ijon	& \ra &	C$_4$		+ C$_3$	&	1.84 {\rm x} 10$^{-6}$	&	-0.50	&	0	&	2\\
C$_8$\jon~		+ e\ijon	& \ra &	C$_7$		+ C 		&	6.00 {\rm x} 10$^{-8}$	&	-0.30	&	0	&	2\\
C$_8$\jon~		+ e\ijon	& \ra &	C$_6$ 	+ C\dtwo		&	2.00 {\rm x} 10$^{-8}$	&	-0.30	&	0	&	2\\
C$_8$\jon~ 		+ e\ijon	& \ra &	C$_5$ 	+ C\dthree	&	1.80 {\rm x} 10$^{-6}$	&	-0.30	&	0	&	2\\
C$_8$\jon~ 		+ e\ijon	& \ra &	C$_4$	+ C$_4$		&	1.20 {\rm x} 10$^{-7}$	&	-0.30	&	0	&	2\\
C$_9$\jon~		+ e\ijon	& \ra &	C$_7$		+ C\dtwo	&	1.20 {\rm x} 10$^{-7}$	&	-0.30	&	0	&	2\\
C$_9$\jon~ 		+ e\ijon 	& \ra &	C$_6$	+ C\dthree	&	1.32 {\rm x} 10$^{6}$	&	-0.30	&	0	&	2\\
C$_9$\jon~ 		+ e\ijon	& \ra &	C$_5$ 	+ C$_4$		&	5.60 {\rm x} 10$^{-7}$	&	-0.30	&	0	&	2\\
C$_{10}$\jon~ 		+ e\ijon	& \ra &	C$_9$		+ C 		&	2.00 {\rm x} 10$^{-8}$	&	-0.30	&	0	&	2\\
C$_{10}$\jon~ 		+ e\ijon	& \ra &	C$_8$		+ C\dtwo	&	2.00 {\rm x} 10$^{-8}$	&	-0.30	&	0	&	2\\
C$_{10}$\jon~ 		+ e\ijon	& \ra &	C$_7$ 	+ C\dthree	&	1.40 {\rm x} 10$^{-6}$	&	-0.30	&	0	&	2\\
C$_{10}$\jon~ 		+ e\ijon	& \ra &	C$_6$	+ C$_4$		&	6.00 {\rm x} 10$^{-8}$	&	-0.30	&	0	&	2\\
C$_{10}$\jon~ 		+ e\ijon	& \ra &	C$_5$ 	+ C$_5$		&	5.00 {\rm x} 10$^{-7}$	&	-0.30	&	0	&	2\\
CNC\jon~			+ e\ijon	& \ra &	C\dtwo~	+ 	N 		&	2.00 {\rm x} 10$^{-8}$	&	-0.60	&	0	&	1\\
CNC\jon~			+ e\ijon	& \ra &	CN		+ C 			&	3.80 {\rm x} 10$^{-7}$	&	-0.60	&	0	&	1\\
H\dthree\jon~ 		+ e\ijon	& \ra &	H	+	H 	+ H 		&	5.44 {\rm x} 10$^{-8}$	&	-0.50	&	0	&	3\\
H\dthree\jon~ 		+ e\ijon	& \ra &	H\dtwo~ 	+ H			&	1.36 {\rm x} 10$^{-8}$	&	-0.50	&	0	&	3\\
H\dtwo CO\jon~ 	+ e\ijon 	& \ra &	CH\dtwo~	+ 	O		&	2.50 {\rm x} 10$^{-8}$	&	-0.70	&	0	&	1\\
H\dtwo CO\jon~ 	+ e\ijon 	& \ra &	CO 		+ H 	+ H		&	2.50 {\rm x} 10$^{-7}$	&	-0.70	&	0	&	1\\
H\dtwo CO\jon~ 	+ e\ijon 	& \ra &	CO 		+	H\dtwo	&	7.50 {\rm x} 10$^{-8}$	&	-0.70	&	0	&	1\\
H\dtwo CO\jon~ 	+ e\ijon	& \ra &	HCO 	+ H			&	1.50 {\rm x} 10$^{-7}$	&	-0.70	&	0	&	1\\
HC$_5$NH\jon~ 	+ e\ijon	& \ra &	C$_5$N 	+ 	H\dtwo	&	8.00 {\rm x} 10$^{-8}$	&	-0.70	&	0	&	1\\
HC$_5$NH\jon~ 	+ e\ijon	& \ra &	HC\dthree N + 	C\dtwo H	&	1.20 {\rm x} 10$^{-7}$	&	-0.70	&	0	&	1\\
HC$_5$NH\jon~ 	+ e\ijon	& \ra &	HC$_5$N + 	H 		&	9.20 {\rm x} 10$^{-7}$	&	-0.70	&	0	&	1\\
HC$_5$NH\jon~ 	+ e\ijon	& \ra &	HCN 	+ 	C$_4$H	&	4.40 {\rm x} 10$^{-7}$	&	-0.70	&	0	&	1\\
HC$_5$NH\jon~ 	+ e\ijon	& \ra &	HNC 	+ 	C$_4$H	&	4.40 {\rm x} 10$^{-7}$	&	-0.70	&	0	&	1\\
HCNH\jon	~		+ e\ijon	& \ra &	CN		+	H 	+ H	&	9.06 {\rm x} 10$^{-8}$	&	-0.65	&	0	&	1\\
HCNH\jon	~		+ e\ijon	& \ra &	HCN		+	H 		&	9.62 {\rm x} 10$^{-8}$	&	-0.65	&	0	&	1\\
HCNH\jon	~		+ e\ijon	& \ra &	HNC		+	H 		&	9.62 {\rm x} 10$^{-8}$	&	-0.65	&	0	&	1\\
%
C\dtwo H 			+ CRPHOT	& \ra &	C\dtwo~ 	+ H 			&	5.27 {\rm x} 10$^{-14}$	&	0.00	&	0	&	2\\
C\dtwo H	 		+ CRPHOT 	& \ra &	C 		+ CH			&	1.24 {\rm x} 10$^{-14}$	&	0.00	&	0	&	2\\
C\dthree H 		+ CRPHOT	& \ra &	C\dtwo H 	+ C 			&	2.15 {\rm x} 10$^{-14}$	&	0.00	&	0	&	2\\
C\dthree H 		+ CRPHOT 	& \ra &	C\dtwo~ 	+ CH			&	1.30 {\rm x} 10$^{-15}$	&	0.00	&	0	&	2\\
C\dthree H 		+ CRPHOT	& \ra &	C\dthree~ + H			&	4.23 {\rm x} 10$^{-14}$	&	0.00	&	0	&	2\\
C\dthree H\dtwo~ 	+ CRPHOT 	& \ra &	C\dthree H 	+ H 		&	2.99 {\rm x} 10$^{-14}$	&	0.00	&	0	&	2\\
C\dthree H\dtwo~	+ CRPHOT 	& \ra &	C\dthree~ 	+ H\dtwo	&	1.56 {\rm x} 10$^{-14}$	&	0.00	&	0	&	2\\
C\dthree H\dtwo~ 	+ CRPHOT 	& \ra &	C\dtwo H\dtwo~ + C		&	1.50 {\rm x} 10$^{-14}$	&	0.00	&	0	&	2\\
C\dthree H\dtwo~ 	+ CRPHOT 	& \ra &	C\dtwo H 	+ CH 		&	2.60 {\rm x} 10$^{-15}$	&	0.00	&	0	&	2\\
C\dthree H\dtwo~ 	+ CRPHOT 	& \ra &	C\dtwo~ 	+ CH\dtwo	&	1.95 {\rm x} 10$^{-15}$	&	0.00	&	0	&	2\\
C$_4$ 			+ CRPHOT	& \ra &	C\dthree~ + C 			&	1.00 {\rm x} 10$^{-14}$	&	0.00	&	0	&	2\\
C$_4$ 			+ CRPHOT 	& \ra &	C\dtwo~ 	+ C\dtwo		&	2.99 {\rm x} 10$^{-15}$	&	0.00	&	0	&	2\\
C$_4$H 			+ CRPHOT 	& \ra &	C$_4$ 	+ H			&	3.77 {\rm x} 10$^{-14}$	&	0.00	&	0	&	2\\
C$_4$H 			+ CRPHOT 	& \ra &	C\dthree H + C 			&	1.69 {\rm x} 10$^{-14}$	&	0.00	&	0	&	2\\
C$_4$H 			+ CRPHOT 	& \ra &	C\dtwo H 	+ C\dtwo		&	1.04 {\rm x} 10$^{-14}$	&	0.00	&	0	&	2\\
C$_5$ 			+ CRPHOT 	& \ra &	C$_4$ 	+ C			&	1.69 {\rm x} 10$^{-15}$	&	0.00	&	0	&	2\\
C$_5$ 			+ CRPHOT 	& \ra &	C\dthree~ + C\dtwo		&	1.13 {\rm x} 10$^{-14}$	&	0.00	&	0	&	2\\
C$_6$ 			+ CRPHOT 	& \ra &	C$_5$ 	+ C			&	1.17 {\rm x} 10$^{-15}$	&	0.00	&	0	&	2\\
C$_6$ 			+ CRPHOT 	& \ra &	C$_4$ 	+ C\dtwo 		&	1.43 {\rm x} 10$^{-15}$	&	0.00	&	0	&	2\\
C$_6$ 			+ CRPHOT	& \ra &	C\dthree~ + C\dthree	&	1.04 {\rm x} 10$^{-14}$	&	0.00	&	0	&	2\\
C$_7$ 			+ CRPHOT	& \ra &	C$_6$ 	+ C 			&	1.30 {\rm x} 10$^{-16}$	&	0.00	&	0	&	2\\
C$_7$ 			+ CRPHOT 	& \ra &	C$_5$	+ C\dtwo		&	2.47 {\rm x} 10$^{-15}$	&	0.00	&	0	&	2\\
C$_7$			+ CRPHOT 	& \ra &	C$_4$ 	+ C\dthree	&	1.04 {\rm x} 10$^{-14}$	&	0.00	&	0	&	2\\
C$_8$ 			+ CRPHOT	& \ra &	C$_7$ 	+ C 			&	3.9 {\rm x} 10$^{-16}$	&	0.00	&	0	&	2\\
C$_8$			+ CRPHOT 	& \ra &	C$_6$ 	+ C\dtwo		&	1.30 {\rm x} 10$^{-16}$	&	0.00	&	0	&	2\\
C$_8$ 			+ CRPHOT 	& \ra &	C$_5$ 	+ C\dthree	&	1.17 {\rm x} 10$^{-15}$	&	0.00	&	0	&	2\\
C$_8$ 			+ CRPHOT 	& \ra &	C$_4$ 	+ C$_4$		&	7.80 {\rm x} 10$^{-16}$	&	0.00	&	0	&	2\\
C$_9$			+ CRPHOT 	& \ra &	C$_7$ 	+ C\dtwo		&	7.80 {\rm x} 10$^{-16}$	&	0.00	&	0	&	2\\
C$_9$			+ CRPHOT 	& \ra &	C$_6$ 	+ C\dthree	&	8.58 {\rm x} 10$^{-15}$	&	0.00	&	0	&	2\\
C$_9$			+ CRPHOT 	& \ra &	C$_5$ 	+ C$_4$		&	3.64 {\rm x} 10$^{-15}$	&	0.00	&	0	&	2\\
CH				+ PHOTON 	&\ra &	CH\jon~	+	e\ijon&		7.60{\rm x} 10$^{-10}$	&	0.00	&	3.80	&	4\\
CH				+ PHOTON 	&\ra &	C		+	H	&		9.20{\rm x} 10$^{-10}$	&	0.00	&	1.72	&	4\\
C\dtwo H 			+ PHOTON	& \ra &	C\dtwo~	+ H 			&	8.10 {\rm x} 10$^{-10}$	&	0.00	&	1.7	&	2\\
C\dtwo H 			+ PHOTON 	& \ra &	CH		+ C			&	1.90 {\rm x} 10$^{-10}$	&	0.00	&	1.7	&	2\\
C\dthree H 		+ PHOTON 	& \ra &	C\dthree~ + H 			&	6.50 {\rm x} 10$^{-10}$	&	0.00	&	1.7	&	2\\
C\dthree H 		+ PHOTON 	& \ra &	C\dtwo H 	+ C			&	3.30 {\rm x} 10$^{-10}$	&	0.00	&	1.7	&	2\\
C\dthree H 		+ PHOTON 	& \ra &	C\dtwo~ 	+ CH 		&	2.00 {\rm x} 10$^{-11}$	&	0.00	&	1.7	&	2\\
C\dthree H\dtwo~ 	+ PHOTON 	& \ra &	C\dtwo H\dtwo + C		&	1.33 {\rm x} 10$^{-9}$	&	0.00	&	1.7	&	2\\
C\dthree H\dtwo~ 	+ PHOTON 	& \ra &	C\dtwo H	+ CH			&	1.16 {\rm x} 10$^{-10}$	&	0.00	&	1.7	&	2\\
C\dthree H\dtwo~ 	+ PHOTON 	& \ra &	CH\dtwo~ + C\dtwo		&	1.16 {\rm x} 10$^{-10}$	&	0.00	&	1.7	&	2\\
C\dthree H\dtwo~ 	+ PHOTON 	& \ra &	C\dthree H	+ H		&	1.33 {\rm x} 10$^{-9}$	&	0.00	&	1.7	&	2\\
C\dthree H\dtwo~ 	+ PHOTON 	& \ra &	C\dthree~ + H\dtwo 		&	6.67 {\rm x} 10$^{-10}$	&	0.00	&	1.7	&	2\\
C$_4$ 			+ PHOTON 	& \ra &	C\dthree~	+ C 			&	3.08 {\rm x} 10$^{-10}$	&	0.00	&	1.7	&	2\\
C$_4$ 			+ PHOTON 	& \ra &	C\dtwo~ 	+ C\dtwo		&	9.02 {\rm x} 10$^{-11}$	&	0.00	&	1.7	&	2\\
C$_4$H 			+ PHOTON 	& \ra &	C\dthree H 	+ C		&	5.20 {\rm x} 10$^{-10}$	&	0.00	&	1.7	&	2\\
C$_4$H 			+ PHOTON	& \ra &	C$_4$ 	+ H			&	1.16 {\rm x} 10$^{-9}$	&	0.00	&	1.7	&	2\\
C$_4$H 			+ PHOTON 	& \ra &	C\dtwo H 	+ C\dtwo		&	3.20 {\rm x} 10$^{-9}$	&	0.00	&	1.7	&	2\\
C$_5$ 			+ PHOTON 	& \ra &	C\dthree~ + C\dtwo		&	8.70 {\rm x} 10$^{-12}$	&	0.00	&	1.7	&	2\\
C$_5$ 			+ PHOTON	& \ra &	C$_4$ 	+ C			&	1.30 {\rm x} 10$^{-12}$	&	0.00	&	1.7	&	2\\
C$_6$ 			+ PHOTON 	& \ra &	C$_5$	+ C			&	9.00 {\rm x} 10$^{-11}$	&	0.00	&	1.7	&	2\\
C$_6$			+ PHOTON	& \ra &	C$_4$ 	+ C\dtwo		&	1.10 {\rm x} 10$^{-10}$	&	0.00	&	1.7	&	2\\
C$_6$ 			+ PHOTON	& \ra &	C\dthree~ + C\dthree	&	8.00 {\rm x} 10$^{-10}$	&	0.00	&	1.7	&	2\\
C$_7$ 			+ PHOTON 	& \ra &	C$_6$ 	+ C			&	1.00 {\rm x} 10$^{-11}$	&	0.00	&	1.7	&	2\\
C$_7$ 			+ PHOTON 	& \ra &	C$_5$ 	+ C\dtwo		&	1.90 {\rm x} 10$^{-10}$	&	0.00	&	1.7	&	2\\
C$_7$ 			+ PHOTON	& \ra &	C$_4$ 	+ C\dthree	&	8.00 {\rm x} 10$^{-10}$	&	0.00	&	1.7	&	2\\
C$_8$			+ PHOTON 	& \ra &	C$_7$ 	+ C 			&	3.00 {\rm x} 10$^{-11}$	&	0.00	&	1.7	&	2\\
C$_8$ 			+ PHOTON 	& \ra &	C$_6$ 	+ C\dtwo		&	1.00 {\rm x} 10$^{-11}$	&	0.00	&	1.7	&	2\\
C$_8$ 			+ PHOTON 	& \ra &	C$_5$ 	+ C\dthree	&	9.00 {\rm x} 10$^{-10}$	&	0.00	&	1.7	&	2\\
C$_8$ 			+ PHOTON	& \ra &	C$_4$	+ C$_4$		&	6.00 {\rm x} 10$^{-11}$	&	0.00	&	1.7	&	2\\
C$_9$			+ PHOTON	& \ra &	C$_7$	+ C\dtwo		&	6.00 {\rm x} 10$^{-11}$	&	0.00	&	1.7	&	2\\
C$_9$			+ PHOTON	& \ra &	C$_6$ 	+ C$_3$		&	6.60 {\rm x} 10$^{-10}$	&	0.00	&	1.7	&	2\\
C$_9$ 			+ PHOTON	& \ra &	C$_5$ 	+ C$_4$		&	2.80 {\rm x} 10$^{-10}$	&	0.00	&	1.7	&	2\\
C$_{10}$			+ PHOTON	& \ra &	C$_9$	+ C			&	1.14 {\rm x} 10$^{-11}$	&	0.00	&	1.7	&	2\\
C$_{10}$			+ PHOTON	& \ra &	C$_8$ 	+ C\dtwo		&	1.14 {\rm x} 10$^{-11}$	&	0.00	&	1.7	&	2\\
C$_{10}$			+ PHOTON	& \ra &	C$_7$	+ C\dthree	&	7.98 {\rm x} 10$^{-10}$	&	0.00	&	1.7	&	2\\
C$_{10}$			+ PHOTON	& \ra &	C$_6$ 	+ C$_4$		&	3.42 {\rm x} 10$^{-11}$	&	0.00	&	1.7	&	2\\
C$_{10}$			+ PHOTON	& \ra &	C$_5$	+ C$_5$		&	2.50 {\rm x} 10$^{-10}$	&	0.00	&	1.7	&	2\\
\multicolumn{5}{c}{Removed reactions} \\
CH\dthree OH\dtwo\jon~ +H\dtwo CO&\ra&	H$_7$C\dtwo O\dtwo\jon	&		$- -$ 			&	$- -$ &	$- -$ 	&	7\\
H\dthree CO\jon~	+ CH$_4$ 	& \ra & 	CH\dthree OCH$_4$\jon 	&		$- -$ 			&	$- -$ 	&	$- -$ 	&	4\\
HCO\jon~			+ CH$_4$ 	& \ra & 	CH\dthree CH\dtwo O\jon & 		$- -$ 			&	$- -$ 	&	$- -$ 	&	4\\
C$_4$H\jon~ 		+ e\ijon		& \ra &	C\dthree~	+ CH 		&		$- -$ 			&	$- -$ &	$- -$ 	&	2\\
C$_9$\jon~		+ e\ijon		& \ra &	C$_8$		+ C		&		$- -$ 			&	$- -$ &	$- -$ 	&	2\\
C$_9$			+ PHOTON	& \ra &	C$_8$		+ C		&		$- -$ 			&	$- -$ &	$- -$	&	2\\
C$_9$ 			+ CRPHOT 	& \ra &	C$_8$		+ C 		&		$- -$ 			&	$- -$	&	$- -$ 	&	2\\
HNO				+ O			& \ra &	NO\dtwo~	+ H 			&		$- -$ 			&	$- -$ &	$- -$ 	&	1\\
O				+ NH 		& \ra &	OH		+ N 			&		$- -$ 			&	$- -$ &	$- -$ 	&	1\\
NH\dtwo~			+ O			& \ra &	NO		+ H\dtwo		&		$- -$ 			&	$- -$ &	$- -$ 	&	1\\
NH\dthree~ 		+ CN 		& \ra &	NH\dtwo CN	+ H		&		$- -$ 			&	$- -$ &	$- -$ 	&	1\\[1.0pt]
\hline\\[1.0pt]
\multicolumn{7}{l}{Cosmic ray-induced photoionization: $k = \alpha \zeta_{\rm CR}$ } \\
\multicolumn{7}{l}{Photoreactions: $k = \alpha $e$^{-\gamma A_{\rm V}}$ }	\\
\multicolumn{7}{l}{Ion-neutral reactions (Kooji formula): $k = \alpha (T/300)^\beta $e$^{-\gamma / T}$ }	\\
\multicolumn{7}{l}{(1) \citet{2010SSRv..156...13W}; (2) \citet{2010A&A...524A..39C}; (3) \citet{2004A&A...424..905R} }\\
\multicolumn{7}{l}{(4) KIDA database; (5) \citet{2011ApJ...728...71L}; (6) \citet{2006A&A...457..927G}}\\
\multicolumn{7}{l}{(7) \citet{2004ApJ...611..605H} }
\end{longtable}
\end{center}

\clearpage
\begin{center}
\begin{longtable}{lclcccc}
\caption[Added and updated deuterium reactions. ]{Added and updated deuterium reactions. } \label{tab:netwDeut} \\
\tablewidth{0.9\textwidth}
Reaction				&	&							&	$\alpha$				& $\beta$	& $\gamma$&	Refs.\\
\endhead
\hline
%
C\dtwo D 			+ H 			& \ra &	C\dtwo H 			+ D 			& 	5.00 {\rm x} 10$^{-11}$ 	&	0.50	&	832	&	9	\\
C\dtwo H 			+ D 			& \ra & 	C\dtwo D 			+ H 			&	5.00 {\rm x} 10$^{-11}$	& 	0.50	&	250	&	9	\\
C\dtwo H\dtwo\jon~	+ HD			& \ra	&	C\dtwo HD\jon~	+ H\dtwo		&	1.00 {\rm x} 10$^{-9}$	&	0.00	&	0	&	5	\\
C\dtwo HD\jon~	+ H\dtwo		& \ra	&	C\dtwo H\dtwo\jon~	+ HD			&	2.50 {\rm x} 10$^{-9}$	&	0.00	&	550	&	5	\\
CH\dthree\jon~ 	+ D\dtwo		& \ra &	CH\dtwo D\jon~	+ HD			&	4.40 {\rm x} 10$^{-10}$ 	&	0.00	&	0	&	3	\\
CH\dthree\jon~ 	+ D\dtwo		& \ra & 	CHD\dtwo\jon~		+ H\dtwo		&	6.60 {\rm x} 10$^{-10}$ 	&	0.00	&	0	&	3	\\
CH\dthree\jon~		+ HD			& \ra	&	CH\dtwo D\jon~	+	H\dtwo	&	1.30 {\rm x} 10$^{-9}$	&	0.00	&	0	&	3	\\
CH\dtwo D\jon~	+ H\dtwo		& \ra	&	CH\dthree\jon~ 	+	HD		&	8.70 {\rm x} 10$^{-10}$	&	0.00	&	370	&	3	\\
CH\dtwo D\jon~	+ HD			& \ra 	&	CHD\dtwo\jon~		+ H\dtwo		&	1.60 {\rm x} 10$^{-9}$ 	&	0.00	&	0	&	3	\\
CH\dtwo D\jon~	+ HD			& \ra &	CH\dthree\jon~		+ D\dtwo		&	4.40 {\rm x} 10$^{-10}$ 	&	0.00	&	400	&	3	\\
CH\dtwo D\jon~	+ D\dtwo		& \ra &	CHD\dtwo\jon~		+ HD			&	1.20 {\rm x} 10$^{-9}$ 	&	0.00	&	0	&	3	\\
CHD\dtwo\jon~ 	+ H\dtwo		& \ra &	CH\dtwo D\jon~	+ HD			&	1.60 {\rm x} 10$^{-9}$ 	&	0.00	&	370	&	3	\\
CHD\dtwo\jon~ 	+ H\dtwo		& \ra &	CH\dthree\jon~		+ D\dtwo		&	6.60 {\rm x} 10$^{-10}$ 	&	0.00	&	400	&	3	\\
CHD\dtwo\jon~ 	+ HD			& \ra 	&	CD\dthree\jon~		+ H\dtwo		&	1.50 {\rm x} 10$^{-9}$ 	&	0.00	&	0	&	3	\\
CHD\dtwo\jon~ 	+ HD			& \ra &	CH\dtwo D\jon~	+ D\dtwo		&	1.20 {\rm x} 10$^{-9}$ 	&	0.00	&	400	&	3	\\
CD\dthree\jon~ 	+ H\dtwo		& \ra &	CHD\dtwo\jon~ 	+ HD			&	1.50 {\rm x} 10$^{-9}$ 	&	0.00	&	370	&	3	\\
D\jon~			+ H 			& \ra	&	H\jon			+ D 			&	1.00 {\rm x} 10$^{-9}$	&	0.00	&	0	&	6	\\
D\jon~			+ H\dtwo		& \ra	&	H\jon~			+ HD			&	2.10 {\rm x} 10$^{-9}$	&	0.00	&	0	&	3	\\
D\dthree\jon~		+ H 			& \ra &	HD\dtwo\jon~		+ D 			&	2.00 {\rm x} 10$^{-9}$ 	&	0.00	&	655	&	4	\\
D\dthree\jon~ 		+ H\dtwo		& \ra &	H\dtwo D\jon~		+ D\dtwo		&	7.00 {\rm x} 10$^{-10}$ 	&	0.00	&	341	&	2	\\
D\dthree\jon~		+ H\dtwo		& \ra & 	HD\dtwo\jon~		+ HD			&	2.00 {\rm x} 10$^{-10}$ 	&	0.00	&	234	&	15	\\
D\dthree\jon~		+ HD			& \ra &	HD\dtwo\jon~		+ D\dtwo		&	8.70 {\rm x} 10$^{-10}$ 	&	0.00	&	159	&	2	\\
DCN				+ H 			& \ra & 	HCN 			+ D 			&	1.00 {\rm x} 10$^{-10}$ 	&	0.50	&	500	&	9$^{*}$ \\
DCO\jon~			+ H 			& \ra &	HCO\jon~			+ D 			&	2.20 {\rm x} 10$^{-9}$	&	0.00	&	796	&	7	\\
H\jon~			+ D 			& \ra	&	D\jon			+ H 			&	1.00 {\rm x} 10$^{-9}$	&	0.00	&	41	&	6	\\
H\jon~			+ HD			& \ra	&	D\jon~			+ H\dtwo		&	1.00 {\rm x} 10$^{-9}$	&	0.00	&	464	&	3	\\
H\dtwo D\jon~ 		+ H 			& \ra &	H\dthree\jon~		+ D 			&	1.00 {\rm x} 10$^{-9}$	&	0.00	&	632	&	7	\\
H\dtwo D\jon~		+ H\dtwo		& \ra	& 	H\dthree \jon+	HD				&	3.50 {\rm x} 10$^{-9}$	&	0.00	&	220	&	15	\\
H\dthree\jon~		+ D 			& \ra & 	H\dtwo D\jon~		+ H 			&	1.00 {\rm x} 10$^{-9}$	&	0.00	&	0	&	7$^{*}$	\\
H\dthree\jon~		+ D\dtwo		& \ra &	H\dtwo D\jon~		+ HD			&	3.50 {\rm x} 10$^{-10}$ 	&	0.00	&	0	&	2	\\
H\dthree\jon~ 		+ D\dtwo		& \ra &	HD\dtwo\jon~		+ H\dtwo		&	1.10 {\rm x} 10$^{-9}$ 	& 	0.00	&	0	&	2	\\
H\dthree\jon~		+ HD			& \ra 	&	H\dtwo D\jon~+ 	H\dtwo		&	3.50 {\rm x} 10$^{-10}$	&	0.00 	&	0	&	15	\\
H\dtwo D\jon~ 		+ D 			& \ra &	HD\dtwo\jon~ 		+ H 			&	2.00 {\rm x} 10$^{-9}$ 	&	0.00	&	0	&	4	\\
H\dtwo D\jon~ 		+ HD			& \ra &	H\dthree\jon~		+ D\dtwo		& 	3.50 {\rm x} 10$^{-10}$ 	&	0.00	&	63	&	2	\\
H\dtwo D\jon~ 		+ HD			& \ra &	HD\dtwo\jon~		+ H\dtwo		& 	2.60 {\rm x} 10$^{-10}$ 	& 	0.00	& 	0	&	15	\\
H\dtwo D\jon~		+ D\dtwo		& \ra &	HD\dtwo\jon~		+ HD			&	7.00 {\rm x} 10$^{-10}$ 	&	0.00	&	0	&	2	\\
H\dtwo D\jon~ 		+ D\dtwo		& \ra &	D\dthree\jon~		+ H\dtwo		&	7.00 {\rm x} 10$^{-10}$ 	&	0.00	&	0	&	2	\\
HCN				+ D 			& \ra & 	DCN 			+ H 			&	1.00 {\rm x} 10$^{-10}$ 	&	0.50	&	500	&	9$^{*}$	\\
HCO\jon~			+ D 			& \ra & 	DCO\jon~			+ H 			&	1.00 {\rm x} 10$^{-9}$	&	0.00	&	0	&	7	\\
HD\dtwo\jon~		+ D 			& \ra &	D\dthree\jon~		+ H 			&	2.00 {\rm x} 10$^{-9}$ 	&	0.00	&	0	&	4	\\
HD\dtwo\jon~		+ H	 		& \ra &	H\dtwo D\jon~		+ D 			&	2.00 {\rm x} 10$^{-9}$ 	&	0.00	&	550	&	4	\\
HD\dtwo\jon~		+ D\dtwo		& \ra &	D\dthree\jon~		+ HD			&	8.70 {\rm x} 10$^{-10}$ 	&	0.00	&	0	&	2	\\
HD\dtwo\jon~		+ H\dtwo		& \ra &	H\dthree\jon~		+ D\dtwo		&	1.10 {\rm x} 10$^{-9}$ 	&	0.00	&	251	&	2	\\
HD\dtwo\jon~ 		+ H\dtwo		& \ra &	H\dtwo D\jon~ 		+ HD			&	2.60 {\rm x} 10$^{-10}$ 	&	0.00	&	187	&	15	\\
HD\dtwo\jon~		+ HD			& \ra &	H\dtwo D\jon~		+ D\dtwo		&	7.00 {\rm x} 10$^{-10}$ 	&	0.00	&	107	&	2	\\
HD\dtwo\jon~		+ HD			& \ra &	D\dthree\jon~		+ H\dtwo		&	2.00 {\rm x} 10$^{-10}$ 	&	0.00	&	0	&	15	\\
N\dtwo H\jon~ 		+ D 			& \ra &	N\dtwo D\jon~		+ H 			&	1.00 {\rm x} 10$^{-9}$	&	0.00	&	0	&	7	\\
N\dtwo D\jon~ 		+ H 			& \ra &	N\dtwo H\jon~		+ D 			&	2.20 {\rm x} 10$^{-9}$	&	0.00	&	550	&	7	\\
OH 		 		+ D 			& \ra &	OD				+ H 			&	1.30 {\rm x} 10$^{-10}$	&	0.50	&	0	&	8	\\
OD 				+ H 			& \ra & 	OH				+ D 			&	1.30 {\rm x} 10$^{-10}$	&	0.50	&	810	&	8	\\
C\dtwo HD\jon~ 	+ H\dtwo		& \ra &	C\dtwo H\dthree D\jon~		&	3.39 {\rm x} 10$^{-14}$ 	&	-1.50	&	0	&	12	\\
CH\dthree\jon~ 	+ D\dtwo 		& \ra &	CH\dthree D\dtwo\jon~		&	3.50 {\rm x} 10$^{-14}$ 	&	-1.00	&	0	&	12	\\
CH\dthree\jon~ 	+ D\dtwo O	& \ra &	CH\dthree OD\dtwo\jon~		&	1.65 {\rm x} 10$^{-11}$ 	&	-1.70	&	0	&	12	\\
CH\dthree\jon~ 	+ HDO		& \ra &	CH\dthree OHD\jon~		&	1.10 {\rm x} 10$^{-11}$ 	&	-1.70	&	0	&	12	\\
CH\dtwo D\jon~ 	+ D\dtwo O	& \ra &	CH\dtwo DOD\dtwo\jon~		&	2.20 {\rm x} 10$^{-11}$ 	&	-1.70	&	0	&	12	\\
CH\dtwo D\jon~ 	+ H\dtwo		& \ra &	CH$_4$D\jon~				&	2.00 {\rm x} 10$^{-14}$ 	&	-1.00	&	0	&	12	\\
CH\dtwo D\jon~ 	+ H\dtwo O	& \ra &	CH\dtwo DOH\dtwo\jon~		&	1.10 {\rm x} 10$^{-11}$ 	&	-1.70	&	0	&	12	\\
CH\dtwo D\jon~ 	+ HDO		& \ra &	CH\dtwo DOHD\jon~		&	1.65 {\rm x} 10$^{-11}$ 	&	-1.70	&	0	&	12	\\
CH\dtwo D\jon~ 	+ HD			& \ra 	&	CH\dthree D\dtwo\jon~		&	3.50 {\rm x} 10$^{-14}$ 	&	-1.00	&	0	&	12	\\
CHD\dtwo\jon~ 	+ D\dtwo O 	& \ra &	CHD\dtwo OD\dtwo\jon~		&	2.75 {\rm x} 10$^{-11}$ 	&	-1.70	&	0	&	12	\\
CHD\dtwo\jon~ 	+ H\dtwo		& \ra &	CH\dthree D\dtwo\jon~		& 	3.50 {\rm x} 10$^{-14}$ 	&	-1.00	&	0	&	12	\\
CHD\dtwo\jon~ 	+ H\dtwo O	& \ra &	CHD\dtwo OH\dtwo\jon~		&	1.65 {\rm x} 10$^{-11}$ 	&	-1.70	&	0	&	12	\\
CHD\dtwo\jon~ 	+ HDO		& \ra &	CHD\dtwo OHD\jon~		&	2.20 {\rm x} 10$^{-11}$ 	&	-1.70	&	0	&	12	\\
CD\dthree\jon~ 	+ D\dtwo 		& \ra &	CD\dthree OD\dtwo\jon~		&	2.75 {\rm x} 10$^{-11}$ 	&	-1.70	&	0	&	12	\\
CD\dthree\jon~ 	+ H\dtwo		& \ra &	CH\dtwo D\dthree\jon~		&	6.30 {\rm x} 10$^{-14}$ 	&	-1.00	&	0	&	12	\\
CD\dthree\jon~ 	+ H\dtwo O 	& \ra &	CD\dthree OH\dtwo\jon~		&	2.20 {\rm x} 10$^{-11}$ 	&	-1.70	&	0	&	12	\\
CD\dthree\jon~ 	+ HDO		& \ra &	CD\dthree OHD\jon~		&	2.75 {\rm x} 10$^{-11}$ 	&	-1.70	&	0	&	12	\\
CD\dthree OCD$_4$\jon~ + e\ijon 	& \ra & 	CD\dthree OCD\dthree~ + D	&	8.50 {\rm x} 10$^{-8}$ 	&	-0.70	&	0	&	14	\\
CD\dthree OCD$_4$\jon~ + e\ijon 	& \ra & 	CD\dthree OD + CD\dthree	&	9.18 {\rm x} 10$^{-7}$ 	&	-0.70	&	0	&	14	\\
CD\dthree OCD$_4$\jon~ + e\ijon 	& \ra & 	CD\dthree~ + CD$_4$ + O 	&	6.97 {\rm x} 10$^{-7}$ 	&	-0.70	&	0	&	14	\\
%
D\dthree\jon~		 + e$^-$		& \ra &	D\dtwo~	+	D 			&	5.40 {\rm x} 10$^{-9}$ 	&	-0.50	&	0	&	12	\\
D\dthree\jon~		 + e$^-$		& \ra &	D 		+ 	D +	D 		&	2.16 {\rm x} 10$^{-8}$ 	&	-0.50	&	0	&	12	\\
H\dtwo D\jon~ 		+ e$^-$		& \ra &	D 		+ H 	+ H 			&	4.38 {\rm x} 10$^{-8}$ 	&	-0.50	&	0	&	10	\\
H\dtwo D\jon~ 		+ e$^-$		& \ra &	H\dtwo~		+ D 			&	4.20 {\rm x} 10$^{-9}$ 	&	-0.50	&	0	&	10	\\
H\dtwo D\jon~ 		+ e$^-$		& \ra &	HD		 	+ H 			&	1.20 {\rm x} 10$^{-8}$ 	&	-0.50	&	0	&	10	\\
HD\dtwo\jon~	 	+ e$^-$		& \ra &	D 		+	D 	+ H 		&	4.38 {\rm x} 10$^{-8}$ 	&	-0.50	&	0	&	11	\\
HD\dtwo\jon~		+ e$^-$		& \ra &	D\dtwo~	+	H 			&	1.20 {\rm x} 10$^{-8}$ 	&	-0.50	&	0	&	11	\\
HD\dtwo\jon~ 		+ e$^-$		& \ra &	HD		+	D 			&	4.20 {\rm x} 10$^{-9}$ 	&	-0.50	&	0	&	11	\\
DCNH\jon~		+ e\ijon		& \ra &	DCN		+ H				&	2.33 {\rm x} 10$^{-7}$ 	&	-0.50	&	0	&	13	\\	
DCNH\jon~		+ e\ijon		& \ra &	HNC		+ D				&	1.16 {\rm x} 10$^{-7}$ 	&	-0.50	&	0	&	13	\\	
HCND\jon~		+ e\ijon		& \ra &	DNC		+ H				&	2.33 {\rm x} 10$^{-7}$ 	&	-0.50	&	0	&	13	\\	
HCND\jon~		+ e\ijon		& \ra &	HCN		+ D				&	1.16 {\rm x} 10$^{-7}$ 	&	-0.50	&	0	&	13	\\	
HDCN\jon~		+ e\ijon		& \ra &	NHD		+ C				&	1.75 {\rm x} 10$^{-7}$ 	&	-0.50	&	0	&	13	\\	
HDNC\jon~		+ e\ijon		& \ra &	HNC		+ D				&	0.58 {\rm x} 10$^{-7}$ 	&	-0.50	&	0	&	13	\\	
HDNC\jon	~		+ e\ijon		& \ra &	DNC		+ H				&	1.16 {\rm x} 10$^{-7}$ 	&	-0.50	&	0	&	13	\\	
%
\\[1.0pt]
\multicolumn{5}{c}{Removed reactions} \\[1.0pt]
C\dtwo H\dtwo\jon~ + HD			& \ra &	C\dtwo H\dthree D\jon		&	$- - $				 	&$- - $	&$- - $	&	11	\\
CH\dthree\jon~ 	+ HD			& \ra &	CH$_4$D\jon 				&	$- - $				 	&$- - $	&$- - $	&	11	\\
CH\dtwo D\jon~ 	+ D\dtwo		& \ra &	CH\dtwo D\dthree\jon		& 	$- - $					&$- - $	&$- - $	&	11	\\
CHD\dtwo\jon~ 	+ HD			& \ra & 	CH\dtwo D\dthree\jon		&	 $- - $				&$- - $	&$- - $	&	11	\\[1.0pt]
\hline\\[1.0pt]
\multicolumn{7}{l}{* Estimate}\\[0.1pt]
\multicolumn{7}{l}{Cosmic ray-induced photoionization: $k = \alpha \zeta_{\rm CR}$}\\
\multicolumn{7}{l}{Photoreactions: $k = \alpha $e$^{-\gamma A_{\rm V}}$ }\\
\multicolumn{7}{l}{Ion-neutral reactions (Kooji formula): $k = \alpha (T/300)^\beta $e$^{-\gamma / T}$}\\
\multicolumn{7}{l}{(1) \citet{2005A&A...440..583C}; (2) \citet{doi:10.1021/j100198a030}; (3) \citet{10.1063/1.444002, 1982ApJ...263..123S}} \\\\[0.1pt]
\multicolumn{7}{l}{(4) \citet{2004A&A...418.1035W}; (5) \citet{1987ApJ...312..351H}; (6) \citet{RevModPhys.48.513}}\\\\[0.1pt]
\multicolumn{7}{l}{(7) \citet{1985ApJ...294L..63A}; (8) \citet{1985ApJ...289..618C}; (9) \citet{1992A&A...256..595S}}\\\\[0.1pt]
\multicolumn{7}{l}{(10) \citet{1994Sci...263..785S}; (11) \citet{2004A&A...424..905R}; (12) \citet{PhysRevLett.79.395}}\\\\[0.1pt]
\multicolumn{7}{l}{(13) \citet{2005A&A...438..585R}; (14) \citet{2010A&A...514A..83H}; (15) \citet{2002P&SS...50.1275G} }
\end{longtable}
\end{center}

\clearpage
\begin{center}
\begin{longtable}{lcll|lcll}
\caption{Problematic reactions that show correlation coefficients $> 0.05$ and strong chemical ties to any of the following species, including their isotopologues and isomers: water, H\dthree\jon, HCO\jon~and HCN. An asterisk alongside the uncertainty signifies a reaction resulting from cloning the network. \label{tab:probreact}}	\\
\tablewidth{0.95\textwidth}
Reaction		&			&									&	Uncertainty& 	Reaction				&		&							&	Uncertainty\\
\endhead
\hline\\[0.1pt]
\multicolumn{4}{l|}{\textit{Problematic reactions connected to all species}}\\
H\dtwo~		+	CRP		&\ra&	H\dtwo\jon~	+	e\ijon	&	2.00		& He			+	CRP		&\ra&	He\jon~		+	e\ijon	&	2.00		\\\\[0.1pt]
\multicolumn{4}{l|}{\textit{H\dthree\jon~isotopologues}}	\\
H\dthree\jon~	+	e\ijon	&\ra&	H	+	H	+	H		&	2.00		&	H\dtwo D\jon~	+	e\ijon	&\ra&	D	+	H	+	H	&	2.00		\\
HD\dtwo\jon~	+	e\ijon	&\ra&	D	+	D	+	H		&	2.00		&	H\dthree\jon~	+	HD		&\ra&	H\dtwo D\jon~	+	H\dtwo	&	1.25	\\
H\dtwo D\jon~	+	HD		&\ra&	HD\dtwo\jon~	+	H\dtwo	&	1.25		&	HD\dtwo\jon~	+	HD		&\ra&	D\dthree\jon~	+	H\dtwo	&	1.25	\\
H\dtwo D\jon~	+	H\dtwo	&\ra&	H\dthree\jon~	+	HD		&	2.00		&	HD\dtwo\jon~	+	H\dtwo	&\ra&	H\dtwo D\jon~	+	HD		&	2.00	\\
D\dthree\jon~	+	H\dtwo	&\ra&	HD\dtwo\jon~	+	HD		&	2.00		&	H\dthree\jon~	+	D\dtwo	&\ra&	HD\dtwo\jon~	+	H\dtwo	&	2.00	\\
H\dtwo D\jon~	+	D\dtwo	&\ra&	D\dthree\jon~	+	H\dtwo	&	2.00		&	H\dthree\jon~	+	D		&\ra&	H\dtwo D\jon~	+	H		&	2.00	\\
H\dtwo D\jon~	+	D		&\ra&	HD\dtwo\jon~	+	H		&	2.00		&	HD\dtwo\jon~	+	D		&\ra&	D\dthree\jon~	+	H		&	2.00	\\
H\dthree\jon~	+	CO		&\ra&	HCO\jon~		+	H\dtwo	&	1.25		&	H\dtwo D\jon~	+	CO		&\ra&	DCO\jon~		+	H\dtwo	&	1.25$^*$	\\	
HD\dtwo\jon~	+	CO		&\ra&	DCO\jon~		+	HD		&	1.25$^*$	&	D\dthree\jon~	+	CO		&\ra&	DCO\jon~		+	D\dtwo	&	1.25$^*$	\\
H\dthree\jon~	+	OH		&\ra&	H\dtwo O\jon~	+H\dtwo		&	2.00		&	H\dtwo D\jon~	+	OH		&\ra&	H\dtwo O\jon~	+	HD		&	2.00		\\ 	%
H\dthree\jon~	+	OD		&\ra&	H\dtwo O\jon~	+	HD		&	2.00		&			&	\\
\hline\\[0.1pt]
\multicolumn{4}{l|}{\textit{HCO\jon~isotopologues}}\\
%
H\dthree\jon~	+	CO		&\ra&	HCO\jon~		+	H\dtwo	&	1.25		&	H\dthree\jon~	+	CO		&\ra&	HOC\jon~		+	H\dtwo	&	1.25		\\	
H\dtwo D\jon~	+	CO		&\ra&	DCO\jon~		+	H\dtwo	&	1.25		&	H\dtwo D\jon~	+	CO		&\ra&	DOC\jon~		+	H\dtwo	&	1.25		\\
HCO\jon~		+	e\ijon	&\ra&	CO			+	H		&	1.25		&	HOC\jon~		+	e\ijon	&\ra&	CO			+	H		&	1.25		\\
DCO\jon~ 	+	e\ijon	&\ra&	CO			+	D$^*$	&	1.25		&	DOC\jon~		+	e\ijon	&\ra&	CO			+	D		&	1.25$^*$	\\
HOC\jon~		+	H\dtwo	&\ra&	HCO\jon~		+	H\dtwo	&	2.00		&	HCO\jon~		+	D		&\ra&	DCO\jon~		+	H		&	2.00		\\
DOC\jon~		+	H\dtwo	&\ra&	DCO\jon~		+	H\dtwo	&	2.00		&	DOC\jon~		+	H\dtwo	&\ra&	HCO\jon~		+	HD		&	2.00		\\
HCO\jon~		+	OH		&\ra&	H\dtwo O\jon~	+	CO		&	2.00		&	H\jon~		+	D		&\ra&	D\jon~		+	H		&	2.00		\\
\hline\\[0.1pt]
\multicolumn{4}{l|}{\textit{HCN~isotopologues}}\\
%
H\dtwo CN\jon~	+	e\ijon	&\ra&	HCN			+	H		&	2.00		&	H\dtwo CN\jon~	+	e\ijon	&\ra&	HNC			+	H		&	2.00		\\
HDCN\jon~		+	e\ijon	&\ra&	HCN			+	D		&	2.00$^*$	&	HDCN\jon	~		+	e\ijon	&\ra&	DCN			+	H		&	2.00		\\
HDCN\jon~		+	e\ijon	&\ra&	HNC			+	D		&	2.00		&	HDCN\jon~		+	e\ijon	&\ra&	DNC			+	H		&	2.00$^*$	\\
H\dtwo CN\jon~	+	e\ijon	&\ra&	CN	+	H	+	H		&	2.00		&	HDCN\jon~		+	e\ijon	&\ra&	CN	+	H	+	D		&	2.00		\\
C\jon~			+	HNC		&\ra&	H\dtwo CN\jon~+	H\dtwo O	&	2.00$^*$	&	C\jon~			+	HCN		&\ra&	C\dtwo N\jon~	+	H		&	2.00		\\
C\jon~			+	DNC		&\ra&	C\dtwo N\jon~	+	D		&	2.00$^*$	&	C\jon~			+	DCN		&\ra&	C\dtwo N\jon~	+	D		&	2.00$^*$	\\
C\jon~			+	HCN		&\ra&	CNC\jon~		+	H		&	2.00		&	C\jon~			+	NH\dtwo	&\ra&	HCN\jon~		+	H		&	2.00		\\
C\jon~			+	NH\dthree&\ra&	H\dtwo CN\jon~+	H		&	2.00		&	NH\dtwo~			+	O		&\ra&	HNO			+	H		&	2.00		\\
N				+	CN		&\ra&	C			+	N\dtwo	&	2.00		&	&	\\
\hline\\[0.1pt]
\multicolumn{4}{l|}{\textit{H\dtwo O~isotopologues}}\\
H\dthree O\jon~	+	e\ijon	&\ra&	H\dtwo O		+	H		&	1.25$^*$	&	H\dtwo DO\jon~	+	e\ijon	&\ra&	HDO			+	H		&	1.25$^*$	\\
HD\dtwo O\jon~	+	e\ijon	&\ra&	D\dtwo O		+	H		&	1.25$^*$	&	H\dthree O\jon~	+	e\ijon	&\ra&	OH	+	H	+	H		&	1.25		\\
H\dtwo DO\jon~	+	e\ijon	&\ra&	OH	+	D	+	H		&	1.25$^*$	&	HD\dtwo O\jon~	+	e\ijon	&\ra&	OD	+	D	+	H		&	1.25$^*$	\\
H\dthree\jon~	+	H\dtwo O	&\ra&	H\dthree O\jon~	+	H\dtwo	&	1.25		&	H\dthree\jon~		+	HDO		&\ra&	H\dthree O\jon~	+	HD		&	1.25$^*$	\\
H\dthree\jon~	+	HDO		&\ra&	H\dtwo DO\jon~	+	H\dtwo	&	1.25$^*$	&	H\dthree\jon~		+	O		&\ra&	H\dtwo O\jon~	+	H		&	1.40		\\
HCO\jon~		+	H\dtwo O	&\ra&	H\dthree O\jon~	+	CO		&	1.50		&	HCO\jon~			+	HDO		&\ra&	H\dtwo DO\jon~	+	CO		&	1.50$^*$	\\
DCO\jon~		+	H\dtwo O	&\ra&	H\dtwo DO\jon~	+	CO		&	1.50$^*$	&	DCO\jon~			+	HDO		&\ra&	HD\dtwo O\jon~	+	CO		&	1.50$^*$	\\
HCO\jon~		+	D\dtwo O	&\ra&	HD\dtwo O\jon~	+	CO		&	1.50$^*$	&	&	\\[1.0pt]
\hline
\end{longtable}
\end{center}

\section{Dominant formation and destruction pathways for deuterated species }
\label{sec:Dominant}
In this Appendix, we list the dominant pathways identified for isotopologues and isomers of the selected species H\dthree\jon, HCO\jon, HCN, H\dtwo O as well as species {involved in the formation of these essential species}. The intention is to help with future comparisons with our model.
\begin{table}
\centering
\tablewidth{350pt}
\caption{Most essential formation and destruction pathways for H\dthree\jon, H\dtwo D\jon, HD\dtwo\jon~and D\dthree\jon. \label{tab:essReactH3+}}
\begin{tabular}{lclcccccl}
\multicolumn{3}{l}{Reaction} 	& {$\alpha$}	& {$\beta$}	& {$\gamma$}	&	{Accuracy}	&	Ref		&	Estimated by\\
\hline
H\dthree\jon~		+	HD			&	\ra 			&	H\dtwo D\jon~		+	H\dtwo			&	1.70E-09	&~0.00E+00&	0.00E+00		&	\textit{factor 2}	&	(1)	&	T	\\
H\dthree\jon~		+	D			&	\ra 			&	H\dtwo D\jon~		+	H 				&	1.00E-09	&~0.00E+00&	0.00E+00		&	\textit{factor 2}	&	(2)	&	T	\\
H\dtwo D\jon~		+	HD			&	\ra 			&	HD\dtwo \jon~		+	H\dtwo			&	8.10E-10	&~0.00E+00&	0.00E+00		&	$\pm 15\%$	&	(3)	&	M	\\
H\dtwo D\jon~		+	D			&	\ra 			&	HD\dtwo \jon~		+	H 				&	2.00E-09	&~0.00E+00&	0.00E+00		&	\textit{factor 2}	&	(4)	&	T	\\
H\dthree\jon~		+	D\dtwo		&	\ra 			&	HD\dtwo\jon~		+	H\dtwo			&	1.10E-09	&~0.00E+00&	0.00E+00		&	$\pm 15\%$ 	&	(3)	&	M	\\
HD\dtwo\jon~		+	HD			&	\ra 			&	D\dthree\jon~		+	H\dtwo			&	6.40E-10	&~0.00E+00&	0.00E+00		&	$\pm 15\%$ 	&	(3)	&	M	\\
HD\dtwo\jon~		+	D			&	\ra 			&	D\dthree\jon~		+	H				&	2.00E-09	&~0.00E+00&	0.00E+00		&	\textit{factor 2}	&	(4)	&	T	\\
H\dtwo D\jon~		+	D\dtwo 		&	\ra 			&	D\dthree\jon~		+	H\dtwo			&	7.00E-10	&~0.00E+00&	0.00E+00		&	$\pm 15\%$ 	&	(3)	&	M	\\
\hline
H\dtwo D\jon~		+	CO			&	\ra 			&	DCO\jon~			+	H\dtwo 			&	5.37E-10	&~0.00E+00&	0.00E+00		&	$< 25\%$		&	(5)	&	C	\\
H\dtwo D\jon~		+	CO			&	\ra 			&	HCO\jon~			+	HD 				&	1.07E-09	&~0.00E+00&	0.00E+00		&	$< 25\%$		&	(5)	&	C	\\
HD\dtwo \jon~		+	CO			&	\ra 			&	DCO\jon~			+	HD				&	1.07E-09	&~0.00E+00&	0.00E+00		&	$< 25\%$ 		&	(5)	&	C	\\
HD\dtwo \jon~		+	CO			&	\ra 			&	HCO\jon~			+	D\dtwo			&	5.37E-10	&~0.00E+00&	0.00E+00		&	$< 25\%$		&	(5)	&	C	\\
D\dthree\jon~		+	CO			&	\ra 			&	DCO\jon~			+	D\dtwo			&	1.61E-09	&~0.00E+00&	0.00E+00		&	$< 25\%$		&	(5)	&	C	\\
D\dthree\jon~		+	e\ijon		&	\ra 			&	D				+	D 	+ D			&	2.16E-08	&-0.50E+00 &	0.00E+00		&	$< 25\%$		&	(5)	&	C	\\
\hline
\multicolumn{9}{l}{T, Theoretical/Calculated}	\\
\multicolumn{9}{l}{C, Cloned} 				\\
\multicolumn{9}{l}{Ion-neutral reactions (Kooji formula): $k = \alpha (T/300)^\beta $e$^{-\gamma / T}$}	\\
\multicolumn{9}{l}{Pathways extracted at 1 Myr from four points with T = 10 or 100 K and densities = 10$^4$ and 10$^8$ cm$^{-3}$ in the ÒPrimordialÓ model}	\\
\multicolumn{9}{l}{(1) \citet{1992A&A...255..453S}; (2) \citet{1985ApJ...294L..63A}; (3) \citet{doi:10.1021/j100198a030}; (4) \citet{2004A&A...418.1035W}}\\
\multicolumn{9}{l}{(5) OSU \url{www.physics.ohio-state.edu/\textasciitilde eric/} }
\end{tabular}
\end{table}

\begin{table}
\centering
\tablewidth{350pt}
\caption{Most essential formation and destruction pathways for HCO\jon, HOC\jon, DCO\jon~and DOC\jon. \label{tab:essReactDCO+}}
\begin{tabular}{lclcccccl}
\multicolumn{3}{l}{Reaction} 	& {$\alpha$}	& {$\beta$}	& {$\gamma$}	&	{Accuracy}	&	Ref		&	Estimated by\\
\hline
H\dthree\jon~		+	CO			&	\ra	&	HCO\jon~		+	H\dtwo			&	1.61E-09	&~0.00E+00&	0.00E+00		&	$< 25\%$ 		&	(1)	&	M	\\
H\dthree\jon~		+	CO			&	\ra	&	HOC\jon~		+	H\dtwo			&	9.44E-11	&~0.00E+00&	0.00E+00		&	$< 25\%$ 		&	(1)	&	M	\\
HOC\jon~			+	H\dtwo		&	\ra	&	HCO\jon~		+	H\dtwo			&	1.00E-11	&~0.00E+00&	0.00E+00		&	factor 2		&	(1)	&	?	\\
CH\dthree\jon~		+	O			&	\ra	&	HCO\jon/HOC\jon~+	H\dtwo			&	2.05E-10	&~0.00E+00&	0.00E+00		&	factor 2 		&	(1)	&	?	\\
\hline
HCO\jon~			+	SO			&	\ra	&	HSO\jon~		+	CO				&	3.30E-09	&-0.50E+00&	0.00E+00		&	factor 2		&	(1)	&	?	\\
HCO\jon~			+	e\ijon		&	\ra	&	CO			+	H				&	2.80E-07	&-0.69E+00&	0.00E+00		&	$< 25\%$ 		&	(1)	&	?	\\
HCO\jon~			+	C			&	\ra	&	CH\jon~		+	CO				&	1.10E-09	&~0.00E+00&	0.00E+00		&	factor 2 		&	(1)	&	?	\\
\hline
H\dtwo D\jon~		+	CO			&	\ra	&	DCO\jon~		+	H\dtwo			&	5.37E-10	&~0.00E+00&	0.00E+00		&	$< 25\%$ 		&	(1)	&	M	\\	
HCO\jon~			+	D 			&	\ra	&	DCO\jon~		+	H 				&	1.00E-09	&~0.00E+00&	0.00E+00		&	\textit{factor 2}	&	(2)	&	M	\\	
D\dthree\jon~		+	CO			&	\ra	&	DCO\jon~		+	D\dtwo			&	1.61E-09	&~0.00E+00&	0.00E+00		&	$< 25\%$ 		&	(1)	&	C	\\
DOC\jon~			+	H\dtwo		&	\ra	&	DCO\jon~		+	H\dtwo			&	3.33E-12	&~0.00E+00&	0.00E+00		&	$< 50\%$ 		&	(1)	&	C	\\
CH\dtwo D\jon~	+	O			&	\ra	&	DCO\jon~		+	H\dtwo			&	6.83E-11	&~0.00E+00&	0.00E+00		&	$< 25\%$ 		&	(1)	&	C	\\
\hline
DCO\jon~			+	e\ijon		&	\ra	&	CO			+	D 				&	2.40E-07	&-0.69E+00&	0.00E+00		&	$< 25\%$ 		&	(1)	&	C	\\
DCO\jon~			+	SO			&	\ra	&	DSO\jon~		+	CO 				&	3.30E-09	&-0.50E+00&	0.00E+00		&	factor 2 		&	(1)	&	C	\\
DCO\jon~			+	H 			&	\ra	&	HCO\jon~		+	D 				&	2.20E-09	&~0.00E+00&	7.69E+02		&	\textit{factor 2}	&	(2) 	&	M	\\	
DCO\jon~			+	C 			&	\ra	&	CD\jon~		+	CO 				&	1.10E-09	&~0.00E+00&	0.00E+00		&	factor 2 		&	(1)	&	C	\\
DCO\jon~			+	HCN			&	\ra	&	HDCN\jon~	+	CO 				&	7.30E-09	&-0.50E+00&	0.00E+00		&	factor 2 		&	(1)	&	C	\\
DCO\jon~			+	HNC			&	\ra	&	HDCN\jon~	+	CO 				&	6.63E-09	&-0.50E+00&	0.00E+00		&	factor 2		&	(1)	&	C	\\
\hline
\multicolumn{9}{l}{T, Theoretical/Calculated}	\\
\multicolumn{9}{l}{M, Laboratory measurements}	\\
\multicolumn{9}{l}{C, Cloned} 	\\
\multicolumn{9}{l}{?, No listing for estimation method, most likely theoretical} 	\\
\multicolumn{9}{l}{Ion-neutral reactions (Kooji formula): $k = \alpha (T/300)^\beta $e$^{-\gamma / T}$}	\\
\multicolumn{9}{l}{Pathways extracted at 1 Myr from four points with T = 10 or 100 K and densities = 10$^4$ and 10$^8$ cm$^{-3}$ in the ÒPrimordialÓ model}	\\
\multicolumn{9}{l}{(1) OSU \url{www.physics.ohio-state.edu/\textasciitilde eric/}; (2) KIDA: \url{http://kida.obs.u-bordeaux1.fr/}}
\end{tabular}
\end{table}

\begin{table}
\centering
\tablewidth{350pt}
\caption{Most essential formation and destruction pathways for HCN, HNC, DCN and DNC. \label{tab:essReactDCN}}
\begin{tabular}{lclcccccl}
\multicolumn{3}{l}{Reaction} 	& {$\alpha$}	& {$\beta$}	& {$\gamma$}	&	{Accuracy}	&	Ref		&	Estimated by\\
\hline
H\dtwo CN\jon~	+	e\ijon		&	\ra	&	HCN			+	H					&	9.62E-08	&-0.65E+00&	0.00E+00		&	factor 2		&	(1)	&	M	\\
H\dtwo CN\jon~	+	e\ijon		&	\ra	&	HNC			+	H					&	1.85E-07	&-0.65E+00&	0.00E+00		&	$< 25\%$ 		&	(2)	&	?	\\
H\dtwo NC\jon~	+	e\ijon		&	\ra	&	HNC			+	H					&	1.80E-07	&-0.50E+00&	0.00E+00		&	$< 25\%$ 		&	(2)	&	?	\\
CH\dtwo~			+	N			&	\ra	&	HCN	/HNC	+	H					&	3.95E-11	&-0.17E+00&	0.00E+00		&	$< 25\%$ 		&	(2)	&	T	\\
HNC				+	H\jon		&	\ra	&	HCN			+	H\jon				&	2.78E-08	&-0.50E+00&	0.00E+00		&	factor 2 		&	(2)	&	?	\\
\hline
NH (ice)			+	C (ice)		&	\ra	&	HCN (ice)								&	1.00		&	0.00		&	0.00		&	-- 			&	(3)	&	 	\\
CN (ice)			+	H (ice)		&	\ra	&	HCN (ice)								&	1.00		&	0.00		&	0.00		&	-- 			&	(3)	&	 	\\
ND (ice)			+	C (ice)		&	\ra	&	DCN (ice)								&	1.00		&	0.00		&	0.00		&	-- 			&	(3)	&	C	\\
CN (ice)			+	D (ice)		&	\ra	&	DCN (ice)								&	1.00		&	0.00		&	0.00		&	-- 			&	(3)	&	C	\\
\hline
HDCN\jon~		+	e\ijon		&	\ra	&	DCN			+	H 					&	2.33E-07	&-0.50E+00&	0.00E+00		&	\textit{factor 2}	&	(4)	&	T	\\	
D\dtwo CN\jon~	+	e\ijon		&	\ra	&	DCN/DNC	+	D 					&	1.85E-07	&-0.65E+00&	0.00E+00		&	\textit{factor 2}	&	(2)	&	C	\\
HDNC\jon~		+	e\ijon		&	\ra	&	DNC			+	H 					&	1.66E-07	&-0.50E+00&	0.00E+00		&	\textit{factor 2}	&	(4)	&	T	\\	
HCN				+	D 			&	\ra	&	DCN			+	H 					&	1.00E-10	&~0.50E+00&	5.00E+02		&	\textit{factor 2}	&	(5)	&	T	\\	
CHD 			+	N 			&	\ra	&	DCN			+	H 					&	1.98E-11	&~1.67E-01&	0.00E+00		&	$< 50\%$		&	(5)	&	T	\\	
\hline
DCN				+	H\dthree\jon	&	\ra	&	HDCN\jon~	+	H\dtwo				&	8.50E-09	&-0.50E+00&	0.00E+00		&	$< 25\%$ 		&	(2)	&	C	\\
DCN				+	H\dthree\jon 	&	\ra	&	H\dtwo CN\jon~+ 	HD 					&	8.50E-09	&-0.50E+00&	0.00E+00		&	$< 25\%$ 		&	(2)	&	C	\\
DCN				+	HCO\jon		&	\ra	&	HDCN\jon	~	+	CO					&	7.30E-09	&-0.50E+00&	0.00E+00		&	factor 2 		&	(2)	&	C	\\
DCN				+	H\jon		&	\ra	&	DCN\jon~		+	H 					&	1.39E-08	&-0.50E+00&	0.00E+00		&	factor 2 		&	(2)	&	C	\\
DCN				+	H 			&	\ra	&	HCN			+	D 					&	1.00E-10	&~0.50E+00&	5.00E+02		&	\textit{factor 2}	&	(5)	&	C	\\
DCN				+	H\dthree O\jon	&	\ra	&	HDCN\jon~	+	H\dtwo O				&	4.10E-09	&-0.50E+00&	0.00E+00		&	$< 25\%$ 		&	(2)	&	C	\\
DCN				+	H\dthree O\jon	&	\ra	&	H\dtwo CN\jon~ +	HDO					&	4.10E-09	&-0.50E+00&	0.00E+00		&	$< 25\%$ 		&	(2)	&	C	\\
DCN				+	D\dthree O\jon	&	\ra	&	D\dtwo CN\jon~ +	D\dtwo				&	8.20E-09	&-0.50E+00&	0.00E+00		&	$< 25\%$ 		&	(2)	&	C	\\
\hline
\multicolumn{9}{l}{T, Theoretical/Calculated}	\\
\multicolumn{9}{l}{C, Cloned} 			\\
\multicolumn{9}{l}{?, No listing for estimation method, most likely theoretical} 	\\
\multicolumn{9}{l}{Ion-neutral reactions (Kooji formula): $k = \alpha (T/300)^\beta $e$^{-\gamma / T}$}	\\
\multicolumn{9}{l}{Pathways extracted at 1 Myr from four points with T = 10 or 100 K and densities = 10$^4$ and 10$^8$ cm$^{-3}$ in the ÒPrimordialÓ model}	\\
\multicolumn{9}{l}{(1) \citet{2010SSRv..156...13W}; (2) OSU \url{www.physics.ohio-state.edu/\textasciitilde eric/}; (3) \citet{2006A&A...457..927G}; (4) \citet{2005A&A...438..585R}}\\
\multicolumn{9}{l}{(5) \citet{1992A&A...256..595S} }
\end{tabular}
\end{table}

\begin{table}
\centering
\tablewidth{350pt}
\caption{Most essential formation and destruction pathways for H\dtwo O, HDO and D\dtwo O. \label{tab:essReactWater}}
\begin{tabular}{lclcccccl}
\multicolumn{3}{l}{Reaction} 	& {$\alpha$}	& {$\beta$}	& {$\gamma$}	&	{Accuracy}	&	Ref		&	Estimated by\\
\hline
H\dthree O\jon~	+	e\ijon		&	\ra	&	H\dtwo O			+	H				&	1.10E-07& -0.50E+00	&	0.00E+00		&	$< 25\%$ &	(1)	&	M\\ %
H\dthree O\jon~	+	HCN			&	\ra	&	H\dtwo O			+	H\dtwo CN\jon		&	8.20E-09& -0.50E+00	&	0.00E+00		&	$< 25\%$ &	(2)	&	?\\ %
H\dthree O\jon~	+	HNC			&	\ra	&	H\dtwo O			+	H\dtwo NC\jon		&	7.42E-09& -0.50E+00	&	0.00E+00		&	factor 2 	&	(2)	&	? \\ %
\hline
OH (ice)			+	H (ice)		&	\ra	&	H\dtwo O	(ice)							&	1.00		&	0.00		&	0.00			&	- - 		&	(3)	&		\\
OD (ice)			+	H (ice)		&	\ra	&	HDO	(ice)								&	1.00		&	0.00		&	0.00			&	- - 		&	(3)	&	C	\\
OH (ice)			+	D (ice)		&	\ra	&	HDO	(ice)								&	1.00		&	0.00		&	0.00			&	- - 		&	(3)	&	C	\\
\hline
H\dtwo O			+	H\jon		&	\ra	&	H\dtwo O\jon~		+	H				&	7.30E-09 &-0.50E+00	&	0.00E+00		&	$< 25\%$	&	(2)	&	?	\\ %
H\dtwo O			+	H\dthree\jon	&	\ra	&	H\dthree O\jon~	+	H\dtwo			&	4.50E-09 &-0.50E+00	&	0.00E+00		&	$< 25\%$	&	(2)	&	?	\\ %
\hline
H$_2$DO\jon~		+ 	e\ijon		&	\ra	&	HDO				+	H 				&	7.33E-08 &-0.50E+00	&	0.00E+00		&	$< 25\%$ &	(2)	&	C	\\ 
HD$_2$O\jon~		+ 	e\ijon		&	\ra	&	HDO				+	D 				&	7.33E-08&-0.50E+00	&	0.00E+00		&	$< 25\%$ &	(2)	&	C	\\
H$_2$DO\jon~		+ 	HCN			&	\ra	&	HDO				+	H$_2$CN$^+$ 	&	4.10E-09&-0.50E+00	&	0.00E+00		&	$< 25\%$ &	(2)	&	C	\\
H$_2$DO\jon~		+ 	HNC			&	\ra	&	HDO				+	H$_2$CN$^+$ 	&	3.71E-09&-0.50E+00	&	0.00E+00		&	factor 2 	&	(2)	&	C	\\
H\dtwo~ 			+	OD			&	\ra	&	HDO				+	H 	 			&	5.60E-13&~0.00E+00	&	1.04E+03		&	factor 2 	&	(2)	&	C	\\
\hline
HDO				+	HCO\jon		&	\ra	&	H\dtwo DO\jon~	+	CO					&	2.10E-09&	-0.50E+00&	0.00E+00		&	$< 50\%$	&	(2)	&	C	\\ 
HDO				+	H\dthree\jon	&	\ra	&	H\dtwo DO\jon~	+	H\dtwo				&	2.70E-09&	-0.50E+00&	0.00E+00		&	$< 25\%$ &	(2)	&	C	\\
HDO				+	H\dthree\jon	&	\ra	&	H\dthree O\jon~	+	HD					&	1.80E-09&	-0.50E+00&	0.00E+00		&	$< 25\%$ &	(2)	&	C	\\
HDO				+	C\jon		&	\ra	&	HOC\jon/DOC\jon~	+	H					&	9.00E-10&	-0.50E+00&	0.00E+00		&	$< 25\%$ &	(2)	&	C	\\
HDO				+	C\jon		&	\ra	&	HCO\jon/DCO\jon~	+	H					&	4.45E-10&	-0.50E+00&	0.00E+00		&	$< 25\%$ &	(2)	&	C	\\
\hline
HD$_2$O$^+$		+ 	e\ijon		&	\ra	&	D$_2$O		+	H 					&	3.67E-08&	-0.50E+00&	0.00E+00		&	$< 25\%$ &	(2)	&	C	\\
D\dthree O\jon~	+	e\ijon		&	\ra	&	D$_2$O		+	D 					&	1.10E-07&	-0.50E+00&	0.00E+00		&	$< 25\%$ &	(2)	&	C	\\ 
HD\dtwo O\jon~	+	HCN			&	\ra	&	D$_2$O		+	H\dtwo CN\jon			&	1.37E-09&	-0.50E+00&	0.00E+00		&	$< 25\%$ &	(2)	&	C	\\
HD\dtwo O\jon~	+	HNC			&	\ra	&	D$_2$O		+	H\dtwo CN\jon			&	1.24E-09&	-0.50E+00&	0.00E+00		&	factor 2 	&	(2)	&	C	\\
HD 				+	OD			&	\ra	&	D\dtwo O		+	H 	 				&	2.80E-13&	~1.00E+00&	1.04E+03		&	factor 2 	&	(2)	&	C	\\
\hline
OD (ice)			+	D (ice)		&	\ra	&	D$_2$O	(ice)							&	1.00		&	0.00		&	0.00			&	- - 		&	(3)	&	C	\\
\hline
D\dtwo O			+	C\jon		&	\ra	&	DOC\jon~		+	D 					&	1.80E-09&	-0.50E+00&	0.00E+00		&	$< 25\%$ &	(2)	&	C	\\
D\dtwo O			+	C\jon		&	\ra	&	DCO\jon~		+	D 					&	8.90E-10&	-0.50E+00&	0.00E+00		&	$< 25\%$ &	(2)	&	C	\\
D\dtwo O			+	H\dthree\jon	&	\ra	&	HD\dtwo O\jon~+	H\dtwo				&	1.35E-09&	-0.50E+00&	0.00E+00		&	$< 25\%$ &	(2)	&	C	\\
D\dtwo O			+	H\dthree\jon	&	\ra	&	H\dtwo DO\jon~+	HD					&	2.70E-09&	-0.50E+00&	0.00E+00		&	$< 25\%$ &	(2)	&	C	\\
D\dtwo O			+	H\dthree\jon	&	\ra	&	H\dthree O\jon~+ 	D\dtwo				&	4.50E-10&	-0.50E+00&	0.00E+00		&	$< 25\%$ &	(2)	&	C	\\
D\dtwo O			+	H\jon		&	\ra	&	D\dtwo O\jon~	+ 	H 					&	2.43E-09&	-0.50E+00&	0.00E+00		&	$< 25\%$ &	(2)	&	C	\\
D\dtwo O 			+	H\jon		&	\ra	&	HDO\jon~		+ 	D					&	4.87E-09&	-0.50E+00&	0.00E+00		&	$< 25\%$ &	(2)	&	C	\\
\hline
\multicolumn{9}{l}{M, Laboratory measurement}	\\
\multicolumn{9}{l}{C, Cloned}		\\
\multicolumn{9}{l}{?, No listing for estimation method, but most likely theoretical} 	\\
\multicolumn{9}{l}{Ion-neutral reactions (Kooji formula): $k = \alpha (T/300)^\beta $e$^{-\gamma / T}$}	\\
\multicolumn{9}{l}{Pathways extracted at 1 Myr from four points with T = 10 or 100 K and densities = 10$^4$ and 10$^8$ cm$^{-3}$ in the ÒPrimordialÓ model}	\\
\multicolumn{9}{l}{(1) \citet{2000ApJ...543..764J}; (2) OSU \url{www.physics.ohio-state.edu/\textasciitilde eric/}; (3) \citet{2006A&A...457..927G}}
\end{tabular}
\end{table}

\begin{table}
\centering
\tablewidth{0.95\textwidth}
\caption{Important reactions for species involved in the main pathways of the assorted deuterated species; H\dtwo D\jon, HD\dtwo\jon, D\dthree\jon, HDO, D\dtwo O, DCO\jon~and DCN. \label{tab:essReactives}} 
\begin{tabular}{clclcccccl}
Formation of species:	&	\multicolumn{3}{l}{Reaction} 		& $\alpha$ & {$\beta$}	& {$\gamma$}	&	{Accuracy}	&	Ref		&	Estimated by\\
\hline
H\dtwo\jon	&	H\dtwo~			+	h$\nu_{\rm CR}$&	\ra 			&	H\dtwo\jon~		+	e\ijon			&	9.30E-01&	~0.00E+00&	0.00E+00		&	factor 2 		&	(1)	&	T	\\
CH\dthree	\jon	&	H\dthree\jon~		+	C 	 		&	\ra 			&	CH\jon~			+	H\dtwo 			&	2.00E-09&	~0.00E+00&	0.00E+00		&	factor 2 		&	(1)	&	?	\\
			&	CH\jon~			+	H\dtwo 		&	\ra 			&	CH\dtwo\jon~		+	H 				&	1.20E-09&	~0.00E+00&	0.00E+00		&	$< 25\%$		&	(1)	&	?	\\
			&	CH\dtwo\jon~		+	H\dtwo 		&	\ra 			&	CH\dthree\jon~		+	H 				&	1.20E-09&	~0.00E+00&	0.00E+00		&	$< 25\%$		&	(1)	&	?	\\
isotopologues	&	CH\dthree\jon~		+	HD 			&	\ra 			&	CH\dtwo D\jon~	+	H\dtwo 			&	1.30E-09&	~0.00E+00&	0.00E+00		&	\textit{factor 2}	&	(2)	&	M	\\
			&	CH\dtwo D\jon~	+	HD			&	\ra 			&	CHD\dtwo\jon~		+	H\dtwo			&	6.60E-10&	~0.00E+00&	0.00E+00		&	\textit{factor 2}	&	(2)	&	M	\\
			&	CHD\dtwo \jon~	+	HD			&	\ra 			&	CD\dthree\jon~		+	H\dtwo			&	6.60E-10&	~0.00E+00&	0.00E+00		&	\textit{factor 2}	&	(2)	&	M	\\
			&	D\dthree\jon~		+	CH\dtwo		&	\ra 			&	CD\dthree\jon~		+	H\dtwo			&	5.19E-11&	-0.50E+00&	0.00E+00		&	factor 2 		&	(1)	&	C	\\
\hline
H\dtwo CN\jon	&	H\dthree\jon~		+	HCN			&	\ra 			&	H\dtwo CN\jon~	+	H\dtwo			&	1.70E-08&	-0.50E+00&	0.00E+00		&	$< 25\%$ 		&	(1)	&	?	\\ 
HDCN\jon		&	DCO\jon~			+	HCN			&	\ra 			&	HDCN\jon	~		+	CO				&	7.30E-09&	-0.50E+00&	0.00E+00		&	factor 2 		&	(1)	&	C	\\
			&	DCO\jon~			+	HNC			&	\ra 			&	HDCN\jon	~		+	CO				&	6.63E-09&	-0.50E+00&	0.00E+00		&	factor 2 		&	(1)	&	C	\\
HDNC\jon		&	CH\dtwo D\jon~	+	N			&	\ra 			&	HDNC\jon	~		+	H				&	4.47E-11&	~0.00E+00&	0.00E+00		&	$< 25\%$		&	(1)	&	C	\\
D\dtwo CN\jon	&	HD\dtwo\jon~		+ 	HCN			&	\ra 			&	D\dtwo CN\jon~	+	H\dtwo 			&	2.83E-09&	-0.50E+00&	0.00E+00		&	factor 2 		&	(1)	&	C	\\
			&	HD\dtwo\jon~		+ 	HNC			&	\ra 			&	D\dtwo CN\jon~	+	H\dtwo 			&	2.50E-09&	-0.50E+00&	0.00E+00		&	factor 2 		&	(1)	&	C	\\
CH\dtwo		&	C				+	H\dtwo		&	\ra 			&	CH\dtwo								&	1.00E-17&	~0.00E+00&	0.00E+00		&	factor 10 		&	(1)	&	?	\\
			&	CH 				+	H\dtwo 		& 	\ra 			&	CH\dtwo~			+	H 				&	1.20E-09&	~0.00E+00&	0.00E+00		&	factor 10 		&	(1)	&	?	\\
CHD			&	HD				+	C			&	\ra 			&	CHD									&	1.20E-17&	~0.00E+00&	0.00E+00		&	factor 10 		&	(1)	&	C	\\
\hline
OH			&	H\dtwo DO\jon~	+ 	e\ijon		&	\ra 			&	OD				+ 	H	+ 	H		&	2.60E-10&	-5.00E-01	&	~0.00E+00	&	factor 2 		&	(1)	&	C	\\
OD\jon		&	H\dtwo D\jon~		+	O			&	\ra 			&	OD\jon~			+	H\dtwo			&	2.67E-10&	~0.00E+00&	~0.00E+00	&	\textit{factor 2}	&	(3)	&	C	\\ 
			&	H\jon~			+	OD			&	\ra 			&	OD\jon~			+	H				&	8.00E-09&	-0.50E+00&	~0.00E+00	&	factor 2 		&	(1)	&	C	\\ 
HDO\jon		&	OD\jon~			+	H\dtwo		&	\ra 			&	HDO\jon~			+	H				&	7.33E-10&	~0.00E+00&	~0.00E+00	&	$< 25\%$ 		&	(1)	&	C	\\
			&	H\dthree\jon~		+	OD			&	\ra 			&	HDO\jon~			+	H\dtwo			&	4.75E-09&	-0.50E+00&	~0.00E+00	&	factor 2 		&	(1)	&	C	\\ 
			&	H\dtwo D\jon~		+	O			&	\ra 			&	HDO\jon~			+	H 				&	2.28E-10&	-1.56E-01	&	-1.41E+00	&	\textit{factor 2}	&	(3)	&	C	\\
H\dtwo DO\jon	&	HDO\jon~			+	H\dtwo		&	\ra 			&	H\dtwo DO\jon~	+	H 				&	4.57E-10&	~0.00E+00&	~0.00E+00	&	$< 25\%$		&	(1)	&	C	\\
			&	HD\dtwo\jon~		+	O			&	\ra 			&	D\dtwo O\jon~		+	H 				&	1.14E-10&	-1.56E+00&	-1.41E+00	&	\textit{factor 2}	&	(3)	&	C	\\
			&	D\dtwo O\jon~		+	H\dtwo		&	\ra 			&	HD\dtwo O\jon~	+	H 				&	3.05E-10&	~0.00E+00&	~0.00E+00	&	$< 25\%$		&	(1)	&	C	\\
			\hline
\multicolumn{10}{l}{M, Laboratory measurement}	\\
\multicolumn{10}{l}{T, Theoretical/Calculated}		\\
\multicolumn{10}{l}{(L) Literature}			\\
\multicolumn{10}{l}{C, Cloned}		\\
\multicolumn{9}{l}{?, No listing for estimation method, but most likely theoretical} 	\\
\multicolumn{10}{l}{Ion-neutral reactions (Kooji formula): $k = \alpha (T/300)^\beta $e$^{-\gamma / T}$}	\\
\multicolumn{10}{l}{Pathways extracted at 1 Myr from four points with T = 10 or 100 K and densities = 10$^4$ and 10$^8$ cm$^{-3}$ in the ÒPrimordialÓ model}	\\
\multicolumn{10}{l}{(1) OSU \url{www.physics.ohio-state.edu/\textasciitilde eric/}; (2) \citet{10.1063/1.444002, 1982ApJ...263..123S}; (3) \citet{2010SSRv..156...13W}}
\end{tabular}
\end{table}

\clearpage
\section{Listing of observed D/H ratios in dense interstellar environments}
{We have collected and generated an updated listing of observed deuterated species and D/H ratios. }
\begin{center}
\begin{longtable}{c|c|c|c|c|c}
\caption[Listings of observated interstellar deuterated species.]{Listings of observated interstellar deuterated species.} \label{tab:obs}\\
\tablewidth{0.9\textwidth}
Species 				& 	\multicolumn{2}{c}{Sources} 				& Spacial scale		&Refs 		& 	Model	\\
					&	Class -I				&	Class O/I		& Beam size ['']		&	 		&			\\[0.05pt]
			\hline
\endfirsthead
Species 				& 	\multicolumn{2}{c}{Sources} 				& Spacial scale		&Refs 		& 	Model	\\
					&	Class -I				&	Class O/I		& Beam size ['']		&	 		&			\\[0.05pt]
			\hline
\endhead
D 		/ H 			&	$<$ 4 $\times 10^{-4}$		&					&	1.8$^\circ$		&	1			& $10^{-3}-10^{-2}$	\\
- - 					&	$\leq$ 2.7 $\times 10^{-5}$	&					&	 				&	2			& - - 				\\
- - 					&	2.2 $\times$10$^{-5}$		&					&	 14$^\circ$		&	3			& - - 				\\
- - 					&	$\leq$ 0.14				&					&	 				&	4			&				\\
HD		/ H\dtwo		&	0.74 - 8.6 $\times$10$^{-6}$	&					& 30x30 				&	5			& $<10^{-4} $		\\
- - 					&	1.32 - 14.83 $\times$10$^{-6}$	&					&	 				&	6			& - - 				\\
ND		/ NH			&							&	0.3 - 1.0			&	41(ND), 22(NH)	&	7			& 10$^{-4}-10^{0}$	\\
OD		/ OH			&	$\leq$ 2.5 $\times$10$^{-3}$	& 					&	1.8$^\circ$ 		&	8			& 10$^{-2}-10^{0}$	\\[0.05pt]
\hline\\[0.1pt]
C$_2$D	/ C$_2$H 		&	0.01						&					&		33			& 	9			& $10^{-3}-10^{-2}$	\\[0.1pt]
- - 					&	0.01 						&	0.18				&		20			& 	10			&	- - 			\\[0.01pt]
D$_2$O	/ H$_2$O		&							&	5 $\times$10$^{-5}$	&	1.5x1.5			&	11			& $<10^{-5}-10^{-3}$\\
DCN		/ HCN		&	0.008-0.015				&					&	 				&	12			&	$10^{-3}-10^{-1}$	\\[0.1pt]
- - 					&	0.012 - 0.11				&					&	 				&	13			&	- - 				\\[0.1pt]
- - 					&	0.013					&					&	20				&	10			&	- - 				\\[0.1pt]
- - 					&	0.023					&					&					&	14			&	- - 				\\[0.1pt] 
- - 					&							&	0.005 - 0.02		&					&	15			&	$<10^{-5} - 10^{-3}$	\\[0.1pt] 
DCO$^+$	/ HCO$^+$	&	0.007 - 0.081				&					&	 				&	13			&	10$^{-2}-10^{0}$ 	\\[0.1pt]
- - 					&	0.02 - 0.18				&					&	20				&	16			&	 - - 				\\[0.1pt]
- - 					&	0.006 - 0.04				&					&	25-57			&	17			&	- - 				\\[0.1pt]
- - 					&	0.0086 - 0.015				&					&	20				& 	10			&	 - - 				\\[0.1pt] 
- - 					&	0.031 - 0.059				&					&	30-96			&	18			&	- - 				\\[0.1pt]
- - 					&	$\leq $ 0.03				&					&	 				&	4			&	- - 				\\[-0.1pt]
- - 					&							&	0.04 - 1 $\times$10$^{-2}$	&	13\tablenotemark{b}&	19	&	$10^{-4}-10^{-2}$	\\[-0.1pt] 
DNC 		/ HNC	&	$<$ 0.014				&					&	10 - 30			&	20			&	10$^{-3}-10^{-1}$	\\[0.1pt] 
- - 					&	0.02 - 0.09			&					&	$\sim 20$			&	21			& 	- - 				\\[0.1pt]
- - 					&	0.008 - 0.122			&					&	17-20\tablenotemark{a}&	22			&	- - 				\\[0.1pt]
- - 					&	0.015 - 0.03			&					&	20				&	10			&	- - 				\\[0.1pt]
H\dtwo D\jon/ H\dthree\jon &						&	$<$ 3 $\times$10$^{-3}$	&	13\tablenotemark{b}	&	19		& 10$^{-4}-10^{-2}$		\\[0.1pt]
HDO / H\dtwo O		&						&	0.014 - 0.058		&					&	23			& 10$^{-3}-10^{-2}$ 		\\[0.1pt] 
- - 					&						&	$\geq$ 0.01		&	10-30			&	24			&	- - 				\\[0.1pt] 
- - 					&						&	$\geq$ 6 $\times$10$^{-4}$&	3.1 $\times$2.5	&	25			&	- - 				\\[0.1pt]
- - 					&						&	2.94 $\times$10$^{-5}$	&	1.5x1.5		&	11			&	- - 				\\[0.1pt]
- - 					&						&	0.03				&	10-33			&	26			&	- - 				\\[0.1pt]
- - 					&						&	0.6 - 5 $\times$10$^{-4}$	&				&	27			&	- - 				\\[0.1pt] 
- - 					&						&	$\gtrsim$ 0.01		&	20				&	28			&	- - 				\\[0.1pt]
HDO / H\dtwo O (solid)	&						&	0.005 - 0.02		&	 				&	29			&	$10^{-3}-10^{-1}$	\\[0.1pt]
- - 					&						&	8 $\times$$10^{-4}-10^{-2}$&	 			&	30			&	- - 				\\[0.1pt]
HDS		/ H$_2$S		&	0.05 - 0.15			&					&	20				& 	10			& 10$^{-2}-10^{-1}$		\\[0.1pt]
N$_2$D$^+$ 	/ N$_2$H$^+$&	0.016	- 0.051	&					&	$\sim$30			&	31			& 10$^{-2}-10^{0}$		\\[0.1pt]
- - 					&	0.01 - 0.16			&					&	10-20			&	32			&	- - 				\\[0.1pt]
- - 					&	0.03 - 0.04 			&					&	26.4 \tablenotemark{a}&	 33			&	- - 				\\[0.1pt]
- - 					&	0.11					&					&	44				&	34			&	- - 				\\[0.1pt]
- - 					&	0.08 - 0.14			&					&	18 \tablenotemark{a}&	35			&	- - 				\\[0.1pt]
- - 					&	0.02 - 0.52 			&					&	11 \tablenotemark{a}&	36			&	- - 				\\[0.1pt]
- - 					&	$\sim 0.1$				&					&	9-26\tablenotemark{a} &	37			&	- - 				\\[0.1pt]
- - 					&	0.08 - 0.35			&					&	20				&	16			& 	- - 				\\[0.1pt]
- - 					&						&	0.005 - 0.014		&	11-26 \tablenotemark{a}	& 	38		&	$10^{-4}-10^{-2}$	\\[0.1pt]
- - 					&						&	0.033 - 0.271		&	11-16			&	39			&	- - 				\\[0.1pt]
- - 					&						&	0.003 - 0.027		&	9-26\tablenotemark{a}	&	40		&	- - 				\\[0.1pt]
%
%
\hline\\[0.1pt]
D$_2$CO	/ H$_2$CO	&	0.11 - 0.19			&					&	27 \tablenotemark{a}&	41			& 10$^{-3}-10^{-2}$		\\[0.1pt]
- - 					&	0.40					&					&					&	42			&	- - 				\\[0.1pt]
- - 					&	0.01 - 0.1				&					&	17				&	43			&	- - 				\\[0.1pt]
- - 					&	$\leq$ 0.07			&					&	22				&	44			& 	- - 				\\[0.1pt]
- - 					&	2.05 - 3.3 $\times$10$^{-2}$	&				&	20-60			&	45			&	- - 				\\[0.1pt]
- - 					&						&	0.01 - 0.04		&	20-60			&	46			&	- - 				\\[0.1pt]
- - 					&						&	0.022 - 1.04		&	10-30			&	47			&	$10^{-5}-10^{-3}$	\\[0.1pt]
- - 					&						&	0.02 - 0.4			&					&	48			&	- - 				\\[0.1pt]
- - 					&						&	0.03 - 0.16		&					&	49			&	- - 				\\[0.1pt] 
D$_2$CS	/ H$_2$CS	&	0.333				&					&					&	50			& 10$^{-3}-10^{-1}$		\\[0.1pt]
HDCO	/ H$_2$CO	&	0.092 - 0.122			&					&	27 \tablenotemark{a} &	41			& 10$^{-3}-10^{-1}$		\\[0.1pt]
- - 					&	0.015				&					&	20				& 	10			& 	- - 				\\[0.1pt]
- - 					&						&	0.07 - 4.3			&	10-30			&	47			& $10^{-4}-10^{-2}$		\\[0.1pt] 
- - 					&						&	0.09 - 2.6			&	20-60			&	45			&	- - 				\\[0.1pt]
HDCS	/ H$_2$CS	&	0.333				&					&	 				&	42			& $10^{-2}-10^{-1}$		\\[0.1pt]
- - 					&	0.015 - 0.025			&					&	$>$60			&	51			& 10$^{-2}-10^{-1}$		\\[0.1pt]
ND$_3$	/ NH$_3$		&	1.1 - 65 $\times$10$^{-4}$&					&	22 \tablenotemark{b}&	52			& $<10^{-5}-10^{-3}$	\\[0.1pt]
- - 					&	8 $\times$10$^{-4}$		&					&	25				&	53			&	- - 				\\[0.1pt]
- - 					&						&	9.35 $\times$10$^{-4}$	&	25			&	54			& $<10^{-5}-10^{-4}$	\\[0.1pt]
NH$_2$D / NH$_3$		&	0.1 - 0.8				&	 				&	7				&	55			& $10^{-3}-10^{-1}$		\\[0.1pt]
- - 					&	0.07 - 0.42			&					&	22 \tablenotemark{b} &	52			& - - 					\\[0.1pt]
- - 					&	0.02 - 0.1				&					&	20				&	16			&	- - 				\\[0.1pt]
- - 					&	0.025 - 0.18			&					&	18				&	56			&	- - 				\\[0.1pt]
- - 					&	$<$ 0.02				&					&	20				& 	10			&	- - 				\\[0.1pt]
- - 					&	 					&	$<$ 0.1			&	7				&	55			& 	$10^{-4}-10^{-2}$	\\[0.1pt]
- - 					&						&	0.071			&	$\sim$ 20			&	57			&	 - - 				\\[0.1pt]
- - 					&						&	0.06 - 0.1			& 	22				&	44			&	- - 				\\[0.1pt]
- - 					&						&	0.04 - 0.33		&	37				&	58			&	 - - 				\\[0.1pt]
- - 					&						&	0.06- 0.1			&	20				&	10			&	- - 				\\[0.1pt]
- - 					&						&	0.06				&	20				&	28			&	- - 				\\[0.1pt]
- - 					&						&	2.6 - 17.3 $\times$10$^{-2}$&	20-60		&	45			& 	- - 				\\[0.1pt]
NHD$_2$	/ NH$_3$		&	0.03 - 0.27 			&					&	22 \tablenotemark{b} &	52			&	$10^{-4}-10^{-1}$	\\[0.1pt]
- - 					&	$\sim$5 $\times$10$^{-3}$&					&	22				&	44			&	- - 				\\[0.1pt]
- - 					&						&	0.02 - 0.4			&					&	59			&	$<10^{-5}-10^{-3}$	\\[0.1pt]
%
%
\hline\\[0.1pt]
C$_3$HD	/ C$_3$H$_2$	&	0.05 - 0.15			&					&	1.7' \tablenotemark{b} &	60			& $10^{-3}-10^{-1}$		\\[0.1pt]
C$_4$D 	/ C$_4$H		&	4.30 $\times$10$^{-3}$	&					&	1.7' \tablenotemark{b} &	61			& $10^{-3}-10^{-2}$		\\[0.1pt]
- - 					&						&	0.0043-0.023		&	17-28			&	62			& - - 					\\[0.1pt]
C$_4$HD	/ C$_4$H$_2$	&						&	0.013 - 0.051		&	17-28			&	62			& $10^{-3}-10^{-1}$		\\[0.1pt]
CD$_3$OH/ CH$_3$OH	&						&	0.001 - 0.028		&	15				&	63			& $<10^{-5}-10^{-2}$ 	\\[0.1pt]
C3H$_3$D/ C3H$_4$	&	0.04 - 0.18			&					&	40-60			&	64			&	 $10^{-3}-10^{-1}$	\\[0.1pt]
- - 					&	0.1 - 0.22				&					&	40				&	65			&	- - 				\\[0.1pt]
- - 					&	0.05 - 0.06			&					&	27\tablenotemark{a}&	66			&	 - - 				\\[0.1pt]
CH$_2$DCN/ CH$_3$CN&						&	$\gtrsim$ 0.01		&	27 \tablenotemark{a} &	67			& $10^{-2}-10^{-1}$		\\[0.1pt]
CH$_2$DOH / CH$_3$OH&						&	0.01$\pm$ 0.73 	&	17-28			&	62			& $10^{-3}-10^{-2}$		\\[0.1pt]
- - 					&						&	0.05 - 0.30		&	 				&	68			&	- -				\\[0.1pt]
- - 					&						&	0.05	- 0.95		&	10-30			&	47			& - - 					\\[0.1pt]
- - 					&						&	0.60 - 1.2			&	11-30			&	69			& 	- - 				\\[0.1pt]
CH$_3$OD/ CH$_3$OH	&		 				&	0.008 - 0.076		&	10-30			&	44			&	 $10^{-4}-10^{-2}$	\\[0.1pt]
- - 					&						&	0.02 - 0.06		&	11-30			&	69			&	 - - 				\\[0.1pt]
- - 					&						&	$\leq$ 0.1			&	20				& 	10			&	 - - 				\\[0.1pt]
CHD$_2$OH/ CH$_3$OH&						&	0.06 - 0.0.51		&	10-30			&	47			& $10^{-3}-10^{0}$		\\[0.1pt]
- - 					&						&	0.1 - 0.3			&	11-30			&	69			& - - 					\\[0.1pt]

DC$_3$N	/HC$_3$N	&	0.010 - 0.020			&					&					&	70			&	$10^{-3}-10^{-2}$	\\[0.1pt]
DC$_3$N	/HC$_3$N	&						&	0.02 - 0.045		&	17-28			&	62			&	$10^{-4}-10^{-3}$	\\[0.1pt]
DC$_5$N/HC$_5$N	&	0.006 - 0.016			&					&					&	71			&	$10^{-3}-10^{-2}$	\\[0.1pt]
DC$_5$N/HC$_5$N	&						&	0.018 - 0.036		&	17-28			&	62			&	$10^{-5}-10^{-3}$	\\[0.1pt]
- - 					&	0.013 - 0.019			&					&					&	72			&	- - 				\\[0.1pt]
DCOOCH$_3$ / HCOOCH$_3$ &	0.02 - 0.06		&					&	 				&	73			& $10^{-3}-10^{-1}$		\\[0.1pt]
- - 					&	$\sim $ 0.15			&					&	9-33 				&	71			&	- - 				\\[0.1pt]
\hline\\[0.1pt]
\multicolumn{6}{l}{{Young stellar object IR classification -I: Prestellar objects, O/I: Embedded/revealed protostellar sources}}\\[0.1pt]
\multicolumn{6}{l}{(a) Half-power beam width (HPBW)}	\\[0.1pt]
\multicolumn{6}{l}{(b) Full-width half-maximum (FWHM)}	\\[0.1pt]
\multicolumn{6}{l}{(c) LWRS}						\\[2.0pt]
\multicolumn{6}{l}{(1) \citet{1973ApJ...180L...1C}; (2) \citet{2006ASPC..348...47H}; (3) \citet{2007AJ....133.1625R}; (4) \citet{1993ApJS...89..271H}; (5) \citet{2005A&A...430..967L}}\\[0.1pt]
\multicolumn{6}{l}{(6) \citet{2008ApJ...688.1124S}; (7) \citet{2010A&A...521L..42B}; (8) \citet{1974ApJ...188...33A}; (9) \citet{1986HiA.....7..513F}; (10) \citet{1995ApJ...447..760V}}\\[0.1pt]
\multicolumn{6}{l}{(11) \citet{2007ApJ...659L.137B}; (12) \citet{2007msl..confE..12P}; (13) \citet{2002ApJ...569..322L}; (14) \citet{1987IAUS..120..311W}; (15) \citep{1991ApJ...369..169M}}\\[0.1pt]
\multicolumn{6}{l}{(16) \citet{2000A&A...356.1039T}; (17) \citet{1999A&A...347..983A}; (18) \citet{1995ApJ...448..207B}; (19) \citet{1999ApJ...521L..67S}; (20) \citet{2009A&A...498..771G}}\\[0.1pt]
\multicolumn{6}{l}{(21) \citet{2003ApJ...594..859H}; (22) \citet{2001ApJ...547..814H}; (23) \citet{2012A&A...539A.132C}; (24) \citet{2011A&A...527A..19L}; (25) \citet{2010ApJ...725L.172J}}\\[0.1pt]
\multicolumn{6}{l}{(26) \citet{2005A&A...431..547P}; (27) \citet{1996A&A...314..281G}; (28) \citet{1990A&A...228..447J}; (29) \citet{2003A&A...410..897P}; (30) \citet{1999A&A...347L..19T}}\\[0.1pt]
\multicolumn{6}{l}{(31) \citet{2010ApJ...713L..50C}; (32) \citet{2010ApJ...718..666F}; (33) \citet{2009A&A...500..845M}; (34) \citet{2008A&A...477L..45F}; (35) \citet{2006A&A...454L..51B}}\\[0.1pt]
\multicolumn{6}{l}{(36) \citet{2005ApJ...619..379C}; (37) \citet{2004A&A...420..957C}; (38) \citet{2010A&A...518A..52A}; (39) \citet{2009A&A...493...89E}; (40) \citet{2006A&A...460..709F}}\\[0.1pt]
\multicolumn{6}{l}{(41) \citet{2011A&A...527A..39B}; (42) \citet{2005ApJ...620..308M}; (43) \citet{2003ApJ...585L..55B}; (44) \citet{2000A&A...354L..63R}; (45) \citet{1990ApJ...362L..29T}}\\[0.1pt]
\multicolumn{6}{l}{(46) \citet{2007A&A...471..849R}; (47) \citet{2006A&A...453..949P}; (48) \citet{2002P&SS...50.1205L}; (49) \citet{2001A&A...372..998C}; (50) \citet{2005ApJ...620..308M}}\\[0.1pt] 
\multicolumn{6}{l}{(51) \citet{1997ApJ...491L..63M}; (52) \citet{2005A&A...438..585R}; (53) \citet{2002ApJ...571L..55L}; (54) \citet{2002A&A...388L..53V}; (55) \citet{2010A&A...517L...6B}}\\[0.1pt]
\multicolumn{6}{l}{(56) \citet{2000ApJ...535..227S}; (57) \citet{2001ApJ...554..933S}; (58) \citet{2003A&A...403L..25H}; (59) \citet{2003cdsf.conf..351L}; (60) \citet{1988ApJ...326..924B}}\\[0.1pt]
\multicolumn{6}{l}{(61) \citet{2009ApJ...702.1025S}; (62) \citet{1987IAUS..120..199S}; (63) \citet{2004A&A...416..159P}; (64) \citet{2005ApJ...627L.117M}; (65) \citet{2002A&A...381..560M}}\\[0.1pt] 
\multicolumn{6}{l}{(66) \citet{1992A&A...253L..29G}; (67) \citet{1992A&A...259L..35G}; (68) \citet{2007msl..confE...9B}; (69) \citet{2002A&A...393L..49P}; (70) \citet{1987IAUS..120..199S}}\\[0.1pt]
\multicolumn{6}{l}{(71) \citet{1981ApJ...251L..33M}; (72) \citet{1981ApJ...251L..37S}; (73) \citet{2010ApJ...714.1120M}}
\end{longtable}
\end{center}

\end{document}